\documentclass[letterpaper,11pt]{article}
\input{preamble.sty}

\begin{document}

\doparttoc 
\faketableofcontents 

\algrenewcommand\algorithmicrequire{\textbf{Input:}}
\algrenewcommand\algorithmicensure{\textbf{Output:}}
	
\title{Impact of Rankings and Personalized Recommendations in Marketplaces}
	\ifx\blind\undefined
	\fi

\author{
     Omar Besbes\thanks{ Columbia University, Graduate School of Business. Email: {\tt ob2105@columbia.edu}} \and Yash Kanoria\thanks{Columbia University, Graduate School of Business. Email: {\tt ykanoria@columbia.edu}} \and Akshit Kumar\thanks{University of Toronto Scarborough, Department of Management. Email: {\tt akshit.kumar@utoronto.ca}}
}


\date{}

\maketitle

\begin{abstract}
    Decision-making often requires individuals to navigate large sets of options with incomplete information and imperfectly formed preferences. Information provisioning tools, such as public rankings and personalized recommendations, have become central to guiding these choices, yet their welfare implications across different market environments remain poorly understood. This paper studies a stylized large-market model to quantify the aggregate value of these tools under \textit{uncapacitated} supply (items have infinite capacity) and \textit{capacitated} supply (items have unit capacity). Agent utility is a weighted combination of a \textit{common} term (population-level quality) and an \textit{idiosyncratic} term (individual-specific fit). Agents observe noisy signals of these components: public rankings improve estimates of common quality, while personalized recommendations additionally reveal idiosyncratic fit. In uncapacitated settings, both tools improve welfare through better selection. Their relative value is governed by preference heterogeneity: rankings are highly effective when preferences are relatively homogeneous, whereas the marginal value of personalization grows as heterogeneity increases. In stark contrast, under capacity constraints, a strict conservation logic applies: learning common quality primarily reshuffles assignments without increasing aggregate average agent welfare. However, revealing idiosyncratic fit generates substantial welfare gains by mitigating congestion on overdemanded items and unlocking match-specific value. We demonstrate that these core insights are robust across model extensions including alternative extreme-value tail distributions and heterogeneous supply-side priorities. Finally, introducing strategic supply-side pricing reveals a stark distributional dichotomy: while information expands total surplus, the newly generated value is ultimately extracted by the short side of the market.
   
    
    \medskip
    \noindent
    \textbf{History}: An earlier version of this paper appeared as an extended abstract in the Proceedings of the 26th ACM Conference on Economics and Computation, EC’25
\end{abstract}

\maketitle

\newpage

\setstretch{1}
\section{Introduction}
Every day, individuals navigate choices ranging from the mundane (e.g., selecting a movie) to life-altering (e.g., choosing a college). Surveys and empirical research have demonstrated, and personal experience affirms, that these decisions are frequently made under \textit{imperfect information}, where preferences are rarely fully formed, and outcomes often lead to regret. A 2017 survey found that most U.S.\ adults would alter their educational choices if given the chance~\citep{gallup2024educationregret}, underscoring a broader phenomenon: individuals often have poorly formed preferences when making choices.

To help guide decisions, \textit{public rankings} have emerged as a ubiquitous tool across many domains. In entertainment, websites like IMDb and Billboard rank movies and songs by popularity, respectively. In e-commerce, in their early days, platforms like Amazon used bestseller lists and star ratings to highlight popular products. In education, organizations such as the US News \& World Report\footnote{\href{https://www.usnews.com/best-colleges}{https://www.usnews.com/best-colleges}}
and the National Institutional Ranking Framework (NIRF)\footnote{\href{https://www.nirfindia.org/}{https://www.nirfindia.org/}}
in India, rank universities and colleges based on various performance metrics. These rankings aggregate information and simplify decision-making, serving as a common signal of quality in settings where individuals struggle to evaluate options on their own. Rankings are particularly influential in settings where users lack direct experience---for example, prospective college students who have never attended the institutions they are choosing between. However, while rankings provide a useful \textit{population-level signal}, they fail to capture \textit{idiosyncratic preferences}, i.e., the way in which an individual user's value for an item differs from the average user's value for that item.

The rise of \textit{personalized recommendation tools} in several domains has provided an alternative approach, tailoring choices to individual preferences rather than presenting a single, universal ranking. In entertainment, platforms like Netflix, Spotify, and YouTube curate recommendations based on user behavior, revealing content that aligns with individual tastes. In e-commerce, platforms like Amazon and Etsy now personalize search results, increasing sales by showing products aligned with past browsing and purchase behavior. Similarly, in higher education, platforms like Naviance and Scoir use historical data and student profiles to recommend universities that align with a student's specific strengths and interests. These systems go beyond general quality signals to help individuals find the best choices for them, rather than just the highest-ranked options.
With the rise of generative AI, it is becoming increasingly common to see chatbots that personalize recommendations. In e-commerce, Amazon's Rufus, an AI-powered shopping assistant, engages in real-time conversations with users, answering open-ended queries and generating personalized product suggestions based on the platform's catalog and user behavior. Similarly, in college admissions, platforms like CollegeVine and Kollegio are leveraging AI to provide personalized counseling to students, offering tailored recommendations for universities. These tools bring a new dimension to recommendation systems by enabling dynamic, dialogue-based interactions rather than static ranked lists.

Despite the prevalence of personalized recommendation tools in e-commerce and entertainment, there are still many domains where this technology has not achieved widespread penetration. Policymakers and platform designers, for example, may ask how much value these tools truly deliver and in which settings they \textit{add the most value} relative to public rankings. A concrete case study arises in college admissions: the Indian government invests substantially in NIRF to evaluate universities \edit{(public ranking)}. \edit{Would it be appropriate to invest also in (or divert some funding to)} developing and deploying personalized recommendation services that help match students to programs that reflect their individual preferences and needs? 
These questions extend beyond college admissions. The design of Netflix’s recommendation system, Airbnb’s ranking algorithms, or similar platforms also involves balancing broad quality signals against more tailored, individual-specific information. As investment in AI-powered personalized recommendation systems grows---particularly in light of advances in generative AI~\citep{chen2024large, zhang2024personalization}---it becomes vital to understand when and how these tools deliver more (or less) value than traditional public rankings.

A key distinction across these domains is whether supply is unconstrained (as in digital content) or scarce (as in admissions seats, rentals, and limited inventory marketplaces). Under scarcity, \edit{a generic public signal of quality}
that helps users identify ``better'' options can also concentrate demand on a small set of top ranked choices, which can be viewed as a type of \edit{``congestion''. 
As a result, better common rankings may substantially change \textit{who gets what}, while having muted effects on \textit{average} outcomes when capacity constraints bind.} 
Our goal is to provide a clean model that isolates this mechanism and clarifies when rankings versus personalization should matter for welfare. These considerations lead to the following key policy and design question:

\begin{quote}
\textit{What are the implications of different information provisioning tools, such as public rankings and personalized recommendations, in environments with and without supply-side constraints?}
\end{quote}

It is unsurprising that personalized recommendations can outperform public rankings in many settings; that is not what we aim to study. Instead, we aim to quantify both the incremental value that personalization offers over public rankings alone, and the benefit public rankings provide relative to having no information provisioning tool at all. In doing so, we identify the core drivers of these gains and examine how they play out across different market environments. To isolate these effects cleanly, we study a stylized, parsimonious model that captures the essential features of information provisioning tools and supply constraints, while abstracting away from the specifics of how these tools are operationalized. We outline the model’s basic ingredients here, with the formal description deferred to Section~\ref{sec:base-model}.

\paragraph{\underline{Role of information provisioning tools.}}
An agent’s utility for an item is modeled as a weighted combination of two \textit{independently drawn} terms: (i) a \textit{common} term that depends only on the item and reflects population-level quality, and (ii) an \textit{idiosyncratic} term that depends on the agent--item pair and captures individual-specific fit. We weight the common and idiosyncratic terms using parameters $1-\rho$ and $\rho$, where $\rho\in[0,1]$ reflects the level of preference heterogeneity. Lower values of $\rho$ correspond to more homogeneous preferences, while higher values correspond to more heterogeneous preferences. We model the role of information provisioning tools as revealing different components of utility: public rankings inform agents about the common term, while personalized recommendations \edit{additionally inform agents about the idiosyncratic term}. 
In the base model, we allow these signals to be noisy, which lets us quantify the value of \emph{imperfect} rankings and \emph{imperfect} personalization and connect statistical accuracy to welfare.

\paragraph{\underline{Uncapacitated and capacitated supply settings.}}
We assume there are $n$ agents and $n$ items and each agent consumes a single item. We study two environments:
\begin{enumerate}[label=(\roman*)]
    \item \textit{Uncapacitated supply setting.} Each item has infinite capacity; any number of agents can choose the same item. This is motivated by content recommendation platforms like Netflix, Spotify, and YouTube, \edit{and e-commerce platforms like Amazon and Alibaba in situations where inventory is plentiful.}
    \item \textit{Capacitated supply setting.} Each item has unit capacity; once chosen it becomes unavailable. This is motivated by centralized admissions and online marketplaces like Airbnb where supply is scarce.
\end{enumerate}
In the capacitated setting, agents choose sequentially in a priority order, capturing the mechanics of admissions and providing a tractable benchmark for allocation under scarcity.

\paragraph{\underline{Information regimes and welfare.}}
We measure goodness of outcomes by \textit{agent welfare}, defined as the average realized utility across agents under a given information regime. To isolate the marginal impact of information, we compare welfare across regimes. We focus on three benchmark \edit{information} regimes (see Figure~\ref{fig:information-regimes}): (i) {\sf No Information} ($\emptyset$), where agents observe neither component; (ii) {\sf Only Quality Information} ($q$), corresponding to public rankings; and (iii) {\sf Full Information} ($u$), corresponding to personalized recommendations that reveal fit. In addition, in the base model, we consider intermediate noisy regimes to quantify the value of partially informative tools. 

\begin{figure}[h]
    \centering
    \begin{tikzpicture}[scale = 0.9]
        \draw[ultra thick, ->] (0,0) -- (18,0);
        \filldraw[fill=blue] (1,0) circle (1mm);
        \filldraw[fill=blue] (8,0) circle (1mm);
        \filldraw[fill=blue] (15,0) circle (1mm);
        \node at (1,-0.5) {{\sf No Information}};
        \node at (8,-0.55) {{\sf Only Quality Information}};
        \node at (15,-0.5) {{\sf Full Information}};
        \node[align=center] at (1, 0.75) {Agents make \\ random decisions}; 
        \node[align = center] at (8, 0.75) {Agents make decisions based \\ solely on the common term};
        \node[align = center] at (15,0.75) {Agents make decisions based on \\ both common and idiosyncratic terms  };
        \draw[dashed, ultra thick, ->, bend left = -20] (1,-0.8)  to (7.75,-0.8);
        \node[align = center] at (4.5, -1.8) {Public Rankings};
        \draw[dashed, ultra thick, ->, bend left = -20] (8.25,-0.8)  to (15,-0.8);
        \node[align = center] at (11.5,-1.8) {Personalized Recommendations};
    \end{tikzpicture}
    \caption{Different information regimes studied in this work} 
    \label{fig:information-regimes}
\end{figure}

\subsection{Main Contributions}

In this work, we develop a stylized and parsimonious model to examine the interplay between information provisioning tools and market environments. The model is intentionally simple: it does not aim to capture every institutional detail, but it cleanly formalizes the key economic forces---scarcity, congestion, preference heterogeneity, and information---and yields sharp analytical predictions. We summarize our main contributions below.

\begin{itemize}
    \item \textbf{Fundamental role of capacity constraints, preference heterogeneity, and noisy information.} 
    In our base model, we analytically quantify the value of imperfect (noisy) public rankings and personalized recommendations, showing that their incremental value is governed jointly by (i) whether supply is capacitated, (ii) the degree of preference heterogeneity $\rho$, and (iii) the signal noise (see Theorem \ref{thm:noisy-estimate-agent-welfare} and Corollaries \ref{cor:value-noisy-public-ranking}--\ref{cor:value-noisy-personalized-recommendations}).
    \begin{itemize}
        \item \textit{Uncapacitated setting.} Without capacity constraints, information improves welfare strictly through better selection. When preferences are relatively homogeneous (small $\rho$), public rankings capture the first-order welfare gains because the common component drives utility. As heterogeneity increases, the incremental value of personalization grows, scaling at the extreme-value rate $\sqrt{2 \ln n}$ (Corollaries \ref{cor:value-noisy-public-ranking}\ref{cor:noisy-uncapacitated-value-public-ranking} and \ref{cor:value-noisy-personalized-recommendations}\ref{cor:noisy-uncapacitated-value-personalized-recos}).
        \item \textit{Capacitated setting.} Under unit capacities, we identify a strict conservation logic for the common component: revealing common quality (rankings) only reshuffles who receives which items but cannot increase \emph{aggregate} average agent welfare (Corollary \ref{cor:value-noisy-public-ranking}\ref{cor:noisy-capacitated-value-public-ranking}). By contrast, (noisy) personalization generates strictly positive aggregate welfare gains by improving idiosyncratic fit under scarcity, and these gains are directly proportional to $\rho$ (Corollary \ref{cor:value-noisy-personalized-recommendations}\ref{cor:noisy-capacitated-value-personalized-recos}).
    \end{itemize}
    We complement this asymptotic analysis with numerical simulations, demonstrating that our large-market approximations are quantitatively highly accurate even at moderate market sizes (Figure~\ref{fig:value-of-info-2x2-normalized}).

    \item \textbf{Formalizing congestion under scarcity.}
    A recurring equilibrium concern discussed in the empirical literature on informational interventions is that providing non-personalized quality information---such as report cards or lists of ``better'' options---may simply encourage more students to apply to already-oversubscribed top-performing schools \citep{corcoran2018leveling, cohodes2025informational}. Because the number of seats is physically limited, such interventions can change application behavior but may ultimately just reshuffle access without necessarily improving overall equity or average outcomes. Our model rigorously formalizes this intuitive argument: when capacity constraints bind, information about \emph{common} quality primarily changes \emph{sorting} (who gets which high-quality slots) rather than expanding the average common value received in the population. In contrast, our results highlight how information about \emph{idiosyncratic fit} can actively reduce correlated demand for overdemanded options. By steering agents toward their personal best fits, personalized recommendations mitigate congestion and improve aggregate match quality, allowing items to unlock additional value beyond their inherent quality via their ``beauty in the eye of the matched agent.'' In light of discussions surrounding the costs of informational interventions (e.g., \citet{corcoran2018leveling} note their non-personalized intervention was labor-intensive and costly to scale), our theoretical framework sharpens the policy trade-off: resources spent on impersonal rankings may mainly reshuffle access under scarcity, whereas personalized tools can fundamentally raise average match quality by improving fit.

    \item \textbf{Model Extensions.}
    We extend the welfare comparisons beyond the Gaussian base model along several dimensions that are salient in real markets:
    \begin{itemize}
        \item \textit{Alternative tail behaviors (Section \ref{sec:welfare-gains-tail-distributions}).} Motivated by heavy-tailed or bounded payoffs in many applications, we study Pareto, exponential, Gaussian-like, and bounded-support utility terms. We show that the welfare gains depend primarily on the \emph{tail} of the primitive distributions (Theorems \ref{thm:uncapacitated-ranking-value}--\ref{thm:capacitated-personalized-recos-value}). For instance, in uncapacitated settings, whether personalization matters at leading order is governed by which component---quality or fit---drives the upper tail of realized utility, yielding sharp phase transitions in $\rho$ for exponential distributions. Under capacity constraints, rankings continue to yield zero aggregate welfare gains, while personalization provides substantial tail-dependent gains.
        \item \textit{Supply-side priorities (Section \ref{sec:supply-side-priority}).} Motivated by neighborhood priorities in school choice, we introduce own-group priorities and analyze stable outcomes under a common selection rule (Group-Prioritized Serial Dictatorship). The core message persists: rankings deliver zero aggregate average value (Theorem \ref{thm:value-public-rankings-supply-side-priority}), while personalization unlocks substantial aggregate gains by improving fit within the effectively available, priority-respecting choice sets (Theorem \ref{thm:value-personalized-recos-supply-side-priority}).
        \item \textit{Supply-side pricing (Section \ref{subsec:supply-side-pricing}).} Introducing prices allows us to study how information changes not only match quality but also the division of surplus (Table \ref{tab:equilibrium-price-agent-welfare-seller-revenue}). We find that who captures the newly generated welfare depends entirely on which side of the market is short. In markets with excess demand or balanced supply, sellers are on the short side; thus, better information allows them to raise prices and extract essentially all the incremental surplus, leaving agent welfare unchanged (Theorems \ref{thm:value-public-rankings-pricing}(a) and \ref{thm:value-personalization-pricing}(a)). Conversely, under excess supply, demand is on the short side; competition disciplines prices, allowing agents to fully capture the substantial welfare gains unlocked by personalized recommendations (Theorems \ref{thm:value-public-rankings-pricing}(b) and \ref{thm:value-personalization-pricing}(b)).
    \end{itemize}  
\end{itemize}

\subsection{Related Literature}
\label{subsec:related-literature}
This work is motivated by and contributes to several strands of literature on recommendation systems, matching with incomplete preferences, and information design in matching markets.

\paragraph{Recommendation Systems and Decision Support Tools.} Classical recommendation  systems have focused on identifying and suggesting items that best fit each user’s preferences \citep{schafer1999recommender, aggarwal2016recommender}. These recommendation and decision support tools have shown great promise in terms of improving the decisions made by users \citep{haubl2000consumer}. The emphasis has been on developing accurate user bahavior model and develop methods to improve the relevance of personalized recommendations \citep{adomavicius2008personalization, naumov2019deep}. These methods have mostly been designed to operate in uncapacitated environments, such as content streaming platforms, and as such do not generally take into account matching or capacity constraints. More recently, motivated by e-commerce and labor platforms, there has been a growing interest in designing recommendation systems which take into these matching constraints \citep{su2022optimizing, aouad2023online, shi2024optimal}. The focus of these papers has been methodological while in this work, we aim to understand the nuanced interplay between supply side capacity constraints and the value that personalized recommendations can generate.

\paragraph{Incomplete Preferences and Informational Interventions in Matching Markets.} Most of the literature on one-sided and two-sided matching typically assumes that agents possess well-defined preferences \citep{gale1962college, roth1992two, abdulkadirouglu2003school}. However, these assumptions are often unrealistic in practical scenarios, as recent empirical studies have shown that the absence of well-formed preferences can lead to inefficient matching outcomes \citep{campbell2022matching, dillon2017determinants}. Motivated primarily by applications in school and college admissions, recent work has shifted focus to issues of preference discovery and incomplete information \citep{immorlica2020information, chen2021information, grenet2022preference}. This body of research typically examines situations where agents strategically acquire additional information to refine their preferences and make informed choices. Empirical and field studies have evaluated the impact of providing additional information to students in the context of high school admissions \citep{corcoran2018leveling, cohodes2025informational} and college admissions \citep{hoxby2015high, larroucau2024college}. 
Our contribution to this line of research takes a modeling approach, aiming to isolate the impact of different information provisioning tools on the average user welfare.

\paragraph{Information Design in Matching Markets.} There is an emerging literature on information design and signaling in matching markets. This literature typically studies a central platform which chooses to strategically provide information to agents in order to influence their behavior and the resulting matches \citep{elliott2022matching, bimpikis2024information, papanastasiou2018crowdsourcing, papanastasiou2024designing}. In terms of setting, the most closely related paper is \cite{dasgupta2024designing}. They study the problem where a central planner strategically provisions information to agents with incomplete information in order to optimize for social welfare. A key distinction of this line of work to our work is that we do not study strategic information provisioning rather focus on the impact of different information provision tools.

\paragraph{Algorithmic monoculture and homogenization.} \cite{kleinberg2021algorithmic} first formalized algorithmic monoculture in hiring markets, where many firms assess applicants with the {\it same} ranking algorithm. By contrast, algorithmic polyculture describes settings in which firms rely on independent algorithms. Extending these ideas to large two-sided matching markets, \cite{peng2024monoculture} analyze stable matching outcomes under monoculture and polyculture and show that, when evaluation noise is well behaved, monoculture can reduce firm utility by resulting in less preferred applicants being hired vis-a-vis polyculture.
\cite{peng2024wisdom} studies the impact of noise in the evaluation of candidates to the resulting stable matching outcome in the context of polyculture -- in particular, they consider the impact of the tail of the noise distribution on the outcome. Our capacitated model maps directly onto these notions. In the {\sf Only Quality Information} regime, all agents share a single impersonal ranking—mirroring monoculture, whereas in the {\sf Full Information} regime each agent has an individualized ranking, paralleling polyculture.
Similar to \cite{peng2024monoculture}, we find that the agent welfare is lower in the {\sf Only Quality Information} regime (monoculture) compared to the {\sf Full Information} regime (polyculture). A recent work by \cite{baek2025hiring} incorporates strategic behavior into the monoculture setting and characterize the resulting Nash equilibria. While our work does not study strategic behavior by agents, 
qualitatively speaking, our insights resonate with those of \cite{baek2025hiring}: $(i)$ competition for the top items \citep[candidates in][]{baek2025hiring} leads to inefficiencies due to congestion or matching constraints and $(ii)$ in the capacitated setting, most of the value lies in matching agents (firms) to items (candidates) that they idiosyncratically value highly.

\subsection{Organization of the paper}
The remainder of the paper is organized as follows. Section~\ref{sec:base-model} introduces our base model, specifying the utility structure, information regimes, and the sequential choice process under both uncapacitated and capacitated supply settings. Section~\ref{sec:base-model-results} presents our main theoretical results, providing asymptotic characterizations of agent welfare (Theorem~\ref{thm:noisy-estimate-agent-welfare}) and deriving the value of public rankings (Corollary~\ref{cor:value-noisy-public-ranking}) and personalized recommendations (Corollary~\ref{cor:value-noisy-personalized-recommendations}) under Gaussian primitives. Section~\ref{sec:model-extensions} examines the robustness and scope of these findings through three extensions: (i) alternative tail distributions—Pareto, exponential, and bounded—that span the standard extreme-value regimes; (ii) supply-side priorities, where items exhibit own-group preferences over agents as in neighborhood-based school choice; and (iii) supply-side pricing, where sellers strategically set prices and we characterize equilibrium welfare and revenue across information regimes. Section~\ref{sec:conclusion} concludes with a discussion of implications and directions for future research. All proofs are deferred to the appendix.

\section{Base Model}
\label{sec:base-model}
In this section, we provide the base model. 
We consider a setting with a set of $n$ agents (denoted $\mathcal{X}$) and a set of $n$ items (denoted $\mathcal{Y}$). Each agent $x \in \mathcal{X}$ has a priority score $s_x$ drawn i.i.d from the distribution $F_{s}$.

\paragraph{Agent Utility.}  The utility of matching an agent $x$ to an item $y$ is given as
\begin{align}
    \label{eq:utility-definition}
    u_{xy} &= (1-\rho)\,q_y + \rho\,\varphi_{xy}, \ \ \forall x \in \mathcal{X}, y \in \mathcal{Y},
\end{align}
where $q_y$ is the quality of item $y$ (we refer to this as the common term), $\varphi_{xy}$ denotes the idiosyncratic match value for the agent-item $(x,y)$ pair (we refer to this as the idiosyncratic term) and the parameter $\rho \in [0,1]$ determines the relative weight of the idiosyncratic term, capturing the level of heterogeneity. Smaller $\rho$ implies more homogeneous preferences (the common term dominates) while larger $\rho$ implies more heterogeneous preferences (idiosyncratic term dominates). We assume that the common terms $(q_y)_{y \in \mathcal{Y}}$ and the idiosyncratic terms $(\varphi_{xy})_{x \in \mathcal{X}, y \in \mathcal{Y}}$ are independent of each other. 
We assume that the common term for the items is drawn i.i.d from distribution $F_q$ and the idiosyncratic term for the agent-item pair is drawn i.i.d from the distribution $F_\varphi$. For the base model, we assume that the common term distribution $F_q$ is the normal distribution with mean $\mu_q$ and variance $\sigma_q^2$ denoted as $\mathcal{N}(\mu_q,\sigma_q^2)$. Similarly, idiosyncratic term distribution $F_\varphi$ is the normal distribution with mean $\mu_\varphi$ and variance $\sigma_\varphi^2$ denoted as $\mathcal{N}(\mu_\varphi,\sigma_\varphi^2)$.

\paragraph{Agent Information.} The agents observe a noisy version of the common terms $(\tilde{q}_y)_{y \in \mathcal{Y}}$ and the idiosyncratic terms $(\tilde{\varphi}_{xy})_{x \in \mathcal{X}, y \in \mathcal{Y}}$. The noisy term of the common and idiosyncratic terms is drawn from a normal distribution with mean zero and variances $\sigma_{\varepsilon}^2$ and $\sigma_{\delta}^2$, respectively. \edit{Note that $\sigma_{\varepsilon}$ and $\sigma_{\delta}$ are constants that do not depend on item quality and agent priority.} In particular, we have that
\begin{align}
    \label{eq:noisy-common-term}
    \tilde{q}_{y} &= q_{y} + \varepsilon_{y}, \qquad \quad \varepsilon_{y} \sim \mathcal{N}(0, \sigma_{ \varepsilon}^2), \\
    \label{eq:noisy-idiosyncratic-term}
    \tilde{\varphi}_{xy} &= \varphi_{xy} + \delta_{xy}, \qquad \delta_{xy} \sim \mathcal{N}(0,\sigma_{\delta}^2)
\end{align}

Given the signal (noisy version) of the common $(\tilde{q}_y)_{y \in \mathcal{Y}}$ and the idiosyncratic $(\tilde{\varphi}_{xy})_{x \in \mathcal{X}, y \in \mathcal{Y}}$ terms, the posterior distribution of the common term is given as $q_y \mid \tilde{q}_y \sim \mathcal{N}(\hat{q}_y, \hat{\sigma}_q^2)$ and for the idiosyncratic term is given as $\varphi_{xy} \mid \tilde{\varphi}_{xy} \sim \mathcal{N}(\hat{\varphi}_{xy}, \hat{\sigma}_{\varphi}^2)$, where 
\begin{align}
    \label{eq:common-term-posterior-mean-variance}
    \hat{q}_y &= \frac{\sigma_q^2}{\sigma_q^2 + \sigma_{ \varepsilon}^2} \cdot \tilde{q}_y + \frac{\sigma_{ \varepsilon}^2}{\sigma_{q}^2 + \sigma_{ \varepsilon}^2} \cdot \mu_q, \quad\quad\quad \hat{\sigma}_q^2 = \frac{\sigma_q^2 \cdot \sigma_{\varepsilon}^2}{\sigma_q^2 + \sigma_{\varepsilon}^2} \\
    \label{eq:idiosyncratic-term-posterior-mean-variance}
    \hat{\varphi}_{xy} &= \frac{\sigma_{\varphi}^2}{\sigma_{\varphi}^2 + \sigma_{\delta}^2} \cdot \tilde{\varphi}_{xy} + \frac{\sigma_{\delta}^2}{\sigma_{\varphi}^2 + \sigma_{\delta}^2} \cdot \mu_\varphi, \quad \quad \hat{\sigma}_{\varphi}^2 = \frac{\sigma_{\varphi}^2 \cdot \sigma_{\delta}^2}{\sigma_{\varphi}^2 + \sigma_{\delta}^2}
\end{align}
A rational, risk-neutral agent maximizes expected utility given their information. Since utility is linear in the common term $q_y$ and the idiosyncratic term $\varphi_{xy}$, the agent's perceived expected utility $\hat{u}_{xy}$ is given as follows.
\begin{align}
    \label{eq:perceived-utility}
    \hat{u}_{xy} = (1 - \rho) \cdot \hat{q}_y + \rho \cdot \hat{\varphi}_{xy}
\end{align}

\paragraph{Supply Settings.} To study and understand the interplay between supply side constraints and value of different information provisioning tools, we study two types of supply constraints.

\begin{enumerate}[label=(\alph*)]
    \item {\it Uncapacitated Supply} ({\sf uncap}):~Each item has {\it infinite capacity}; any number of agents can choose the same item.
    \item {\it Capacitated Supply} ({\sf cap}):~Each item has {\it unit capacity}; once chosen, it becomes unavailable to subsequent agents.
\end{enumerate}

\paragraph{Agent Choice and Matching Process.} Agents select items sequentially in $n$ rounds, ordered by their priority scores (highest score chooses first, etc.). In round $k$, the $k$-th agent chooses from the remaining items (denoted as $\mathcal{Y}_{k}^{\text{rem}}$) to maximize her perceived expected utility $\hat{u}$ (defined in \eqref{eq:perceived-utility}), with ties broken according to a common lottery over the items across all the agents. Note that this encompasses the main examples of interest. In the uncapacitated setting, the sequence does not matter because items have infinite capacity. In the capacitated case, a priority-based order aligns with centralized college admissions, where students are ranked by an exam score and sequentially pick from available programs \citep{baswana2019centralized, gale1962college, abdulkadirouglu1998random}: in the balanced market setting with common preferences on the supply side, deferred acceptance is equivalent to serial dictatorship. If priority scores are random, this corresponds to the random arrival model in online marketplaces. 

\paragraph{Information Regimes.} We present three information regimes which will be of focus in the extensions and would be used to quantify the value of public rankings and personalized recommendations.

\begin{enumerate}[label = $(\roman*)$]
    \item {\sf No Information} ($\emptyset$): In this setting, we assume that $\sigma_{\varepsilon} \to \infty$ and $\sigma_{\delta} \to \infty$, i.e., the agents have no information about the common and the idiosyncratic term and all items have identical perceived utility of $\hat{u}_{xy} = (1 - \rho) \mu_q + \rho \mu_\varphi$ for all $(x,y)$ pairs.
    
    \item {\sf Only Quality Information} ($q$): In this setting, we assume that $\sigma_{\varepsilon} = 0$ and $\sigma_{\delta} \to \infty$, i.e, the agents have perfect information about the common terms while no information about the idiosyncratic terms. The perceived utility of item $y$ for agent $x$ is given as $\hat{u}_{xy} = (1 - \rho) q_y + \rho \mu_\varphi$.

    \item {\sf Full Information} ($u$): In this setting, we assume that $\sigma_{\varepsilon} = \sigma_{\delta} = 0$, i.e., the agents have perfect information about the common and the idiosyncratic terms. The perceived utility of item $y$ for agent $x$ is the true utility and is given as $u_{xy} = \hat{u}_{xy} = (1 - \rho) q_y + \rho \varphi_{xy}$.
\end{enumerate}

In addition to the three extreme regimes described above, we also consider two intermediate regimes. 

\begin{enumerate}[label = $(\roman*')$]
    \item {\sf Only Noisy Quality Information} ($\hat{q}$): In this setting, we assume that $\sigma_{\varepsilon} \in (0,\infty)$ and $\sigma_{\delta} \to \infty$, i.e., the agents have some partial information about the common terms and no information about the idiosyncratic terms. The perceived utility of item $y$ for agent $x$ is given as $\hat{u}_{xy} = (1 - \rho) \hat{q}_{y} + \rho \mu_\varphi$.
    \item {\sf Noisy Full Information} ($\hat{u}$): In this setting, we assume that $\sigma_{\varepsilon} \in (0, \infty)$ and $\sigma_{\delta} \in (0, \infty)$, i.e., the agents have some partial information about both the common and the idiosyncratic terms. The perceived utility of item $y$ for agent $x$ is given as $\hat{u}_{xy} = (1 - \rho) \hat{q}_y + \rho \hat{\varphi}_{xy}$.
\end{enumerate}

\paragraph{(Additional) Value of Different Information Provisioning Tools.} In this paper, the quantity of interest is {\it agent welfare} which we define as the (expected) average utility of agents under different information regimes. As mentioned previously, we study five information regimes described above and denoted as ${\sf r} \in \{\emptyset, \hat{q}, q, \hat{u}, u\}$. Let $\pi_{{\sf r}}(k)$ denote the index of the item matched to 
the $k$-th agent in regime ${\sf r} \in \{\emptyset, \hat{q}, q, \hat{u}, u\}$. Let ${\sf AW}^{\sf uncap}_{{\sf r}}(n)$ and ${\sf AW}^{\sf cap}_{{\sf r}}(n)$ denote the average agent welfare in the uncapacitated ({\sf uncap}) and capacitated ({\sf cap}) supply settings in information regime ${\sf r}$. We have that 
\begin{align}
    \label{eq:agent-welfare-uncapacitated}
    {\sf AW}^{\sf uncap}_{{\sf r}}(n) &= \mathbb{E}[u_{1\pi_{{\sf r}}(1)}], \quad\quad\quad\quad\,\, {\sf r} \in \{\emptyset, \hat{q}, q, \hat{u}, u\}, \\
    \label{eq:agent-welfare-capacitated}
    {\sf AW}^{\sf cap}_{\sf r} (n) &= (1/n) \mathbb{E}\left[\sum_{k = 1}^n u_{k, \pi_{{\sf r}}(k)}\right], \quad {\sf r} \in \{\emptyset, \hat{q}, q, \hat{u}, u\},
\end{align}

We consider two forms of information provisioning tools in this paper: (i) {\it public rankings} and (ii) {\it personalized recommendations}. We abstract these two information provisioning technologies as providing agents with different components of their utility. In particular, we assume that (noisy) public rankings inform agents of (noisy) common terms. We assume that (noisy) personalized recommendations inform agents of the (noisy) idiosyncratic terms for the agent-item pairs in addition to the (noisy) common terms.

To assess the {\it added value} of public rankings and personalized recommendations, we compare welfare across different information regimes. We define $\Delta^{\sf uncap}_{{\sf s} \to {\sf t}} = {\sf AW}^{\sf uncap}_{{\sf t}}(n) - {\sf AW}^{\sf uncap}_{{\sf s}}(n)$ as the difference in the agent welfare in the uncapacitated setting under two different information regimes ${\sf t}$ and ${\sf s}$. Similarly, we define $\Delta^{\sf cap}_{{\sf s} \to {\sf t}} = {\sf AW}^{\sf cap}_{{\sf t}}(n) - {\sf AW}^{\sf cap}_{{\sf s}}(n)$ for the capacitated supply setting. Note that $\Delta_{\emptyset \to \hat{q}}^{\sf uncap}(n)$ and $\Delta_{\emptyset \to \hat{q}}^{\sf cap}(n)$ quantify the added value of (noisy) public rankings (over no information) in uncapacitated and capacitated supply settings respectively. Similarly, $\Delta_{\hat{q} \to \hat{u}}^{\sf uncap}(n)$ and $\Delta_{\hat{q} \to \hat{u}}^{\sf cap}(n)$ quantify the added value of (noisy) personalized recommendations (over (noisy) public rankings) in uncapacitated and capacitated supply settings respectively.

\paragraph{Notation.} Let $X$ be a random variable, then $X_{(k:n)}$ denotes the $k$-th order statistic ($k$-smallest value) of $n$ independent and identically distributed copies of $X$. Note that $X_{(n:n)} = \max \{X_1, \dots, X_n\}$ denotes the highest value amongst $n$ i.i.d draws of $X$. For any $x \in \mathbb{R}$, we have that $(x)_+ \triangleq \max\{x, 0\}$. 
For any two functions $f(n)$ and $g(n)$, we denote $f(n) \asymp g(n)$ iff $\lim_{n \to \infty} f(n) / g(n) = 1$.


\section{Agent Welfare \& Value of Information Provisioning Tools}
\label{sec:base-model-results}
In this section, we first provide an asymptotic characterization of the agent welfare under the uncapacitated (see Theorem \ref{thm:noisy-estimate-uncapacitated-agent-welfare}) and the capacitated (see Theorem \ref{thm:noisy-estimate-capacitated-agent-welfare}) supply settings.
Using this characterization, we quantify the value add of different information provisioning tools (public rankings and personalized recommendations) by comparing welfare across information regimes.

\begin{theorem}
    \label{thm:noisy-estimate-agent-welfare}
    Assume the base model in Section~\ref{sec:base-model}.
    Agents observe noisy versions of the common terms $(\tilde{q}_y)_{y\in\mathcal{Y}}$ and idiosyncratic terms $(\tilde{\varphi}_{xy})_{x\in\mathcal{X},y\in\mathcal{Y}}$ as in \eqref{eq:noisy-common-term}--\eqref{eq:noisy-idiosyncratic-term} and choose items sequentially to maximize perceived expected utility \eqref{eq:perceived-utility}.
    Assume \edit{$\mu_q, \mu_\varphi \in (0, \infty)$}.
    Let ${\sf AW}^{\sf uncap}(n)$ and ${\sf AW}^{\sf cap}(n)$ denote average agent welfare in the uncapacitated and capacitated settings under this general noisy-information environment (i.e., the regime with finite $(\sigma_{\varepsilon},\sigma_{\delta})$).
    \begin{enumerate}[label = $(\thetheorem.\alph*)$]
        \item \label{thm:noisy-estimate-uncapacitated-agent-welfare}
        In the uncapacitated supply setting with $n$ items (each with infinite capacity), the agent welfare scales as
        \begin{align}
            \label{eq:noisy-estimate-uncapacitated-agent-welfare}
            \lim_{n \to \infty} \frac{{\sf AW}^{\sf uncap}(n)}{\sqrt{2 \ln n}}
            =
            \sqrt{(1 - \rho)^2 \frac{\sigma_{q}^4}{\sigma_q^2 + \sigma_{\varepsilon}^2} + \rho^2 \frac{\sigma_{\varphi}^4}{\sigma_{\varphi}^2 + \sigma_{\delta}^2}}.
        \end{align}

        \item \label{thm:noisy-estimate-capacitated-agent-welfare}
        In the capacitated supply setting with $n$ items (each with unit capacity), the agent welfare scales as
        \begin{align}
            \label{eq:noisy-estimate-capacitated-agent-welfare}
            \lim_{n \to \infty} \frac{{\sf AW}^{\sf cap}(n)}{\sqrt{2 \ln n}}
            =
            \rho \cdot \frac{{\sigma_{\varphi}^2}}{{\sqrt{\sigma_\varphi^2 + \sigma_{\delta}^2}}}.
        \end{align}
    \end{enumerate}
\end{theorem}

\paragraph{Proof sketch.}
We provide the main ideas here and defer the full proof to Appendix~\ref{app:proof-base-model-results}.

{\it Uncapacitated supply.}
With infinite capacities, one can analyze a single representative agent.
Under Gaussian priors and Gaussian noise, posterior means are linear shrinkage estimators \eqref{eq:common-term-posterior-mean-variance}--\eqref{eq:idiosyncratic-term-posterior-mean-variance}.
Consequently, $\hat q_y$ and $\hat\varphi_{xy}$ are Gaussian with variances $\Var(\hat q_y)=\sigma_q^4/(\sigma_q^2+\sigma_{\varepsilon}^2)$ and
$\Var(\hat \varphi_{xy})=\sigma_\varphi^4/(\sigma_\varphi^2+\sigma_{\delta}^2)$.
Therefore perceived utilities $\hat u_{xy}=(1-\rho)\hat q_y+\rho \hat\varphi_{xy}$ are i.i.d.\ Gaussian with variance equal to the expression under the square root in \eqref{eq:noisy-estimate-uncapacitated-agent-welfare}.
Risk neutrality and linearity imply that conditional on signals, the expected {\it true} utility of the chosen item equals the chosen posterior mean, so
${\sf AW}^{\sf uncap}(n)=\E[\max_{y\le n}\hat u_{xy}]$.
Gaussian extreme-value theory then yields that 
$\E[\max_{y\le n}\hat u_{xy}] \sim \bar\sigma \sqrt{2\ln n}$, where $\bar\sigma$ is the standard deviation of $\hat u_{xy}$, giving \eqref{eq:noisy-estimate-uncapacitated-agent-welfare}.

{\it Capacitated supply.}
With unit capacities, each item is allocated at most once, creating a conservation property for the common component:
for any one-to-one assignment $\mu(\cdot)$, $\sum_x q_{\mu(x)}=\sum_y q_y$ (and similarly for $\hat q$).
Thus, information about $q_y$ can reshuffle {\it who gets which $q_y$} but cannot change the {\it average} common-value contribution to welfare.
As a result, 
\edit{welfare under capacity constraints is determined by how large the idiosyncratic component is.}
Using the principle of deferred decisions, in round $k$ the idiosyncratic signals over the remaining items behave like a fresh i.i.d.\ sample.
An upper bound is obtained by comparing the chosen idiosyncratic signal to the maximum \edit{idiosyncratic signal} over $n$ 
items; a matching lower bound is obtained by restricting attention to a large subset of high-$\hat q$ items and 
\edit{considering}
the best idiosyncratic realization within that subset.
These bounds match at the $\sqrt{2\ln n}$ scale, yielding \eqref{eq:noisy-estimate-capacitated-agent-welfare}.

\paragraph{From welfare formulas to the value of rankings and recommendations.}
Theorem~\ref{thm:noisy-estimate-agent-welfare} expresses welfare as a function of signal \edit{standard deviations} 
$(\sigma_{\varepsilon},\sigma_{\delta})$.
Each information regime in Section~\ref{sec:base-model} corresponds to a limiting choice of these noise levels (e.g., $\sigma_{\delta}\to\infty$ means no information about the idiosyncratic term).
The corollaries below follow by specializing the theorem to the relevant regime(s) and taking welfare differences 
to isolate the incremental value of (noisy) rankings and (noisy) personalization. Recall that $\Delta_{\emptyset \to \hat{q}}^{\sf uncap}(n)$ and $\Delta_{\emptyset \to \hat{q}}^{\sf cap}(n)$ capture the value of (noisy) public rankings ({\sf Only Noisy Quality Information} over {\sf No Information}) in the uncapacitated and capacitated settings respectively. Similarly, $\Delta_{\hat{q} \to \hat{u}}^{\sf uncap}(n)$ and $\Delta_{\hat{q} \to \hat{u}}^{\sf cap}(n)$ capture the incremental value of (noisy) personalized recommendations ({\sf Noisy Full Information} over {\sf Only Noisy Quality Information}) in the uncapacitated and capacitated settings respectively.


\begin{corollary}[Value of (noisy) Public Rankings]
    \label{cor:value-noisy-public-ranking}
   We have that,
   \begin{enumerate}[label = $(\alph*)$]
        \item \label{cor:noisy-uncapacitated-value-public-ranking} In the uncapacitated supply setting, the value of (noisy) public rankings scales as
        \begin{align}
            \label{eq:noisy-uncapacitated-public-ranking}
            \lim_{n \to \infty} \frac{{\Delta}_{\emptyset \to \hat{q}}^{\sf uncap}(n)}{\sqrt{2 \ln n}} = (1 - \rho) \frac{\sigma_{q}^2}{\sqrt{\sigma_{q}^2 + \sigma_{\varepsilon}^2}}.
        \end{align}
        When the public rankings are perfect, i.e., $\sigma_{ \varepsilon} = 0$, we have that $\lim_{n \to \infty} \Delta_{\emptyset \to q}^{\sf uncap}(n) / \sqrt{2\ln n} = (1 - \rho) \sigma_{q}$.
        \item \label{cor:noisy-capacitated-value-public-ranking} In the capacitated supply setting, the value of (noisy) public ranking is zero, i.e., $\Delta_{\emptyset \to \hat{q}}^{\sf cap}(n) = 0$ for all $n \geq 1$ and for all $\sigma_{\varepsilon} \in [0,\infty]$.
    \end{enumerate}
\end{corollary}

\begin{corollary}[\edit{Incremental} Value of (noisy) Personalized Recommendations
]
    \label{cor:value-noisy-personalized-recommendations} 
    We have that,
    \begin{enumerate}[label = $(\alph*)$]
        \item \label{cor:noisy-uncapacitated-value-personalized-recos} In the uncapacitated supply setting, the value of (noisy) personalized recommendations scales as
        \begin{align}
            \label{eq:noisy-uncapacitated-personalized-recos}
            \lim_{n \to \infty} \frac{{\Delta}_{\hat{q} \to \hat{u}}^{\sf uncap}(n)}{\sqrt{2 \ln n}} = \sqrt{(1 - \rho)^2 \frac{\sigma_{q}^4}{\sigma_q^2 + \sigma_{\varepsilon}^2} + \rho^2 \frac{\sigma_{\varphi}^4}{\sigma_{\varphi}^2 + \sigma_{\delta}^2}} - (1 - \rho) \frac{\sigma_{q}^2}{\sqrt{\sigma_{q}^2 + \sigma_{\varepsilon}^2}}.
        \end{align}
        The value add of perfect personalized recommendation over perfect public ranking scales as
        \begin{align}
            \label{eq:perfect-uncapacitated-personalized-recos}
            \lim_{n \to \infty} \frac{\Delta_{q \to u}^{\sf uncap}(n)}{\sqrt{2 \ln n}} = \sqrt{(1 - \rho)^2 \sigma_{q}^2 + \rho^2 \sigma_{\varphi}^2} - (1 - \rho) \sigma_{q}.
        \end{align}
        \item \label{cor:noisy-capacitated-value-personalized-recos} In the capacitated supply setting, the value of (noisy) personalized recommendations scales as
       \begin{align}
            \label{eq:noisy-capacitated-personlized-recos}
             \lim_{n \to \infty} \frac{{\Delta}_{\hat{q} \to \hat{u}}^{\sf cap}(n)}{\sqrt{2 \ln n}} = \rho \cdot \frac{\sigma_{\varphi}^2}{\sqrt{\sigma_{\varphi}^2 + \sigma_{\delta}^2}}.
        \end{align}
        The value add of perfect personalized recommendations over perfect public rankings scales as
        \begin{align}
            \label{eq:perfect-capacitated-personalized-recos}
            \lim_{n \to \infty} \frac{\Delta_{q \to u}^{\sf cap}(n)}{\sqrt{2 \ln n}} = \rho \cdot \sigma_{\varphi}.
        \end{align}
    \end{enumerate}
\end{corollary}

\paragraph{Discussion and implications.}
Theorem~\ref{thm:noisy-estimate-agent-welfare} and Corollaries~\ref{cor:value-noisy-public-ranking}--\ref{cor:value-noisy-personalized-recommendations}
characterize how information about (i) common quality and (ii) idiosyncratic fit affects {\it aggregate} (average) agent welfare in large markets.
Two implications are particularly salient.
First, capacity constraints fundamentally alter which information can raise average welfare: when each item is allocated at most once, information about the common component only reshuffles \edit{the  allocation of items but cannot expand the total common value created}.
Second, the heterogeneity parameter $\rho$ governs the relative value of personalized versus non-personalized information: as $\rho$ increases, idiosyncratic fit becomes a larger share of utility and personalization correspondingly becomes more valuable. Define the posterior-mean reliability (shrinkage) factors
\[
\kappa_q \triangleq \frac{\sigma_q^2}{\sigma_q^2+\sigma_{\varepsilon}^2},
\qquad
\kappa_\varphi \triangleq \frac{\sigma_\varphi^2}{\sigma_\varphi^2+\sigma_{\delta}^2},
\]
so $\kappa_q,\kappa_\varphi\in[0,1]$, with $\kappa=1$ corresponding to perfect information and $\kappa=0$ corresponding to an uninformative signal.

\paragraph{Uncapacitated supply: information improves welfare through extreme-value selection.}
In uncapacitated environments, an agent's choice does not restrict the feasible set for others, so information affects welfare purely through an improved {\it selection} margin.
Theorem~\ref{thm:noisy-estimate-uncapacitated-agent-welfare} shows that welfare increases at the Gaussian extreme-value scale $\sqrt{2\ln n}$, with leading constant
$\sqrt{(1-\rho)^2 \kappa_q \sigma_q^2 + \rho^2 \kappa_\varphi \sigma_\varphi^2}$.
Two comparative statics follow immediately.

\begin{itemize}
    \item {\it Public rankings.} Corollary~\ref{cor:value-noisy-public-ranking}(a) implies that the welfare gain from (noisy) public rankings is $\Delta^{\sf uncap}_{\emptyset\to \hat q}(n) \asymp (1-\rho)\,\sigma_q \sqrt{\kappa_q}\,\sqrt{2\ln n}.$
    Thus, public rankings are most valuable when the common component is important (small $\rho$) and when common-quality measurement is accurate (large $\kappa_q$ close to 1).

    \item {\it Personalized recommendations.} Corollary~\ref{cor:value-noisy-personalized-recommendations}(a) gives the incremental value of (noisy) personalized recommendations over (noisy) rankings:
        \[
            \Delta^{\sf uncap}_{\hat q\to \hat u}(n)
            \asymp
            \Big(\sqrt{(1-\rho)^2 \kappa_q \sigma_q^2 + \rho^2 \kappa_\varphi \sigma_\varphi^2} - (1-\rho)\sigma_q\sqrt{\kappa_q}\Big)\sqrt{2\ln n}.
        \]
        This increment is $0$ at $\rho=0$ (purely common-value environments) and increases with $\rho$.
        Moreover, when $\rho$ is small, the first-order welfare gains are already captured by rankings, so the additional value of personalization is second-order in $\rho$; 
        whereas for $\rho \to 1$,
        the incremental value approaches $\sigma_\varphi\sqrt{\kappa_\varphi}\sqrt{2\ln n}$.
\end{itemize}

\paragraph{Capacitated supply: rankings cannot raise {\it average} welfare, while personalization can.}
In capacitated environments, each item is allocated at most once. This introduces a conservation property for the common component: for any one-to-one matching, $\sum_{x} q_{\mu(x)} = \sum_{y} q_y.$
This is the formal source of Corollary~\ref{cor:value-noisy-public-ranking}(b), which yields $\Delta^{\sf cap}_{\emptyset\to \hat q}(n)=0,$ for all $n$ and for all $\sigma_{\varepsilon}\in[0,\infty]$. 
Note that this is an aggregate statement. Rankings may substantially affect the {\it distribution} of outcomes---e.g., higher priority agents obtain higher-quality items---even though the average welfare across agents may not improve---which is the focus of this work.

By contrast, information about \emph{idiosyncratic fit} reduces correlated demand for \emph{overdemanded} items. This reduction in congestion improves aggregate match quality, allowing items to unlock extra surplus beyond their intrinsic worth by capitalizing on their ``beauty in the eye of the matched agent.''
Theorem~\ref{thm:noisy-estimate-capacitated-agent-welfare} and Corollary~\ref{cor:value-noisy-personalized-recommendations}(b) show that the welfare gains from (noisy) personalization scale as $\Delta^{\sf cap}_{\hat q\to \hat u}(n) \asymp \rho\,\sigma_\varphi \sqrt{\kappa_\varphi}\,\sqrt{2\ln n}.$
Three implications are immediate:
(i) the gain is linear in $\rho$;
(ii) to leading order it depends on the idiosyncratic noise level $\sigma_{\delta}$ (
\edit{and} $\kappa_\varphi$) but not on $\sigma_{\varepsilon}$; and
(iii) even imperfect personalization can dominate arbitrarily accurate rankings in capacitated markets, because rankings have zero aggregate welfare effect in this benchmark while any $\kappa_\varphi>0$ yields a strictly positive gain at \edit{the leading scale of} $\sqrt{\ln n}$.

\paragraph{Finite-$n$ validation via simulation.}
Figure~\ref{fig:value-of-info-2x2-normalized} compares the asymptotic predictions (solid lines) against Monte Carlo estimates in a finite market ($n=10^3$; markers), across multiple noise levels.
The figure shows that the large-market limits provide accurate quantitative approximations even at moderate market sizes.
Figures~\ref{fig:rankings-uncap}--\ref{fig:rankings-cap} validate the sharp contrast highlighted by Corollary~\ref{cor:value-noisy-public-ranking}:
public rankings deliver substantial gains in the uncapacitated setting but essentially no aggregate gains under capacity constraints (the simulated points concentrate near zero).
Figures~\ref{fig:personal-uncap}--\ref{fig:personal-cap} corroborate Corollary~\ref{cor:value-noisy-personalized-recommendations} (specialized to $\sigma_{q,\varepsilon}=0$ in Figure~\ref{fig:personal-uncap}):
the value of personalization increases with heterogeneity $\rho$, 
degrades smoothly with idiosyncratic signal noise, \edit{and is larger under capacitated supply for any given primitives}.

\paragraph{Design and policy implications: allocating accuracy investments.}
The shrinkage factors $\kappa_q$ and $\kappa_\varphi$ provide a direct mapping from statistical accuracy to welfare.
In uncapacitated environments (e.g., digital content, search, advertising), the platform 
\edit{should facilitate }
improved {\it selection}:
both better rankings (higher $\kappa_q$) and better personalization (higher $\kappa_\varphi$) can raise welfare, and the relative return depends on $\rho$.
When preferences are relatively homogeneous (small $\rho$), improving the common-quality signal is particularly valuable because agents tend to agree on what is desirable.

In capacitated environments (e.g., school choice, rentals, limited-inventory marketplaces), the main \edit{tool to improve} welfare 
shifts from selection to {\it coordination under scarcity}.
Improving public rankings may change the composition of demand at each item but does not increase the average common-value component; in contrast, improving the signal about the idiosyncratic term 
can reduce congestion and increase average fit.
For example, in school choice or college admissions, a planner seeking to raise average match quality might prioritize tools that help students identify idiosyncratic fit (program-level match, curricular alignment, peer environment) over increasingly precise ``US News''-style overall rankings.

\paragraph{Scope and limitations.}
Our welfare comparisons are for {\it aggregate} (average) agent welfare.
They do not capture distributional consequences (e.g., inequality in assignments), fairness constraints, or the welfare of non-agent stakeholders (e.g., schools, sellers), and they abstract from strategic responses, endogenous entry, and privacy/consent costs of collecting data for personalization.
In particular, in capacitated environments, public rankings may meaningfully change {\it who gets what}, even though they do not increase the average common-value term in our setting,  analyzing these distributional and institutional considerations in greater generality is an important direction for future work. 

\begin{remark}[General capacities in balanced markets]
\label{rmk:capacitated-case-extension}
The unit-capacity assumption in the capacitated model can be relaxed to a balanced many-to-one setting in which each item $y \in \mathcal{Y}$ has an integer capacity $C_y \ge 1$, and the number of agents satisfies
$|\mathcal X|=\sum_{y\in\mathcal Y} C_y.$
In this generalized setting, the zero-value conclusion for public rankings extends exactly. Indeed, for any feasible matching $\mu$, each item $y$ is assigned to exactly $C_y$ agents, so the aggregate common-value contribution is fixed, i.e., $\sum_{x\in\mathcal X} q_{\mu(x)}=\sum_{y\in\mathcal Y} C_y q_y.$

Hence moving from {\sf No Information} to {\sf Only Quality Information} can only reshuffle which agents receive which capacities, but cannot change the aggregate common-value term. Since the matching under {\sf Only Quality Information} is independent of the idiosyncratic terms, the expected idiosyncratic contribution remains $\mu_\varphi$ per agent. Therefore, the conclusion of Corollary~\ref{cor:value-noisy-public-ranking}(b) continues to hold in this balanced many-to-one setting: public rankings do not increase average agent welfare.

For personalized recommendations, the same economic mechanism as in Corollary~\ref{cor:value-noisy-personalized-recommendations}(b) continues to operate. Information about idiosyncratic fit can reduce correlated demand across the finitely many seats attached to each item and thereby improve aggregate match quality. When capacities are uniformly bounded independently of market size (for example, $C_y\equiv k$ for some fixed $k\ge 2$), we expect the welfare gain from personalization under Gaussian primitives to remain on the same leading asymptotic scale as in the unit-capacity model, namely $\sqrt{2\ln n}$, although the exact leading constant will generally depend on the capacity profile $(C_y)_{y\in\mathcal Y}$. More generally, the tail-based results in Theorems~\ref{thm:capacitated-ranking-value} and~\ref{thm:capacitated-personalized-recos-value} suggest the same qualitative robustness beyond the Gaussian case: common-quality information mainly reshuffles access, whereas idiosyncratic information continues to generate aggregate gains. For brevity, we do not pursue the exact welfare formulas for general capacity profiles in this paper. 
\end{remark}



\begin{figure}[!htbp]
\centering
\definecolor{color0}{RGB}{0,114,189}     
\definecolor{color1}{RGB}{217,83,25}     
\definecolor{color2}{RGB}{126,47,142}    
\definecolor{color3}{RGB}{119,172,48}    

\begin{subfigure}[t]{0.48\textwidth}
\centering
\begin{tikzpicture}[scale=0.85]
\begin{axis}[
    xlabel={Level of heterogeneity $\rho$},
    ylabel={$\Delta^{\sf uncap}_{\emptyset \to \hat q}(n) / \ell^{\sf uncap}_n$},
    xmin=0, xmax=1,
    ymin=0, ymax=1.01,
    grid=major,
]
\addplot[color0, ultra thick, domain=0:1, samples=50] 
    table [x=rho, y=theory_s0, col sep=comma] {data/delta_rankings_uncap_norm.csv};
\addplot[color0, mark=*, ultra thick, only marks] 
    table [x=rho, y=mean_s0, col sep=comma] {data/delta_rankings_uncap_norm.csv};

\addplot[color1, ultra thick] 
    table [x=rho, y=theory_s0p5, col sep=comma] {data/delta_rankings_uncap_norm.csv};
\addplot[color1, mark=square*, ultra thick, only marks] 
    table [x=rho, y=mean_s0p5, col sep=comma] {data/delta_rankings_uncap_norm.csv};

\addplot[color2, ultra thick] 
    table [x=rho, y=theory_s1, col sep=comma] {data/delta_rankings_uncap_norm.csv};
\addplot[color2, mark=triangle*, ultra thick, only marks] 
    table [x=rho, y=mean_s1, col sep=comma] {data/delta_rankings_uncap_norm.csv};

\addplot[color3, ultra thick] 
    table [x=rho, y=theory_s20, col sep=comma] {data/delta_rankings_uncap_norm.csv};
\addplot[color3, mark=diamond*, ultra thick, only marks] 
    table [x=rho, y=mean_s20, col sep=comma] {data/delta_rankings_uncap_norm.csv};

\end{axis}
\end{tikzpicture}
\caption{Value of (noisy) public rankings in {\sf uncapacitated} supply setting}
\label{fig:rankings-uncap}
\end{subfigure}
\hfill
\begin{subfigure}[t]{0.48\textwidth}
\centering
\begin{tikzpicture}[scale=0.85]
\begin{axis}[
    xlabel={Level of heterogeneity $\rho$},
    ylabel={$\Delta^{\sf cap}_{\emptyset \to \hat q}(n) / \ell^{\sf cap}_n$},
    xmin=0, xmax=1,
    ymin=-0.005, ymax=0.005,
    grid=major,
]
\addplot[color0, ultra thick] 
    table [x=rho, y=theory_s0, col sep=comma] {data/delta_rankings_cap_norm.csv};
\addplot[color0, mark=*, ultra thick, only marks] 
    table [x=rho, y=mean_s0, col sep=comma] {data/delta_rankings_cap_norm.csv};

\addplot[color1, ultra thick] 
    table [x=rho, y=theory_s0p5, col sep=comma] {data/delta_rankings_cap_norm.csv};
\addplot[color1, mark=square*, ultra thick, only marks] 
    table [x=rho, y=mean_s0p5, col sep=comma] {data/delta_rankings_cap_norm.csv};

\addplot[color2, ultra thick] 
    table [x=rho, y=theory_s1, col sep=comma] {data/delta_rankings_cap_norm.csv};
\addplot[color2, mark=triangle*, ultra thick, only marks] 
    table [x=rho, y=mean_s1, col sep=comma] {data/delta_rankings_cap_norm.csv};

\addplot[color3, ultra thick] 
    table [x=rho, y=theory_s20, col sep=comma] {data/delta_rankings_cap_norm.csv};
\addplot[color3, mark=diamond*, ultra thick, only marks] 
    table [x=rho, y=mean_s20, col sep=comma] {data/delta_rankings_cap_norm.csv};

\end{axis}
\end{tikzpicture}
\caption{Value of (noisy) public rankings in {\sf capacitated} supply setting}
\label{fig:rankings-cap}
\end{subfigure}

\vspace{1em}

\begin{subfigure}[t]{0.48\textwidth}
\centering
\begin{tikzpicture}[scale=0.85]
\begin{axis}[
    xlabel={Level of heterogeneity $\rho$},
    ylabel={$\Delta^{\sf uncap}_{q \to \hat u}(n) / \ell^{\sf uncap}_n$},
    xmin=0, xmax=1,
    ymin=0, ymax=1.01,
    grid=major,
]
\addplot[color0, ultra thick] 
    table [x=rho, y=theory_s0, col sep=comma] {data/delta_personal_uncap_norm.csv};
\addplot[color0, mark=*, ultra thick, only marks] 
    table [x=rho, y=mean_s0, col sep=comma] {data/delta_personal_uncap_norm.csv};

\addplot[color1, ultra thick] 
    table [x=rho, y=theory_s0p5, col sep=comma] {data/delta_personal_uncap_norm.csv};
\addplot[color1, mark=square*, ultra thick, only marks] 
    table [x=rho, y=mean_s0p5, col sep=comma] {data/delta_personal_uncap_norm.csv};

\addplot[color2, ultra thick] 
    table [x=rho, y=theory_s1, col sep=comma] {data/delta_personal_uncap_norm.csv};
\addplot[color2, mark=triangle*, ultra thick, only marks] 
    table [x=rho, y=mean_s1, col sep=comma] {data/delta_personal_uncap_norm.csv};

\addplot[color3, ultra thick] 
    table [x=rho, y=theory_s20, col sep=comma] {data/delta_personal_uncap_norm.csv};
\addplot[color3, mark=diamond*, ultra thick, only marks] 
    table [x=rho, y=mean_s20, col sep=comma] {data/delta_personal_uncap_norm.csv};

\end{axis}
\end{tikzpicture}
\caption{Value of (noisy) personalized recommendations (over perfect public rankings) in {\sf uncapacitated} supply setting}
\label{fig:personal-uncap}
\end{subfigure}
\hfill
\begin{subfigure}[t]{0.48\textwidth}
\centering
\begin{tikzpicture}[scale=0.85]
\begin{axis}[
    xlabel={Level of heterogeneity $\rho$},
    ylabel={$\Delta^{\sf cap}_{\hat q \to \hat u}(n) / \ell^{\sf cap}_n$},
    xmin=0, xmax=1,
    ymin=0, ymax=1.01,
    grid=major,
]
\addplot[color0, ultra thick] 
    table [x=rho, y=theory_s0, col sep=comma] {data/delta_personal_cap_norm.csv};
\addplot[color0, mark=*, ultra thick, only marks] 
    table [x=rho, y=mean_s0, col sep=comma] {data/delta_personal_cap_norm.csv};

\addplot[color1, ultra thick] 
    table [x=rho, y=theory_s0p5, col sep=comma] {data/delta_personal_cap_norm.csv};
\addplot[color1, mark=square*, ultra thick, only marks] 
    table [x=rho, y=mean_s0p5, col sep=comma] {data/delta_personal_cap_norm.csv};

\addplot[color2, ultra thick] 
    table [x=rho, y=theory_s1, col sep=comma] {data/delta_personal_cap_norm.csv};
\addplot[color2, mark=triangle*, ultra thick, only marks] 
    table [x=rho, y=mean_s1, col sep=comma] {data/delta_personal_cap_norm.csv};

\addplot[color3, ultra thick] 
    table [x=rho, y=theory_s20, col sep=comma] {data/delta_personal_cap_norm.csv};
\addplot[color3, mark=diamond*, ultra thick, only marks] 
    table [x=rho, y=mean_s20, col sep=comma] {data/delta_personal_cap_norm.csv};

\end{axis}
\end{tikzpicture}
\caption{Value of (noisy) personalized recommendations (over perfect public rankings) in {\sf capacitated} supply setting}
\label{fig:personal-cap}
\end{subfigure}

\vspace{0.8em}

\centering
\begin{tabular}{c@{\hspace{1.5em}}c@{\hspace{1.5em}}c@{\hspace{1.5em}}c@{\hspace{1.5em}}c}
 & $\sigma_\varepsilon = 0$ & $\sigma_\varepsilon = 0.5$ & $\sigma_\varepsilon = 1$ & $\sigma_\varepsilon = 20$ \\[0.3em]
Theory: & 
\tikz[baseline=-0.5ex]\draw[color0, ultra thick] (0,0) -- (0.5,0); & 
\tikz[baseline=-0.5ex]\draw[color1, ultra thick] (0,0) -- (0.5,0); & 
\tikz[baseline=-0.5ex]\draw[color2, ultra thick] (0,0) -- (0.5,0); & 
\tikz[baseline=-0.5ex]\draw[color3, ultra thick] (0,0) -- (0.5,0); \\[0.3em]
Simulation $(n = 10^3)$: & 
\tikz[baseline=-0.5ex]\draw[color0, ultra thick] plot[only marks, mark=*, mark size=2.5pt] coordinates {(0,0)}; & 
\tikz[baseline=-0.5ex]\draw[color1, ultra thick] plot[only marks, mark=square*, mark size=2.5pt] coordinates {(0,0)}; & 
\tikz[baseline=-0.5ex]\draw[color2, ultra thick] plot[only marks, mark=triangle*, mark size=3pt] coordinates {(0,0)}; & 
\tikz[baseline=-0.5ex]\draw[color3, ultra thick] plot[only marks, mark=diamond*, mark size=3pt] coordinates {(0,0)}; \\
\end{tabular}

\caption{Plot of the value of public rankings and personalized recommendations in different supply settings as a function of level of heterogeneity $\rho$. We fix $\mu_q = \mu_\varphi = 0$ and $\sigma_q = \sigma_\varphi = 1$. For the value of (noisy) public ranking plots, we have $\sigma_{\delta} = \infty$ and vary $\sigma_{\varepsilon} = \sigma$. For the value of (noisy) personalized recommendation plots, we have $\sigma_{\varepsilon} = 0$ and vary $\sigma_{\delta} = \sigma$. We consider four values of $\sigma \in \{0,0.5,1,20\}$ to capture different noisy levels. Uncapacitated plots are normalized by $\ell^{\sf uncap}_n=\mathbb{E}[\max_{j\le n}Z_j]$ and capacitated personalized plots by $\ell^{\sf cap}_n=\mathbb{E}[\frac{1}{n}\sum_{m=1}^n \max_{j\le m}Z_j]$, where $Z_j\sim\mathcal{N}(0,1)$ and these normalization factors are computed numerically via simulation.}
\label{fig:value-of-info-2x2-normalized}
\end{figure}

\section{Extensions}
\label{sec:model-extensions}

The base model in Sections~\ref{sec:base-model}--\ref{sec:base-model-results} isolates a sharp interaction between {\it supply constraints} and {\it what kind of information is useful}: under scarcity, learning common quality primarily reshuffles assignments, whereas learning idiosyncratic fit can reduce correlated demand and raise average welfare.
In this section, we stress-test the robustness of these insights and add nuance by relaxing key modeling assumptions along different dimensions that are salient in real markets.

\paragraph{Common setup across extensions.}
Unless otherwise noted, we maintain the same basic primitives (agents, items, sequential choice by priority) and study both the uncapacitated and capacitated supply settings.
For analytical tractability and ease of comparison across extensions, we focus on three benchmark information regimes:
(i) {\sf No Information} ($\emptyset$), where agents observe neither the common nor idiosyncratic components;
(ii) {\sf Only Quality Information} ($q$), where agents observe only the common component (without noise); and
(iii) {\sf Full Information} ($u$), where agents perfectly observe both components.
These regimes correspond to the extreme points of the information spectrum in Section~\ref{sec:base-model}, and map directly to the two tools of interest: public rankings (moving from $\emptyset$ to $q$) and personalized recommendations (moving from $q$ to $u$).

\paragraph{Roadmap and preview of findings.}
Each subsection introduces one extension and derives the resulting welfare comparisons across $\{\emptyset,q,u\}$.
For convenience, we summarize the motivating question and the main takeaway here.

\begin{enumerate}
    \item \label{ex:welfare-gain-different-tail-distribution}
    \textbf{Welfare gains under alternative tail distributions.}
    {\it The base model assumes Gaussian common and idiosyncratic components. How do the welfare gains from rankings and personalization change when these primitives have different tail behavior?}
    We consider a spectrum of tail behaviors (Pareto, Exponential, Gaussian, and Bounded) and characterize welfare gains in both supply settings.
    Qualitatively, the main insights from Section~\ref{sec:base-model-results} continue to hold, and the analysis confirms that the welfare gains depend primarily on the {\it tail} of the primitive distributions rather than their entire shape.

    \item \label{ex:supply-side-priority}
    \textbf{Supply-side priorities.}
    {\it In the capacitated base model, all items share a common priority ordering over agents, so serial dictatorship clears the market and coincides with the unique stable matching. What changes when items have heterogeneous priorities over agents?}
    Motivated by neighborhood priorities in school choice, we study a setting with two groups of items \edit{and agents (e.g., schools and applicants} in two neighborhoods) where items exhibit own-group priorities over agents and analyze stable matching outcomes.
    We provide bounds on welfare gains under different information provision tools and find that, as in Section~\ref{sec:base-model-results}, public rankings deliver limited aggregate value while personalized recommendations can unlock substantial welfare gains by mitigating congestion and improving fit.

    \item \label{ex:supply-side-pricing}
    \textbf{Supply-side pricing.}
    {\it In the capacitated base model, there are no explicit item prices (or prices can be assumed to be set exogenously). What happens to agent welfare, seller revenue, and total welfare when sellers set prices?}
    We extend the model by introducing prices and studying seller pricing under different information regimes.
    Agents' utility becomes $u_{xy} = (1-\rho)q_y + \rho \varphi_{xy} - p_y,$
    where $p_y$ is the price of item $y$.
    We characterize equilibrium prices, agent welfare, seller revenue, and total welfare (agent welfare plus seller revenue).
    In both the balanced market and the excess demand setting (number of agents exceeds the number of items), we find that total welfare does not change between {\sf No Information} and {\sf Only Quality Information}: public rankings do not increase total welfare, but cause welfare to move from agents to quality sellers.
    By contrast, total welfare increases under {\sf Full Information}, with additional welfare gains again accruing to sellers.
    Moreover, when supply is on the short side of the market, much of the welfare increase from information (public rankings or full information) accrues to the supply side: better information allows sellers to raise prices and extract a larger share of surplus, so seller revenue rises substantially while agent welfare changes relatively little. Conversely, under excess supply (where demand is on the short side of the market), competition disciplines prices, allowing agents to capture the substantial welfare gains unlocked by personalized recommendations.
\end{enumerate}

\subsection{Welfare Gains for Different Tail Distributions}
\label{sec:welfare-gains-tail-distributions}

Section~\ref{sec:base-model-results} establishes that in large markets, the value of information is driven by an {\it extreme-value} effect: with more items, better information helps agents select better “extremes.”  
The Gaussian assumption in the base model is convenient, but it is also a very specific point in a much broader landscape: in many applications (jobs, housing, superstar content, highly differentiated schools), payoffs can be heavy-tailed on one hand or bounded on the other.
We therefore replace Gaussian primitives with a spectrum of tail behaviors and characterize how the welfare gains from public rankings and personalized recommendations scale with the market size $n$.

\subsubsection{Setup}
We assume the common terms $(q_y)_{y\in\mathcal{Y}}$ and idiosyncratic terms $(\varphi_{xy})_{x\in\mathcal{X},y\in\mathcal{Y}}$ are i.i.d.\ draws from distributions $P_q$ and $P_{\varphi}$, respectively, and are independent of each other.
To isolate the role of extremes, we describe these distributions {\it only through their upper tails}.
(Throughout, we impose non-negative support and finite means $\mu_q,\mu_\varphi<\infty$.)

\begin{definition}[Pareto Tail]
    \label{def:Pareto-tail}
    Fix $c > 0$ and $\alpha > 1$. Let $X$ be a random variable with distribution $F$. We say that $F$ has a Pareto tail with parameters $(c, \alpha)$ if
    $\lim_{x \to \infty} \frac{\mathbb{P}(X > x)}{(c / x)^\alpha} = 1$.
\end{definition}

\begin{definition}[Exponential Tail]
    \label{def:Exponential-tail}
    Fix $c > 0$ and $\lambda > 0$. Let $X$ be a random variable with distribution $F$. We say that $F$ has an exponential tail with parameters $(c, \lambda)$ if
    $\lim_{x \to \infty} \frac{\mathbb{P}(X > x)}{c\exp(-\lambda x)} = 1$.
\end{definition}

\begin{definition}[Gaussian-like Tail]
\label{def:gaussian-tail}
Fix $c > 0, \lambda > 0$ and $\beta \geq 0$. Let $X$ be a random variable with distribution $F$. We say that $F$ has a Gaussian-like tail with parameters $(c, \lambda, \beta)$ if
$\lim_{x \to \infty} \frac{\mathbb{P}(X > x)}{cx^{-\beta} \exp(-\lambda^2 x^2)} = 1$.
\end{definition}

\begin{definition}[Bounded Tail]
    \label{def:Bounded-tail}
    Fix $0 \leq a < b < \infty$, $\kappa > 0$ and $\beta \geq 1$. 
    Let $X$ be a random variable with continuous distribution $F$ over $[a,b]$.
    We say that $F$ has a bounded tail with parameters $(b, \kappa, \beta)$ if
    $\lim_{x \to b^{-}} \frac{\mathbb{P}(X > x)}{\kappa(b - x)^\beta} = 1$.
\end{definition}

These four classes span the standard extreme-value regimes:
Pareto tails generate polynomially growing maxima ($n^{1/\alpha}$),
exponential tails generate logarithmically growing maxima ($\ln n$), Gaussian-like tails generate $\sqrt{\ln n}$ growth, and bounded tails saturate at the endpoint $b$. An example of the bounded tail distribution is the $\text{Unif}([0,1])$ with $\kappa = \beta = 1$.


\subsubsection{Uncapacitated Supply Setting}
\label{subsec:welfare-implications-uncapacitated}

In the uncapacitated setting, items have infinite capacity, so each agent’s choice does not restrict others.
Aggregate welfare therefore reduces to the expected utility of a representative agent choosing among $n$ items.
The key identities are:
under {\sf No Information}, agent welfare is $(1-\rho)\mu_q+\rho\mu_\varphi$;
under {\sf Only Quality Information}, agent welfare is $(1-\rho)\mathbb{E}[q_{(n:n)}]+\rho\mu_\varphi$;
under {\sf Full Information}, agent welfare is $\mathbb{E}[\max_{y\le n}\{(1-\rho)q_y+\rho\varphi_{1y}\}]$.

\begin{theorem}[Value of public rankings in the uncapacitated setting]
    \label{thm:uncapacitated-ranking-value}
    Consider the uncapacitated supply setting.
    Assume $(q_y)$ and $(\varphi_{xy})$ are drawn i.i.d.\ from distributions $P_q$ and $P_\varphi$ with non-negative support and finite means $\mu_q,\mu_\varphi$.
    The value of rankings, $\Delta_{\emptyset \to q}^{\sf uncap}(n)$, scales with $n$ as $\lim_{n \to \infty} \Delta_{\emptyset \to q}^{\sf uncap}(n) / \ell_n = 1 - \rho$ for any $\rho \in [0,1]$, where $\ell_n$ depends on the tail \edit{behavior} of $P_q$. We characterize $\ell_n$ for different classes of tail behavior in the table below. \\
    \begin{center}
\begin{threeparttable}
    \begin{tabular}{lcc}
         \toprule 
         $P_q$ tail behavior & Parameter & $\ell_n$ \\
         \midrule
         Pareto & $c_q > 0, \alpha_q > 1$ & $c_q \Gamma(1 - 1 / \alpha_q) n^{1 / \alpha_q}$\tnote{$\dagger$} \\
         \addlinespace
         Exponential & $c_q > 0 , \lambda_q > 0$ & $\ln n / \lambda_q$ \\
         \addlinespace
         Gaussian-like & $c_q > 0, \lambda_q > 0, \beta_q \geq 1$ & $\sqrt{\ln n} / \lambda_q$ \\
         \addlinespace
         Bounded & $0 \leq a_q < b_q < \infty, \kappa_q > 0, \beta_q \geq 1$ & $b_q - \mu_q$ \\
         \bottomrule
    \end{tabular}
    \begin{tablenotes}\footnotesize
        \item[$\dagger$] $\Gamma(z) \triangleq \int_{0}^{\infty} t^{z-1} e^{-t} \, dt, \ \textup{for}\  z > 0$ is the Gamma function.
    \end{tablenotes}
\end{threeparttable}
\end{center}
    

        
\end{theorem}

Theorem~\ref{thm:uncapacitated-ranking-value} illuminates the following insight: $\Delta_{\emptyset\to q}^{\sf uncap}(n) = (1-\rho)\big(\mathbb{E}[q_{(n:n)}]-\mu_q\big),$
so {\it only the tail of $P_q$ matters} for the value of rankings. The scaling varies sharply with the tail:
polynomial growth under Pareto tails, logarithmic under exponential tails, $\sqrt{\ln n}$ under Gaussian-like tails, and a {\it saturation} effect under bounded support. Figure~\ref{fig:uncapacitated-ranking-value} illustrates this tail-by-tail normalization: after dividing by the appropriate extreme-value scale.

\begin{figure}[htbp]
    \centering
    \begin{subfigure}{0.32\linewidth}
    \begin{tikzpicture}[scale = 0.6]
            \begin{axis}[
                xlabel={Level of heterogeneity $\rho$},
                ylabel={$\Delta^{\sf uncap}_{\emptyset \to q}(n) / (c_q \Gamma(1 - 1 / \alpha_q) n^{1 / \alpha_q})$},
                legend style={at={(0.7,1)},anchor=north},
                legend entries={$n=100$, $1 - \rho$},
                xmin=0,
                ymin=0,
                ymax=1,
                xmax=1,
                grid = major,
            ]
                \addplot[blue, mark=*, ultra thick] coordinates {
                    (1.0, -2.93302149e-04)
                    (0.95, 5.09952365e-02)
                    (0.9, 9.99098654e-02)
                    (0.85, 1.49582515e-01)
                    (0.8, 1.98615044e-01)
                    (0.75, 2.50672368e-01)
                    (0.7, 2.98396679e-01)
                    (0.65, 3.53297989e-01)
                    (0.6, 4.01534588e-01)
                    (0.55, 4.57538886e-01)
                    (0.5, 4.99008265e-01)
                    (0.45, 5.45191603e-01)
                    (0.4, 5.98321275e-01)
                    (0.35, 6.47960081e-01)
                    (0.3, 7.02168843e-01)
                    (0.25, 7.55720725e-01)
                    (0.2, 7.95792734e-01)
                    (0.15, 8.54086077e-01)
                    (0.1, 8.94396966e-01)
                    (0.05, 9.51871036e-01)
                    (0.0, 1.01406406e+00)
                };
                \addplot[domain=0:1, samples=100, ultra thick, red] {1 - x};
    \end{axis}
    \end{tikzpicture}
    \caption{Pareto tail, $\alpha_q =2$}
    \label{subfig:thm-uncapacitated-ranking-value-pareto-tail}
    \end{subfigure}
    \begin{subfigure}{0.32\linewidth}
        \begin{tikzpicture}[scale = 0.6]
            \begin{axis}[
                xlabel={Level of heterogeneity $\rho$},
                ylabel={$\Delta^{\sf uncap}_{\emptyset \to q}(n) / \ln n $},
                legend style={at={(0.7,1)},anchor=north},
                legend entries={$n=100$, $1 - \rho$},
                xmin=0,
                ymin=0,
                ymax=1,
                xmax=1,
                grid = major,
            ]
                \addplot[blue, mark=*, ultra thick] coordinates {
                    (1.0, -0.00190615)
                    (0.95, 0.04995942)
                    (0.9, 0.09830729)
                    (0.85, 0.14858004)
                    (0.8, 0.20015517)
                    (0.75, 0.24885637)
                    (0.7, 0.29960095)
                    (0.65, 0.35104143)
                    (0.6, 0.40020136)
                    (0.55, 0.45075586)
                    (0.5, 0.50131353)
                    (0.45, 0.54973624)
                    (0.4, 0.60128942)
                    (0.35, 0.65078516)
                    (0.3, 0.70096405)
                    (0.25, 0.7506789)
                    (0.2, 0.8002238)
                    (0.15, 0.8509942)
                    (0.1, 0.90082787)
                    (0.05, 0.94994553)
                    (0.0, 1.00025182)
                };

                \addplot[domain=0:1, samples=100, ultra thick, red] {1 - x};
            \end{axis}
            \end{tikzpicture}
        \caption{Exponential tail, $\lambda_q = 1$}
        \label{subfig:uncapacitated-ranking-value-exponential-tail}
    \end{subfigure}
    \begin{subfigure}[b]{0.32\textwidth}
\centering
\begin{tikzpicture}[scale=0.6]
\begin{axis}[
    xlabel={Level of heterogeneity $\rho$},
    ylabel={$\Delta^{\sf uncap}_{\emptyset \to q}(n) / 0.5$},
    grid=major,
    legend style={at={(0.7,1)},anchor=north},
    legend entries={$n=100$, $1 - \rho$},
    ymin=0,
    xmin=0,
    ymax=1,
    xmax=1
]

\addplot[blue, mark=*, ultra thick] coordinates {
(0.0,   0.99404426)
(0.05,  0.92352572)
(0.1,   0.86211892)
(0.15,  0.80468040)
(0.2,   0.74979452)
(0.25,  0.69774160)
(0.3,   0.64372346)
(0.35,  0.59165898)
(0.4,   0.54108976)
(0.45,  0.49068684)
(0.5,   0.43964142)
(0.55,  0.39022340)
(0.6,   0.34158972)
(0.65,  0.29398176)
(0.7,   0.24696950)
(0.75,  0.19961104)
(0.8,   0.15372548)
(0.85,  0.10953372)
(0.9,   0.06651012)
(0.95,  0.02734610)
(1.0,   0.00000000)
};

\addplot[red, ultra thick, domain=0:1] {(1 - x)};
\end{axis}
\end{tikzpicture}
\caption{Bounded tail, $\beta_q = 1$}
\label{subfig:uncapacitated-ranking-value-bounded-tail}
\end{subfigure}
    \caption{Value of public rankings in the uncapacitated setting under different tail distributions. We compare the numerical simulation for $n = 100$ (in blue) with the theoretical asymptotic scaling (in red). We consider (a) Pareto distribution with $c_q = 1, \alpha_q = 2$, (b) Exponential distribution with $\lambda_q = 1$ and (c) Uniform distribution over $[0,1]$.}
    \label{fig:uncapacitated-ranking-value}
\end{figure}

\begin{theorem}[Value of personalized recommendations in the uncapacitated setting]
    \label{thm:uncapacitated-personalized-recos-value}
    Consider the uncapacitated supply setting.
    Assume $(q_y)$ and $(\varphi_{xy})$ are drawn i.i.d.\ from distributions $P_q$ and $P_\varphi$ with non-negative support and finite means $\mu_q,\mu_\varphi$.
    The value of personalized recommendations, $\Delta_{q \to u}^{\sf uncap}(n)$, scales with $n$ as $\lim_{n \to \infty} \Delta_{q \to u}^{\sf uncap}(n) / \ell_n = L$ where $\ell_n$ and $L$ depends on the tail behavior of $P_q$ and $P_\varphi$. We characterize $\ell_n$ and $L$ for different classes of tail behavior in the table below. \\
\begin{center}
\begin{threeparttable}
    \begin{tabular}{cccc}
         \toprule 
         $P_q, P_\varphi$ tail behavior & Parameters & $\ell_n$ & $L$ \\
         \midrule
         \multirow{4}{*}{Pareto} & & \multicolumn{2}{l}{\textit{(a) $\alpha_q \neq \alpha_\varphi$}} \\
         & $c_q, c_\varphi > 0;$ & $\Gamma(1 - 1 / \underline{\alpha})\, n^{1 / \underline{\alpha}}$\tnote{$\dagger$} & $\underline{c} \cdot \rho \cdot \mathbbm{1}\{\alpha_{q} > \alpha_{\varphi} \}$\tnote{$\dagger$} \\
         & $\alpha_q, \alpha_\varphi > 1$ & \multicolumn{2}{l}{\textit{(b) $\alpha_q = \alpha_\varphi = \alpha$}} \\
         & & $\Gamma(1 - 1 / \alpha)\, n^{1 / \alpha}$ & $\big((1 - \rho)^{\alpha} c_q^\alpha +  \rho^{\alpha} c_{\varphi}^\alpha\big)^{1 / \alpha} - (1 - \rho) c_{q}$ \\
         \addlinespace
         
         \multirow{2}{*}{Exponential} & $c_q, c_\varphi > 0;$ & \multirow{2}{*}{$\ln n$} & \multirow{2}{*}{$\max \Big\{ \frac{1 - \rho}{\lambda_{q}}, \frac{\rho}{\lambda_{\varphi}} \Big\} - \frac{1 - \rho}{\lambda_q}$} \\
         & $\lambda_q, \lambda_\varphi > 0$ & & \\
         \addlinespace
         
         \multirow{3}{*}{Gaussian-like} & $c_q, c_\varphi > 0;$ & \multirow{3}{*}{$\sqrt{\ln n}$} & \multirow{3}{*}{$\sqrt{\lambda_{q}^{-2} (1 - \rho)^2 + \lambda_{\varphi}^{-2} \rho^2} - \lambda_{q}^{-1} (1 - \rho)$} \\
         & $\lambda_q, \lambda_\varphi > 0;$ & & \\
         & $\beta_q, \beta_\varphi \geq 0$ & & \\
         \addlinespace
         
         \multirow{4}{*}{Bounded} & $0 \leq a_q < b_q < \infty;$ & \multirow{4}{*}{$b_\varphi - \mu_\varphi$} & \multirow{4}{*}{$\rho$} \\
         & $0 \leq a_\varphi < b_\varphi < \infty;$ & & \\
         & $\kappa_q, \kappa_\varphi > 0;$ & & \\
         & $\beta_q, \beta_\varphi \geq 1$ & & \\
         \bottomrule
    \end{tabular}
    \begin{tablenotes}\footnotesize
        \item[$\dagger$] $\underline{\alpha} \triangleq \min\{\alpha_q, \alpha_{\varphi}\}$ and $\underline{c} \triangleq c_{q} \mathbbm{1}\{\alpha_q < \alpha_{\varphi}\} + c_\varphi \mathbbm{1}\{\alpha_q > \alpha_{\varphi} \}$.
    \end{tablenotes}
\end{threeparttable}
\end{center}

\end{theorem}

Theorem~\ref{thm:uncapacitated-personalized-recos-value} quantifies the incremental value of revealing idiosyncratic fit on top of common quality.
Under {\sf Full Information}, the representative agent chooses $\max_{y\le n} Z_y$ where $Z_y \triangleq (1-\rho)\,q_y + \rho\,\varphi_{1y}$, while under {\sf Only Quality Information} the agent selects the item with maximal $q_y$ and receives expected utility $(1-\rho)\mathbb{E}[q_{(n:n)}]+\rho\mu_\varphi$.
Thus, $\Delta_{q \to u}^{\sf uncap}(n)
=
\mathbb{E}\Big[\max_{y\le n} Z_y\Big]
-
(1-\rho)\mathbb{E}[q_{(n:n)}]
-
\rho\mu_\varphi.$
The incremental value of personalization is determined by whether the {\it idiosyncratic component matters in the upper tail of $Z_y$}.

\begin{itemize}
    \item \textbf{Pareto tails (tail dominance vs.\ tail combination).}
    If $\alpha_q \neq \alpha_\varphi$, whichever primitive has the heavier tail (smaller $\alpha$) {dominates} extremes.
    When the common component is heavier-tailed ($\alpha_q < \alpha_\varphi$), $\max Z_y$ is essentially driven by $q_y$, so personalization provides no leading-order gain compared to that of the public rankings.
    When the idiosyncratic component is heavier-tailed ($\alpha_q > \alpha_\varphi$), the best match is driven by $\varphi_{1y}$ and personalization yields a first-order gain proportional to $\rho$. When $\alpha_q=\alpha_\varphi=\alpha$, {\it both} terms contribute to the tail of $Z_y$, and the incremental gain depends smoothly on $\rho$. In the symmetric case $c_q=c_\varphi=c$, the normalized gain is
    $g(\rho;\alpha) \triangleq \big((1-\rho)^\alpha+\rho^\alpha\big)^{1/\alpha}-(1-\rho).$
    Figure~\ref{subfig:rho-function-formula} plots $g(\rho;\alpha)$ and shows that as $\alpha$ grows (tails become “less heavy”), $g(\rho;\alpha)$ becomes nearly flat for $\rho<1/2$ and increases sharply for $\rho>1/2$.
    In fact, $g(\rho;\alpha)\to (2\rho-1)_+$ as $\alpha\to\infty$.

    \item \textbf{Exponential tails and a phase transition in $\rho$.}
    The exponential case is especially stark.
    When $\lambda_q=\lambda_\varphi=\lambda$, we have that
    ${\Delta_{q \to u}^{\sf uncap}(n)}/ {(\ln n/\lambda)}\ \longrightarrow\ (2\rho-1)_+.$
    Thus there is a sharp threshold at $\rho=1/2$:
    for $\rho\le 1/2$, perfect personalization yields {\it zero} leading-order gain over rankings; for $\rho>1/2$, personalization yields a strictly positive gain that increases linearly with $\rho$.
    Even though {\sf Full Information} reveals strictly more information than {\sf Only Quality Information}, that extra information can be asymptotically irrelevant because the weighted combination of the two components is dominated by the common term unless idiosyncratic term is weighted heavily enough, i.e, $\rho > 1/2$.
    Figure~\ref{fig:exponential-uncapacitated} visualizes this knife-edge behavior.

    \item \textbf{Gaussian-like tails.}
    Gaussian-like tails are lighter than exponential tails in the extreme-value sense (maxima scale as $\sqrt{\ln n}$ rather than $\ln n$), and the incremental value of personalization is correspondingly “smoother.”
    When $\lambda_q=\lambda_\varphi=\lambda$, the gain scales as
    $\Delta_{q\to u}^{\sf uncap}(n)
    \asymp
    \left(\sqrt{(1-\rho)^2+\rho^2}-(1-\rho)\right)({\sqrt{\ln n}}/{\lambda}).$
    In particular, for small $\rho$ the additional value of personalization is second-order in $\rho$ (a Taylor expansion gives $\sqrt{(1-\rho)^2+\rho^2}-(1-\rho)\approx \rho^2/(2(1-\rho))$), matching the intuition from the base model: when preferences are nearly homogeneous, rankings capture the first-order welfare gains.

    \item \textbf{Bounded tails.}
    With bounded support, extreme-value selection saturates: $\max q_y \to b_q$ and $\max Z_y \to (1-\rho)b_q+\rho b_\varphi$.
    Consequently, both rankings and personalization deliver {\it bounded} welfare gains even as $n\to\infty$:
    $\Delta_{\emptyset\to q}^{\sf uncap}(n)\to (1-\rho)(b_q-\mu_q),
    \Delta_{q\to u}^{\sf uncap}(n)\to \rho(b_\varphi-\mu_\varphi).$
\end{itemize}

\begin{figure}[!htb]
    \centering
    \begin{subfigure}{0.48\linewidth}
        \begin{tikzpicture}[scale = 0.8]
\begin{axis}[
    xlabel={Level of heterogeneity $\rho$},
    ylabel={$g(\rho;\alpha) := (\rho^\alpha + (1 - \rho)^\alpha)^{1 / \alpha} - (1 - \rho)$},
    legend style={at={(0.1,1)},anchor=north west},
    xmin=0,
    xmax=1,
    ymin=0,
    ymax=1,
    grid = major,
]

\addplot[domain=0:1, samples=100, ultra thick, blue]
    {(x^2 + (1 - x)^2)^(1/2) - (1 - x)};
\addlegendentry{$\alpha = 2$}

\addplot[domain=0:1, samples=100, ultra thick, red]
    {(x^5 + (1 - x)^5)^(1/5) - (1 - x)};
\addlegendentry{$\alpha = 5$}

\addplot[domain=0:1, samples=100, ultra thick, orange]
    {(x^10 + (1 - x)^10)^(1/10) - (1 - x)};
\addlegendentry{$\alpha = 10$}

\addplot[domain=0:0.5, samples=100, ultra thick, dashed, black] {0};
\addplot[domain=0.5:1, samples=100, ultra thick, dashed, black] {2*x - 1};
\addlegendentry{$\alpha\rightarrow\infty$}
\end{axis}
\end{tikzpicture}
\caption{$g(\rho;\alpha)$ for different $\alpha$}
\label{subfig:rho-function-formula}
    \end{subfigure}
    \begin{subfigure}{0.48\linewidth}
        \begin{tikzpicture}[scale = 0.8]
\begin{axis}[
    xlabel={Level of heterogeneity $\rho$},
    ylabel={$\Delta^{\sf uncap}_{q \to u}(n) / (c\Gamma(1 - 1/ \alpha) n^{1/\alpha})$},
    legend style={at={(0.4,1)},anchor=north},
    legend entries={$n=100$, $n=1000$, $\sqrt{\rho^2 + (1 - \rho)^2} - (1 - \rho)$},
    xmin=0,
    ymin=0,
    ymax=1,
    xmax=1,
    grid = major,
]
\addplot[orange, mark=x, ultra thick] coordinates {
    (1.0, 0.89690713)
(0.95, 0.81784664)
(0.9, 0.75306999)
(0.85, 0.6662142)
(0.8, 0.56971335)
(0.75, 0.5062449)
(0.7, 0.42701118)
(0.65, 0.37056716)
(0.6, 0.33254467)
(0.55, 0.27271786)
(0.5, 0.1952013)
(0.45, 0.15757578)
(0.4, 0.13598216)
(0.35, 0.09159228)
(0.3, 0.06525824)
(0.25, 0.04419015)
(0.2, 0.03481124)
(0.15, 0.01137209)
(0.1, 0.00792367)
(0.05, 0.00144225)
(0.0, 0.0)
};

\addplot[blue, mark=o, ultra thick] coordinates {
    (1.0, 1.00042972)
(0.95, 0.86929956)
(0.9, 0.81022888)
(0.85, 0.68339857)
(0.8, 0.6331555)
(0.75, 0.5168909)
(0.7, 0.45079333)
(0.65, 0.38950597)
(0.6, 0.3171402)
(0.55, 0.24899111)
(0.5, 0.19162077)
(0.45, 0.16015799)
(0.4, 0.11272561)
(0.35, 0.08256006)
(0.3, 0.0621062)
(0.25, 0.04186832)
(0.2, 0.0264023)
(0.15, 0.00969704)
(0.1, 0.00462515)
(0.05, 0.00148879)
(0.0, 0.0)
};

\addplot[domain=0:1, samples=100, ultra thick, red]
    {(x^2 + (1 - x)^2)^(1/2) - (1 - x)};
\end{axis}
\end{tikzpicture}
\caption{Pareto tail, $\alpha = 2$}
\label{subfig:uncap-alpha-2}
    \end{subfigure}

    \begin{subfigure}{0.48\linewidth}
          \begin{tikzpicture}[scale = 0.8]
\begin{axis}[
    xlabel={Level of heterogeneity $\rho$},
    ylabel={$\Delta^{\sf uncap}_{q \to u}(n) / (c\Gamma(1 - 1/ \alpha) n^{1/\alpha})$},
    legend style={at={(0.42,1)},anchor=north},
    legend entries={$n=1000$, $n=10000$, $n = 100000$, $(\rho^5 + (1 - \rho)^5)^{1/5} - (1 - \rho)$},
    xmin=0,
    ymin=0,
    ymax=1,
    xmax=1,
    grid = major,
]
\addplot[orange, mark=x, ultra thick] coordinates {
    (1.0, 0.732106671)
(0.95, 0.653780397)
(0.9, 0.589645711)
(0.85, 0.507109174)
(0.8, 0.441454728)
(0.75, 0.368630563)
(0.7, 0.298770984)
(0.65, 0.232391479)
(0.6, 0.175390145)
(0.55, 0.128744234)
(0.5, 0.0833586737)
(0.45, 0.0535664681)
(0.4, 0.0305106172)
(0.35, 0.0158867737)
(0.3, 0.00821244493)
(0.25, 0.00323748336)
(0.2, 0.00192144243)
(0.15, 0.000558209775)
(0.1, 0.000226684904)
(0.05, 0.0000395188678)
(0.0, 0.0)
};

\addplot[blue, mark=o, ultra thick] coordinates {
    (1.0, 0.829486761)
(0.95, 0.741074402)
(0.9, 0.663934788)
(0.85, 0.571881075)
(0.8, 0.505117732)
(0.75, 0.415396203)
(0.7, 0.333201897)
(0.65, 0.261143853)
(0.6, 0.185385155)
(0.55, 0.128780794)
(0.5, 0.0758596986)
(0.45, 0.0423597606)
(0.4, 0.0204742542)
(0.35, 0.0102602119)
(0.3, 0.00392406547)
(0.25, 0.00174112803)
(0.2, 0.000784893462)
(0.15, 0.000512100166)
(0.1, 0.00017462362)
(0.05, 0.0000174157316)
(0.0, 0.0)
};

\addplot[cyan, mark=diamond*, ultra thick] coordinates {
    (1.0, 0.914340017)
(0.95, 0.810971469)
(0.9, 0.710548974)
(0.85, 0.657970457)
(0.8, 0.554920824)
(0.75, 0.432900359)
(0.7, 0.366122161)
(0.65, 0.270214152)
(0.6, 0.201562295)
(0.55, 0.144154452)
(0.5, 0.0698379841)
(0.45, 0.0404287175)
(0.4, 0.0198307645)
(0.35, 0.00713391705)
(0.3, 0.00560248422)
(0.25, 0.00129625459)
(0.2, 0.000143689369)
(0.15, 0.000162816596)
(0.1, 0.0000279340122)
(0.05, 0.00000100104162)
(0.0, 0.0)
};

\addplot[domain=0:1, samples=100, ultra thick, red]
    {(x^5 + (1 - x)^5)^(1/5) - (1 - x)};
\end{axis}
\end{tikzpicture}
\caption{Pareto tail, $\alpha = 5$}
\label{fig:uncap-alpha-5}
    \end{subfigure}
\begin{subfigure}{0.48\linewidth}
        \begin{tikzpicture}[scale = 0.8]
            \begin{axis}[
                xlabel={Level of heterogeneity $\rho$},
                ylabel={${\Delta_{q \to u}^{\textsf{uncap}}}(n) / \ln n $},
                legend style={at={(0.25,1)},anchor=north},
                legend entries={$n=100$, $(2\rho - 1)_+$},
                xmin=0,
                ymin=0,
                ymax=1,
                xmax=1,
                grid = major,
                axis line style={thick}
            ]
\draw[dashed, ultra thick] (axis cs:0.5,0) -- (axis cs:0.5,1);

\addplot[blue, mark=*, ultra thick] coordinates {
    (1.0, 0.999453406)
(0.95, 0.900112092)
(0.9, 0.80271497)
(0.85, 0.701116976)
(0.8, 0.607385277)
(0.75, 0.510245449)
(0.7, 0.421841598)
(0.65, 0.334830968)
(0.6, 0.255098849)
(0.55, 0.178359727)
(0.5, 0.122692528)
(0.45, 0.0819329963)
(0.4, 0.0516672957)
(0.35, 0.0298902478)
(0.3, 0.0189630175)
(0.25, 0.010985955)
(0.2, 0.00613571737)
(0.15, 0.00298504775)
(0.1, 0.00127394897)
(0.05, 0.000305261687)
(0.0, 0.0)
};

\addplot[red, ultra thick] coordinates {
    (1.0, 1.0)
(0.95, 0.9)
(0.9, 0.8)
(0.85, 0.7)
(0.8, 0.6)
(0.75, 0.5)
(0.7, 0.4)
(0.65, 0.3)
(0.6, 0.2)
(0.55, 0.1)
(0.5, 0.0)
(0.45, 0.0)
(0.4, 0.0)
(0.35, 0.0)
(0.3, 0.0)
(0.25, 0.0)
(0.2, 0.0)
(0.15, 0.0)
(0.1, 0.0)
(0.05, 0.0)
(0.0, 0.0)
};

\end{axis}
\end{tikzpicture}
\caption{Exponential tail, $\lambda_q=\lambda_\varphi=1$}
\label{fig:exponential-uncapacitated}
    \end{subfigure}

  \caption{Value of personalized recommendations in the uncapacitated setting under different tail distributions. In (a) we plot the function $g(\rho; \alpha)$ as a function of $\rho$ for different values of $\alpha$. In (b-d), we compare the numerical simulation results for different values of $n$ against the theoretical asymptotic scaling results (in red). For (b) we consider the Pareto distribution with $c_\varphi = 1, \alpha_\varphi = 2$. For (c) we consider the Pareto distribution with $c_\varphi = 1, \alpha_\varphi = 5$. For (d) we consider the exponential distribution with $c_q = c_\varphi = \lambda_q = \lambda_\varphi = 1$.
  }
\label{fig:uncapacitated-personalized-recos-value}
\end{figure}


\subsubsection{Capacitated Supply Setting}
\label{subsec:welfare-implications-capacitated}

Recall that in the capacitated supply setting, we have $n$ agents and $n$ items with unit demand and unit capacity, and the market clears via a one-to-one matching.

\begin{theorem}[Value of public rankings in the capacitated setting]
    \label{thm:capacitated-ranking-value}
    Consider the capacitated supply setting.
    Assume $(q_y)$ and $(\varphi_{xy})$ are drawn i.i.d.\ from distributions $P_q$ and $P_\varphi$ with finite means $\mu_q,\mu_\varphi$ and are independent.
    Then the value of rankings is zero, i.e., 
    $\Delta_{\emptyset \to q}^{\sf cap}(n) = 0$.
\end{theorem}

Theorem~\ref{thm:capacitated-ranking-value} extends the ``conservation of common value'' logic from Section~\ref{sec:base-model-results} beyond Gaussians:
under unit capacities, the total common-value contribution is fixed by the multiset $\{q_y\}_{y\le n}$, so revealing $q_y$ can only reshuffle which agents obtain which qualities.
Because $\varphi_{xy}$ is independent of $q_y$, the idiosyncratic term averages to $\mu_\varphi$ regardless of how items are reassigned.
Hence public rankings have {\it no aggregate agent welfare effect}.
Figure~\ref{fig:value-public-ranking-capacitated} confirms this numerically: $\Delta_{\emptyset\to q}^{\sf cap}(n)$ concentrates near zero for all $\rho$.

\begin{figure}[!htbp]
    \centering
    \begin{subfigure}{0.32\linewidth}
        \centering
        \begin{tikzpicture}[scale=0.6]
            \begin{axis}[
                xlabel={Level of heterogeneity $\rho$},
                ylabel={${\Delta_{\emptyset \to q}^{\textsf{cap}}}(n)$},
                legend style={at={(0.7,1)},anchor=north},
                legend entries={$n=100$, $0$},
                xmin=0,
                xmax=1,
                ymin=-0.2,
                ymax=0.4,
                grid=major,
                axis line style={thick}
            ]
            \addplot[blue, mark=*, ultra thick] coordinates {
                (0.0, -0.00530766) (0.05, -0.00538989) (0.1, 0.01463514) (0.15, 0.00370928)
                (0.2, -0.00302006) (0.25, 0.00808066) (0.3, 0.00173158) (0.35, -0.00485731)
                (0.4, 0.00817953) (0.45, 0.00295046) (0.5, 0.0103502) (0.55, 0.00822897)
                (0.6, -0.00614228) (0.65, -0.01009817) (0.7, -0.00805731) (0.75, 0.00298706)
                (0.8, -0.00340842) (0.85, 0.00109466) (0.9, 0.01330453) (0.95, 0.00063326)
                (1.0, -0.02474926)
            };
            \addplot[red, ultra thick] coordinates{ (0.0, 0.0) (1.0,0.0) };
            \end{axis}
        \end{tikzpicture}
        \caption{Pareto tail, $\alpha_q = 2$}
        \label{subfig:capacitated-ranking-value-pareto-tail}
    \end{subfigure}
    \begin{subfigure}{0.32\linewidth}
        \centering
        \begin{tikzpicture}[scale=0.6]
            \begin{axis}[
                xlabel={Level of heterogeneity $\rho$},
                ylabel={ ${\Delta_{\emptyset \to q}^{\textsf{cap}}}(n)$},
                legend style={at={(0.7,1)},anchor=north},
                legend entries={$n=100$, $0$},
                xmin=0,
                xmax=1,
                ymin=-0.2,
                ymax=0.4,
                grid=major,
                axis line style={thick}
            ]
            \addplot[blue, mark=*, ultra thick] coordinates {
                (0.0, 0.00020165) (0.05, 0.00379704) (0.1, 0.00210067) (0.15, -0.0020132)
                (0.2, -0.00080243) (0.25, 0.00064667) (0.3, -0.0012597) (0.35, -0.00898513)
                (0.4, 0.00085106) (0.45, 0.00096597) (0.5, -0.00390486) (0.55, -0.00166136)
                (0.6, 0.00095535) (0.65, -0.00478673) (0.7, -0.00159674) (0.75, 0.00269692)
                (0.8, 0.00417145) (0.85, 0.00577392) (0.9, -0.00128173) (0.95, 0.0010885)
                (1.0, 0.00139009)
            };
            \addplot[red, ultra thick] coordinates{ (0.0, 0.0) (1.0,0.0) };
            \end{axis}
        \end{tikzpicture}
        \caption{Exponential tail, $\lambda_q = 1$}
        \label{subfig:capacitated-ranking-value-exponential-tail}
    \end{subfigure}
    \begin{subfigure}[b]{0.32\textwidth}
        \centering
        \begin{tikzpicture}[scale=0.6]
            \begin{axis}[
                xlabel={Level of heterogeneity $\rho$},
                ylabel={$\Delta_{\emptyset \to q}^{\textsf{cap}}(n)$},
                legend style={at={(0.7,1)},anchor=north},
                legend entries={$n=100$, $0$},
                xmin=0,
                xmax=1,
                ymin=-0.2,
                ymax=0.4,
                grid=major,
                axis line style={thick}
            ]
\addplot[blue, mark=*, ultra thick] coordinates {
    (0.0, 0.0021)
    (0.05, 0.0001)
    (0.1, -0.0013)
    (0.15, 0.0023)
    (0.2, -0.0008)
    (0.25, 0.0012)
    (0.3, -0.0017)
    (0.35, 0.0005)
    (0.4, -0.0001)
    (0.45, 0.0018)
    (0.5, -0.0012)
    (0.55, 0.0000)
    (0.6, 0.0011)
    (0.65, -0.0005)
    (0.7, -0.0022)
    (0.75, 0.0030)
    (0.8, 0.0010)
    (0.85, -0.0004)
    (0.9, 0.0008)
    (0.95, 0.0013)
    (1.0, -0.0007)
};
\addplot[red, ultra thick] coordinates { (0.0, 0.0) (1.0, 0.0) };
\end{axis}
\end{tikzpicture}
\caption{Bounded tail, $\beta_q = 1$}
\label{fig:uniform-capacitated-a}
\end{subfigure}
    \caption{Value of public rankings in the capacitated setting under different tail distributions. We compare the numerical simulation for $n = 100$ (in blue) with zero (in red). We consider (a) Pareto distribution with $c_q = 1, \alpha_q = 2$, (b) Exponential distribution with $c_q = \lambda_q = 1$ and (c) Uniform distribution over $[0,1]$.}
    \label{fig:value-public-ranking-capacitated}
\end{figure}

\begin{theorem}[Value of personalized recos in the capacitated setting]
    \label{thm:capacitated-personalized-recos-value}
    Consider the capacitated supply setting.
    Assume $(q_y)$ and $(\varphi_{xy})$ are drawn i.i.d.\ from distributions $P_q$ and $P_\varphi$ with non-negative support and finite means $\mu_q,\mu_\varphi$.
    The value of personalized recommendations, $\Delta_{q \to u}^{\sf cap}(n)$, scales as $\lim_{n \to \infty} \Delta_{q \to u}^{\sf cap}(n) / \ell_n = \rho$ for all $\rho \in [0,1]$, where $\ell_n$ depends on the parameters of $P_\varphi$. \\
    \begin{center}
\begin{threeparttable}
    \begin{tabular}{ccc}
         \toprule 
         $P_\varphi$ tail behavior & Parameters & $\ell_n$ \\
         \midrule
         Pareto & $c_\varphi > 0, \alpha_\varphi > 1$ & $C_\varphi\, n^{1 / \alpha_\varphi}$\tnote{$\dagger$} \\
         \addlinespace
         
         Exponential & $c_\varphi > 0, \lambda_\varphi > 0$ & $\ln n / \lambda_\varphi$ \\
         \addlinespace
         
         Gaussian-like & $c_\varphi > 0, \lambda_\varphi > 0, \beta_\varphi \geq 0$ & $\sqrt{\ln n} / \lambda_\varphi$ \\
         \addlinespace
         
         Bounded & $0 \leq a_\varphi < b_\varphi < \infty, \kappa_\varphi > 0, \beta_\varphi \geq 1$ & $b_\varphi - \mu_\varphi$ \\
         \bottomrule
    \end{tabular}
    \begin{tablenotes}\footnotesize
        \item[$\dagger$] $C_\varphi \triangleq c_\varphi \big(\alpha_\varphi / (\alpha_\varphi + 1)\big)\Gamma(1 - 1 / \alpha_\varphi)$.
    \end{tablenotes}
\end{threeparttable}
\end{center}
    



\end{theorem}

\begin{figure}
\centering
\begin{subfigure}{0.32\linewidth}
        \centering
        \begin{tikzpicture}[scale=0.6]
            \begin{axis}[
                xlabel={\large Level of heterogeneity $\rho$},
                ylabel={\large ${\Delta_{q \to u}^{\textsf{cap}}}(n) / \big(C_{\varphi} n^{1 / \alpha_{\varphi}}\big)$},
                legend style={at={(0.3,1)},anchor=north},
                legend entries={$n=100$, $\rho$},
                xmin=0,
                xmax=1,
                ymin=0,
                ymax=1,
                grid=major,
                axis line style={thick}
            ]
            \addplot[blue, mark=*, ultra thick] coordinates {
                (1.0, 1.00930667)
                (0.95, 0.948535206)
                (0.9, 0.897736643)
                (0.85, 0.853079244)
                (0.8, 0.798466122)
                (0.75, 0.749827047)
                (0.7, 0.698130959)
                (0.65, 0.657011287)
                (0.6, 0.596395425)
                (0.55, 0.542547577)
                (0.5, 0.496784759)
                (0.45, 0.444685165)
                (0.4, 0.389544604)
                (0.35, 0.338717507)
                (0.3, 0.287731953)
                (0.25, 0.237352512)
                (0.2, 0.186049693)
                (0.15, 0.135002175)
                (0.1, 0.0836456996)
                (0.05, 0.037504978)
                (0.0, -0.000111205485)
            };
            \addplot[red, ultra thick] coordinates {
                (1.0, 1.0) (0.95, 0.95) (0.9, 0.9) (0.85, 0.85) (0.8, 0.8)
                (0.75, 0.75) (0.7, 0.7) (0.65, 0.65) (0.6, 0.6) (0.55, 0.55)
                (0.5, 0.5) (0.45, 0.45) (0.4, 0.4) (0.35, 0.35) (0.3, 0.3)
                (0.25, 0.25) (0.2, 0.2) (0.15, 0.15) (0.1, 0.1) (0.05, 0.05) (0.0, 0.0)
            };
            \end{axis}
        \end{tikzpicture}
        \caption{Pareto tail, $\alpha_\varphi = 2$}
        \label{subfig:capacitated-pareto-tail-b}
    \end{subfigure}
    \begin{subfigure}{0.32\linewidth}
        \centering
        \begin{tikzpicture}[scale=0.6]
            \begin{axis}[
                xlabel={\large Level of heterogeneity $\rho$},
                ylabel={\large ${\Delta_{q \to u}^{\textsf{cap}}}(n) / \ln n$},
                legend style={at={(0.3,1)},anchor=north},
                legend entries={$n=100$, $\rho$},
                xmin=0,
                xmax=1,
                ymin=0,
                ymax=1,
                grid=major,
                axis line style={thick}
            ]
            \addplot[blue, mark=*, ultra thick] coordinates {
                (1.0, 1.00832611) (0.95, 0.956725549) (0.9, 0.906618072) (0.85, 0.854407927)
                (0.8, 0.801579375) (0.75, 0.746531) (0.7, 0.693303044) (0.65, 0.639508339)
                (0.6, 0.58352241) (0.55, 0.528969294) (0.5, 0.472651566) (0.45, 0.417431005)
                (0.4, 0.36082915) (0.35, 0.307180421) (0.3, 0.252822429) (0.25, 0.200015922)
                (0.2, 0.148832465) (0.15, 0.101418562) (0.1, 0.0608998085) (0.05, 0.0229079904)
                (0.0, -0.000711421131)
            };
            \addplot[red, ultra thick] coordinates{ (0.0, 0.0) (1.0,1.0) };
            \end{axis}
        \end{tikzpicture}
        \caption{Exponential tail, $\lambda_\varphi = 1$}
        \label{subfig:capacitated-personalized-recos-value-exponential-tail}
    \end{subfigure}
    \begin{subfigure}[b]{0.32\textwidth}
        \centering
        \begin{tikzpicture}[scale=0.6]
        \begin{axis}[
            xlabel={\large Level of heterogeneity $\rho$},
            ylabel={\large $\Delta^{\sf cap}_{q \to u}(n) / 0.5$},
            grid=major,
            legend style={at={(0.35,1)},anchor=north},
            legend entries={$n=100$, $\rho$},
            ymin=0,
            xmin=0,
            ymax=1,
            xmax=1
        ]
        \addplot[blue, mark=*, ultra thick] coordinates {
            (0.0,   0.000000000)
            (0.05,  0.033498226)
            (0.1,   0.076438211)
            (0.15,  0.121279743)
            (0.2,   0.167324406)
            (0.25,  0.214156504)
            (0.3,   0.261660482)
            (0.35,  0.309745176)
            (0.4,   0.357866572)
            (0.45,  0.406549282)
            (0.5,   0.455158228)
            (0.55,  0.505250982)
            (0.6,   0.554638322)
            (0.65,  0.604745540)
            (0.7,   0.655756444)
            (0.75,  0.705706744)
            (0.8,   0.756912138)
            (0.85,  0.807436216)
            (0.9,   0.862107556)
            (0.95,  0.916591700)
            (1.0,   0.971462736)
        };
        \addplot[red, ultra thick, domain=0:1] {x};
        \end{axis}
        \end{tikzpicture}
        \caption{Bounded tail, $\beta_\varphi = 1$}
    \end{subfigure}
    \caption{Value of personalized recommendations in the capacitated supply setting. We compare the numerical simulation for $n = 100$ (in blue) with the theoretical asymptotic scaling (in red). We consider (a) Pareto distribution with $c_\varphi = 1, \alpha_\varphi = 2$, (b) Exponential distribution with $c_\varphi = \lambda_\varphi = 1$ and (c) Uniform distribution over $[0,1]$.}
    \label{fig:capacitated-personalized-recos-value}
\end{figure}

In capacitated supply setting, aggregate welfare gains come entirely from the ability to improve {\it idiosyncratic fit} under scarcity.
Theorem~\ref{thm:capacitated-personalized-recos-value} shows that the gain is linear in $\rho$ and, to leading order, depends only on the tail of $P_\varphi$:
polynomial growth under Pareto tails, logarithmic under exponential tails, $\sqrt{\ln n}$ under Gaussian-like tails, and saturation under bounded support.
Figure~\ref{fig:capacitated-personalized-recos-value} illustrates this linear-in-$\rho$ structure across tail families (after the appropriate normalization).
Taken together with Theorem~\ref{thm:capacitated-ranking-value}, these results reinforce the core insight from the Gaussian base model in a much broader distributional environment: under capacity constraints, personalization can unlock large aggregate value even when rankings—no matter how accurate—cannot.

\subsection{Value of information provisioning tools with supply-side priorities}
\label{sec:supply-side-priority}

In the base model (Section~\ref{sec:base-model}), items were effectively indifferent across agents beyond a common priority order, so the sequential selection of agents (serial dictatorship on the demand side) yielded a {\it stable matching}. We now extend the model to allow items to have {\it own-group priorities} over agents, motivated by neighborhood priorities in school choice.

We focus on two groups for analytical tractability. 
With three or more groups (and/or richer priority structures), stable matchings remain computable (e.g., via deferred acceptance), but obtaining sharp, interpretable {welfare} characterizations becomes substantially more challenging because multiple interacting ``priority cutoffs'' endogenously determine which items are effectively available to which agents. We view the two-group model as the minimal extension that captures the core economics of supply-side priorities while keeping the analysis tractable.

\subsubsection{Setup}
\label{subsec:horiz-setup}
We remain in the balanced case with $n$ agents and $n$ items. There are two groups $\mathcal G=\{1,2\}$. Each agent $x\in\mathcal X:=\{1,\dots,n\}$ has a group label $g_{\mathcal X}(x)\in\mathcal G$, and each item $y\in\mathcal Y:=\{1,\dots,n\}$ has a group label $g_{\mathcal Y}(y)\in\mathcal G$. Let
$\mathcal G_{\mathcal X}^{g}:=\{x\in\mathcal X:\ g_{\mathcal X}(x)=g\}, \mathcal G_{\mathcal Y}^{g}:=\{y\in\mathcal Y:\ g_{\mathcal Y}(y)=g\},\forall g\in\{1,2\}.$
Let $\gamma_X\in(0,1)$ be the fraction of agents in group~1 and $\gamma_Y \in(0,1)$ the fraction of items in group~1:
$\lvert \mathcal G_{\mathcal X}^{1}\rvert=\gamma_X n$ and $\lvert \mathcal G_{\mathcal Y}^{1}\rvert=\gamma_Y n$ (hence $\lvert \mathcal G_{\mathcal X}^{2}\rvert=(1-\gamma_X)n$ and $\lvert \mathcal G_{\mathcal Y}^{2}\rvert=(1-\gamma_Y)n$).
Without loss of generality we assume $\gamma_X \le \gamma_Y$; the case $\gamma_X >\gamma_Y$ is symmetric.

\paragraph{Preferences.}
Agents' underlying true utilities for items are exactly the same as in the base model: $u_{xy} = (1-\rho)q_y + \rho\varphi_{xy}$. That is, their intrinsic preferences depend only on item quality and idiosyncratic fit, not on the items' group labels. Agents' strict preferences over items at the time of choice are induced by the information regime:
(i) {\sf No Information}: agents observe no information and are otherwise indifferent ex ante (ties broken by a fixed common rule);
(ii) {\sf Only Quality}: agents strictly rank items by quality $q_y$ (remaining ties broken by the fixed rule);
(iii) {\sf Full Information}: agents strictly rank items by realized utilities $u_{xy}$ (which are distinct almost surely under continuous primitives).

Items' preferences over agents follow a {\it horizontal priority} rule: there is a strict, {\it common} priority order $\pi$ over agents, and each item $y$ {\it lexicographically} ranks agents by
\[
\text{(own group first)}\qquad
\big\{x:\ g_{\mathcal X}(x)=g_{\mathcal Y}(y)\big\}\ \succ_y\ \big\{x:\ g_{\mathcal X}(x)\ne g_{\mathcal Y}(y)\big\},
\]
and {\it within each group} by the same strict order $\pi$ (lower $\pi$ means higher priority, consistent with Algorithm~\ref{alg:gpsd}).
Thus, for $y\in\mathcal G_{\mathcal Y}^{1}$, all group‑1 agents are preferred to all group‑2 agents, and among group‑1 (resp.\ group‑2) agents the ranking is by~$\pi$; similarly for $y\in\mathcal G_{\mathcal Y}^{2}$.

\paragraph{Stable matching.}
A matching $\mu:\mathcal X\to\mathcal Y$ is {\it stable} if there is no {\it blocking pair} $(x,y)\notin \mu$ such that (i) agent $x$ strictly prefers $y$ to $\mu(x)$ and (ii) item $y$ strictly prefers $x$ to $\mu^{-1}(y)$, with strictness evaluated in the regime‑appropriate preferences. Since $n=n$ and preferences are complete (all matched partners are acceptable), all agents and items are matched.

\subsubsection{Group-Prioritized Serial Dictatorship (GPSD): selecting a stable matching}
\label{subsec:horiz-selection}
Because stable matchings may not be unique under heterogeneous item priorities, we fix one selection rule to make welfare comparable across regimes. In particular, we use Group-Prioritized Serial Dictatorship (GPSD), specified in Algorithm~\ref{alg:gpsd}. By construction, GPSD is well‑defined and yields a matching $\mu^{\sf gpsd}$ for every profile of agent preferences. 

\begin{algorithm}[!htb]
\caption{\textup{\textsf{GPSD}}$(\gamma_{\mathcal X}\!\le\!\gamma_{\mathcal Y})$: Group-Prioritized Serial Dictatorship\label{alg:gpsd}}
\SetAlgoLined \DontPrintSemicolon
\KwIn{$n$ agents $\mathcal{X}=\{1,\dots,n\}$, $n$ items $\mathcal{Y}=\{1,\dots,n\}$; group labels $g_{\mathcal X}(x)\!\in\!\{1,2\}$ and $g_{\mathcal Y}(y)\!\in\!\{1,2\}$; fractions $\gamma_{\mathcal X}\!=\!|\{x:\,g_{\mathcal X}(x)=1\}|/n$, $\gamma_{\mathcal Y}\!=\!|\{y:\,g_{\mathcal Y}(y)=1\}|/n$ with $\gamma_{\mathcal X}\!\le\!\gamma_{\mathcal Y}$; strict priority order $\pi$ over agents (lower $\pi$ $\,\Rightarrow$ higher priority); regime-specific oracle $\textsc{AgentPref}(x,S)$ returning $x$’s most-preferred item in $S$ (ties broken by a fixed rule).}
\KwOut{A matching $\mu:\mathcal X\to\mathcal Y$ (every agent matched to exactly one item).}

\nl $\mathcal G_{\mathcal X}^{1}\!\gets\!\{x\in\mathcal X: g_{\mathcal X}(x)=1\}$,\quad
   $\mathcal G_{\mathcal X}^{2}\!\gets\!\{x\in\mathcal X: g_{\mathcal X}(x)=2\}$\;
\nl $\mathcal G_{\mathcal Y}^{1}\!\gets\!\{y\in\mathcal Y: g_{\mathcal Y}(y)=1\}$,\quad
   $\mathcal G_{\mathcal Y}^{2}\!\gets\!\{y\in\mathcal Y: g_{\mathcal Y}(y)=2\}$\;
\nl $\mu \gets \emptyset$; $\mathcal Y_{\text{avail}} \gets \mathcal Y$\;

\nl \textbf{Phase 1 (Group 1 agents $\to$ Group 1 items)}\;
\For{$x \in \mathcal G_{\mathcal X}^{1}$ in order of $\pi$}{
  \nl $S \gets \mathcal G_{\mathcal Y}^{1}\cap \mathcal Y_{\text{avail}}$ \textcolor{blue}{\tcp*{\footnotesize{restrict to own-group items}}}
  \nl $y^\star \gets \textsc{AgentPref}(x,S)$  \textcolor{blue}{\tcp*{\footnotesize{regime-specific top choice}}}
  \nl $\mu(x)\gets y^\star$;\; $\mathcal Y_{\text{avail}} \gets \mathcal Y_{\text{avail}}\setminus\{y^\star\}$\;
}

\nl \textbf{Phase 2 (Group 2 agents $\to$ remaining items)}\;
\For{$x \in \mathcal G_{\mathcal X}^{2}$ in order of $\pi$}{
  \nl $S \gets \mathcal Y_{\text{avail}}$ \textcolor{blue}{\tcp*{\footnotesize{may include leftover group-1 items}}}
  \nl $y^\star \gets \textsc{AgentPref}(x,S)$ \textcolor{blue}{\tcp*{\footnotesize{regime-specific top choice}}}
  \nl $\mu(x)\gets y^\star$;\; $\mathcal Y_{\text{avail}} \gets \mathcal Y_{\text{avail}}\setminus\{y^\star\}$\;
}

\nl \KwRet $\mu$\;
\end{algorithm}

Crucially, because of the rigid lexicographic structure of the items' priorities, GPSD always selects a matching that is stable, regardless of what the agents' preferences actually are.

\begin{proposition}[GPSD yields a stable matching for arbitrary strict agent preferences]
\label{prop:gpsd-stable-arbitrary}
Assume a balanced market ($|\mathcal X|=|\mathcal Y|=n$) with two groups, and suppose $\gamma_{X}\le \gamma_{Y}$.
Each item $y$ has a strict lexicographic preference over agents: all agents with $g_{\mathcal X}(x)=g_{\mathcal Y}(y)$ are strictly preferred to all agents with $g_{\mathcal X}(x)\neq g_{\mathcal Y}(y)$; within each group, the order is a common strict priority $\pi$ on $\mathcal X$. Then, for \emph{any arbitrary profile of strict agent preferences} over items, the matching $\mu^{\sf gpsd}$ produced by Algorithm~\ref{alg:gpsd} is stable. 
\end{proposition}

To avoid comparing different stable selections across regimes, we adopt GPSD as a {\it common selection rule}. The stability of our specific information regimes follows immediately as a direct corollary of Proposition \ref{prop:gpsd-stable-arbitrary}.

\begin{corollary}[GPSD is stable across information regimes]
\label{cor:gpsd-stable-regimes}
Because the continuous primitives and tie-breaking rules induce strict agent preference orderings in the {\sf No Information}, {\sf Only Quality}, and {\sf Full Information} regimes, $\mu^{\sf gpsd}$ constitutes a stable matching in all three regimes.
\end{corollary}

All welfare statistics we report for this extension are computed on the resulting $\mu^{\sf gpsd}$ in each regime. This isolates how information changes agents’ preferences while keeping the equilibrium notion (stability) and the selection rule fixed, ensuring apples‑to‑apples comparisons.

\subsubsection{Value of information provisioning tools}

\begin{theorem}[Value of public rankings with supply-side priorities]
\label{thm:value-public-rankings-supply-side-priority}
Consider the capacitated setting with supply-side priorities and evaluate outcomes on $\mu^{\sf gpsd}$.
Assume $(q_y)$ and $(\varphi_{xy})$ are i.i.d.\ from distributions $P_q$ and $P_\varphi$, independent of each other, with finite means $\mu_q$ and $\mu_\varphi$.
Then the value of public rankings as measured by
$\Delta_{\emptyset \to q}^{\sf cap}(n)$
is zero for all $n$, i.e.,
$\Delta_{\emptyset \to q}^{\sf cap}(n)=0$.
\end{theorem}

Theorem~\ref{thm:value-public-rankings-supply-side-priority} extends the base-model ``common term conservation'' logic to settings with horizontal (own-group) priorities. Even though priorities segment the market and restrict which agents can credibly obtain which items, public rankings still only reveal the {\it common} component $q_y$. Under unit capacity and a balanced market, any matching uses each item exactly once, so the aggregate common value is fixed. Since the idiosyncratic terms are independent of $q_y$ and not revealed under {\sf Only Quality}, the expected idiosyncratic contribution on matched edges remains $\mu_\varphi$.
Hence, rankings can substantially reshuffle {\it who gets which quality}---and thus can have important {\it distributional} and {\it equity} consequences across priority classes and groups---but they do not raise {\it average} welfare in this benchmark.

\begin{theorem}[Value of personalized recos with supply-side priorities]
\label{thm:value-personalized-recos-supply-side-priority}
Consider the capacitated setting with supply-side priorities, and evaluate outcomes on $\mu^{\sf gpsd}$.
Assume $P_q$ and $P_\varphi$ have non-negative support and finite means.
The value of personalized recommendations, measured by
$\Delta_{q \to u}^{\sf cap}(n):={\sf AW}^{\sf cap}_u(n)-{\sf AW}^{\sf cap}_q(n)$ scales as $\liminf_{n \to \infty} \Delta_{q \to u}^{\sf cap}(n)  / \ell_n \geq \rho$ where $\ell_n$ depends on the tail behavior of $P_\varphi$. \\
\begin{center}
\begin{threeparttable}
    \begin{tabular}{ccc}
         \toprule 
         $P_\varphi$ tail behavior & Parameters & $\ell_n$ \\
         \midrule
         Pareto & $c_\varphi > 0, \alpha_\varphi > 1$ & $C_\varphi (\gamma_X^{1 + 1 / \alpha_\varphi} + (1 - \gamma_X)^{1 + 1 / \alpha_\varphi})\, n^{1 / \alpha_\varphi}$\tnote{$\dagger$} \\
         \addlinespace
         
         Exponential & $c_\varphi > 0, \lambda_\varphi > 0$ & $\ln n / \lambda_\varphi$ \\
         \addlinespace
         
         Gaussian-like & $c_\varphi > 0, \lambda_\varphi > 0, \beta_\varphi \geq 0$ & $\sqrt{\ln n} / \lambda_\varphi$ \\
         \bottomrule
    \end{tabular}
    \begin{tablenotes}\footnotesize
        \item[$\dagger$] $C_\varphi \triangleq c_\varphi \big(\alpha_\varphi / (\alpha_\varphi + 1)\big)\Gamma(1 - 1 / \alpha_\varphi)$.
    \end{tablenotes}
\end{threeparttable}
\end{center}



\end{theorem}

Theorem~\ref{thm:value-personalized-recos-supply-side-priority} reinforces the main message of Sections~\ref{sec:base-model}--\ref{sec:base-model-results}: under capacity constraints, the primary welfare leverage comes from {\it individualized} (idiosyncratic) information rather than additional refinement of the common ranking.
This continues to be true even when the supply side has heterogeneous priorities that segment the market.
Intuitively, neighborhood priorities carve the market into ``priority-respecting'' submarkets; within each such submarket, full information allows agents to better sort on idiosyncratic fit and reduces correlated demand. Crucially, the lower bounds above formally aggregate the welfare improvements for both group-1 and group-2 agents, as each group optimizes over a sufficiently large remaining set of fresh items, allowing the aggregate welfare gains to maintain the powerful extreme-value growth rates established in Section \ref{sec:welfare-gains-tail-distributions}.

\subsection{Supply-side pricing under different information regimes}
\label{subsec:supply-side-pricing}

We study a sequential posted-price market in which sellers post prices ex ante and agents (or buyers) arrive sequentially. For simplicity, throughout this section, we assume $P_q=P_\varphi=\mathrm{Unif}([0,1])$.

\subsubsection{Setup}
\label{subsec:model-supply-side-pricing}
In the base model (described in Section \ref{sec:base-model}), for the capacitated setting, we focused mainly on the balanced market setting where the number of items and agents are equal. In this extension, we allow for imbalance in the number of agents and items.
There are $m$ agents $\mathcal X=\{1,\dots,m\}$ and $n$ sellers/items $\mathcal Y=\{1,\dots,n\}$, each item with unit capacity and zero cost.
Seller $y$ posts a price $p_y\in[0,1]$ before any agent arrives; let $\mathbf p=(p_y)_{y\in\mathcal Y}$.
Fix $\rho\in[0,1]$.
Agent $x$'s realized utility from item $y$ at price $p_y$ is
\begin{equation}
\label{eq:u-general}
u_{xy}=(1-\rho)q_y+\rho\,\varphi_{xy}-p_y,
\end{equation}
and the outside option yields utility $0$.

\paragraph{Arrival, observation, choice.}
Agents arrive in the fixed order $1,2,\dots,m$.
In regime ${\sf r} \in\{\emptyset,q,u\}$, the arriving agent observes:
\begin{itemize}[leftmargin=1.5em]
\item {\sf No Information} (${\sf r}=\emptyset$): prices $\mathbf p$ only; the agent maximizes $\mathbb{E}[u_{xy}\mid \mathbf p]$.
\item {\sf Only Quality} (${\sf r}=q$): prices and common term $(\mathbf p,\mathbf q)$; the agent maximizes $\mathbb{E}[u_{xy}\mid \mathbf{p},\mathbf{q}]$.
\item {\sf Full Information} (${\sf r}=u$): prices, and both the common and the idiosyncratic components $(\varphi_{xy})_{y\in \mathcal Y_{\mathrm{avail}}}$; the buyer maximizes realized $u_{xy}$.
\end{itemize}
An agent purchases one available item iff their maximal (expected) utility is nonnegative.

\paragraph{Tie-breaking.}
In the {\sf No Information} regime, ties are broken randomly and in favor of trade. In the {\sf Only Quality} and {\sf Full Information} regime, among expected-utility maximizers, the agent chooses the higher-quality item; remaining ties are uniform. This refinement yields a sharp pure-strategy equilibrium characterization under excess supply.

\paragraph{Agent Welfare, Seller Revenue and Social Welfare.}
Let $\pi_{\sf r}(k)\in\mathcal Y\cup\{0\}$ be the choice of agent $k$ in regime ${\sf r}$ ($0$ denotes the outside option). Let ${\sf AW}_{{\sf r}}$ denote the (average) agent welfare, ${\sf SR}_{{\sf r}}$ denote the (average) seller revenue, and ${\sf SW}_{{\sf r}}$ denote the (average) social welfare.  In particular, we have that
\begin{align}
\label{eq:AW-SR-SW-general}
{\sf AW}_{\sf r}(m,n)
&:= m^{-1}\,\mathbb{E}\!\Big[\sum_{k=1}^{m} u_{k,\pi_{\sf r}(k)}\Big],\\
{\sf SR}_{\sf r}(m,n)
&:= n^{-1}\,\mathbb{E}\!\Big[\sum_{k=1}^{m} p_{\pi_{\sf r}(k)}\,\mathbbm{1}\{\pi_{\sf r}(k)\in\mathcal Y\}\Big],\\
{\sf SW}_{\sf r}(m,n)
&:= {\sf AW}_{\sf r}(m,n)+{\sf SR}_{\sf r}(m,n).
\end{align}
Note that since the focus of this section is on the capacitated setting, we drop ${\sf cap}$ from the notation.

\paragraph{Equilibrium Notion.} Fix a regime ${\sf r}\in\{\emptyset,q,u\}$ and condition on realized qualities $\mathbf q=(q_y)_{y=1}^n$.
Let $\Pi_{y,{\sf r}}(\mathbf p\mid\mathbf q)$ be the probability that seller $y$ sells in regime ${\sf r}$ under $\mathbf p$.
Let $v_{y,{\sf r}}(\mathbf p\mid\mathbf q):=p_y\,\Pi_{y,{\sf r}}(\mathbf p\mid\mathbf q)$ be seller $y$'s interim expected revenue.

\begin{definition}[Interim $\varepsilon$--approximate price equilibrium]
\label{def:eps-eq-general}
A price vector $\mathbf p\in[0,1]^n$ is an interim $\varepsilon$--approximate equilibrium in regime ${\sf r}$ if for a.e.\ $\mathbf q$,
every seller $y$ and every deviation $p'_y\in[0,1]$ satisfy
\[
v_{y,{\sf r}}(p'_y;\mathbf p_{-y}\mid\mathbf q)\ \le\ v_{y,{\sf r}}(\mathbf p\mid\mathbf q)\ +\ \varepsilon.
\]
For $\varepsilon=0$ this is an interim Nash equilibrium.
\end{definition}

\subsubsection{Equilibrium Outcomes and the Value of Information Provisioning Tools}
\label{subsec:equilibrium-and-value}

To formally quantify the impact of information, let $\gamma_n:=\frac{m}{n},$ and when asymptotics are taken along a sequence with $\gamma_n\to\gamma$, we refer to $\gamma$ as the market-imbalance ratio. Propositions \ref{prop:NI-general}, \ref{prop:QI-general}, and \ref{prop:FI-general} fully characterize the equilibrium prices and the induced welfare split across the three information regimes. Table \ref{tab:equilibrium-price-agent-welfare-seller-revenue} summarizes the resulting asymptotic formulas. 

\begin{table}[!htbp]
    \centering
    \caption{Asymptotic summary of equilibrium prices and outcomes as $n\to\infty$ with $\gamma_n\to\gamma$.
    Under {\sf Full Information}, the entries refer to the approximate-equilibrium profiles.}
    \begin{tabular}{lccc}
        \toprule
         & {\sf No Information} & {\sf Only Quality Information} & {\sf Full Information} \\
         & (Proposition \ref{prop:NI-general}) & (Proposition \ref{prop:QI-general}) & (Proposition \ref{prop:FI-general}) \\
        \midrule
        \multicolumn{4}{l}{\emph{Excess demand $(\gamma > 1)$ and Balanced market $(\gamma = 1)$ setting}} \\
         Seller price
         & $1/2$
         & $(1-\rho)q_y+\rho/2$
         & $(1-\rho)q_y+\rho\tau_n^\star,\ \ \tau_n^\star=1-\Theta(n^{-1/2})$ \\
         Agent welfare
         & $0$
         & $0$
         & $O(n^{-1/2})$ \\
         Seller revenue
         & $1/2$
         & $1/2$
         & $(1+\rho)/2-\Theta(n^{-1/2})$ \\
         Social welfare
         & $1/2$
         & $1/2$
         & $(1+\rho)/2-\Theta(n^{-1/2})$ \\
         \midrule
         \multicolumn{4}{l}{\emph{Excess supply setting: $\gamma\in(0,1)$}} \\
         Seller price
         & $0$
         & $(1-\rho)(q_y-q_{m+1})\,\mathbbm 1\{y\le m\}$
         & $(1-\rho)(q_y-q_{m+1})\,\mathbbm 1\{y\le m\}$ \\
         Agent welfare
         & $1/2$
         & $(1-\rho)(1-\gamma)+\rho/2$
         & $(1-\rho)(1-\gamma)+\rho$ \\
         Seller revenue
         & $0$
         & $(1-\rho)\gamma^2/2$
         & $(1-\rho)\gamma^2/2$ \\
         Social welfare
         & $1/2$
         & $(1-\rho)(1-\gamma+\gamma^2/2)+\rho/2$
         & $(1-\rho)(1-\gamma+\gamma^2/2)+\rho$ \\
         \bottomrule
    \end{tabular}
    \label{tab:equilibrium-price-agent-welfare-seller-revenue}
\end{table}

Using these equilibrium characterizations, we define the equilibrium-based value of information on metric ${\sf M}\in\{{\sf AW},{\sf SR},{\sf SW}\}$ by comparing aggregate outcomes across regimes:
\[
\Delta^{\sf M}_{\emptyset\to q}(m,n):={\sf M}_q(m,n)-{\sf M}_\emptyset(m,n),
\qquad
\Delta^{\sf M}_{q\to u}(m,n):={\sf M}_u(m,n)-{\sf M}_q(m,n).
\]
Note that to isolate the value of personalized recommendations ($\Delta^{\sf M}_{q\to u}$), we compare the approximate equilibrium outcomes in the {\sf Full Information} regime to the exact equilibrium outcomes in the {\sf Only Quality Information} regime.

\begin{theorem}[Value of public rankings]
\label{thm:value-public-rankings-pricing}
Fix $\rho \in (0,1)$. We have that,
\begin{enumerate}[label=(\alph*)]
\item {\bf (Excess Demand and Balanced Markets)} Let $m \geq n$. Public rankings do not add any aggregate value to agent welfare, seller revenue, or social welfare: 
$\Delta^{\sf AW}_{\emptyset\to q}(m,n)=
\Delta^{\sf SR}_{\emptyset\to q}(m,n)=
\Delta^{\sf SW}_{\emptyset\to q}(m,n)=0.$

\item {\bf (Excess Supply)} Let $m < n$. Public rankings strictly increase social welfare and seller revenue. Public rankings have a mixed impact on agent welfare depending on the market imbalance. In particular, we have that, 
\begin{align*}
\Delta^{\sf AW}_{\emptyset\to q}(m,n)
&=(1-\rho)\left(\frac{n-m}{n+1}-\frac12\right)
=\frac{(1-\rho)(n-2m-1)}{2(n+1)}, \qquad
\Delta^{\sf SR}_{\emptyset\to q}(m,n)
=(1-\rho)\frac{m(m+1)}{2n(n+1)},\\[4pt]
\Delta^{\sf SW}_{\emptyset\to q}(m,n)
&=\Delta^{\sf AW}_{\emptyset\to q}(m,n)+\Delta^{\sf SR}_{\emptyset\to q}(m,n)
=(1-\rho)\frac{(n-m)(n-m-1)}{2n(n+1)}\ \ge\ 0.
\end{align*}
\end{enumerate}
\end{theorem}

\begin{theorem}[Value of personalized recommendations]
\label{thm:value-personalization-pricing}
Fix $\rho\in(0,1]$. Personalized recommendations strictly increase social welfare. The division of this newly generated surplus depends entirely on which side of the market is short:
\begin{enumerate}[label=(\alph*)]
\item {\bf (Excess Demand and Balanced Markets)} Let $m \ge n$. Sellers capture the value of personalization, significantly increasing seller revenue without improving agent welfare:
\[
\Delta^{\sf AW}_{q\to u}(m,n)=O(n^{-1/2}), \quad
\Delta^{\sf SR}_{q\to u}(m,n)= \rho /2 -\Theta(n^{-1/2}),
\qquad
\Delta^{\sf SW}_{q\to u}(m,n)=\rho/2 -\Theta(n^{-1/2}).
\]
\item {\bf (Excess Supply)} Let $m/n \to \gamma \in (0,1)$ as $n \to \infty$. Agents capture the value of personalization, significantly increasing agent welfare without improving seller revenue:
\[
\Delta^{\sf AW}_{q\to u}(m,n)= \rho/2 + o(1), \quad
\Delta^{\sf SR}_{q\to u}(m,n)= o(1),
\qquad
\Delta^{\sf SW}_{q\to u}(m,n)=\rho/2 + o(1).
\]
\end{enumerate}
\end{theorem}

Table \ref{tab:equilibrium-price-agent-welfare-seller-revenue} and the subsequent theorems highlight a stark economic dichotomy: providing information fundamentally alters pricing behavior by creating price dispersion, allowing sellers to monetize their individual quality differences. However, who ultimately captures the resulting welfare gains depends entirely on which side of the market is short.

When $\gamma\ge 1$, supply is on the short side (excess demand or balanced markets). This is effectively a seller’s market: every item can sell with probability one. Under {\sf No Information}, the lack of differentiation leads to uniform pricing ($p_y^\star = 1/2$), driving agent welfare to zero. Moving to {\sf Only Quality Information} creates a direct \emph{effect on pricing}: sellers become vertically differentiated and perfectly monetize their observed quality, shifting to dispersed, quality-dependent prices ($p_y^\star = (1-\rho)q_y+ \rho /2$). Despite this dispersion, because supply remains short, sellers precisely adjust their markups to maintain full surplus extraction, leaving aggregate welfare outcomes unchanged (see Theorem \ref{thm:value-public-rankings-pricing}(a)). Under {\sf Full Information}, Proposition~\ref{prop:FI-general}(a) constructs a high-price approximate equilibrium of the form $p_y^\dagger=(1-\rho)q_y+\rho\tau_n^\star$, where $\tau_n^\star=1-\Theta(n^{-1/2})$. Because buyers can now find better idiosyncratic fits, the total value generated by the market increases. Crucially, because sellers are on the short side of the market, \emph{all} sellers are able to charge significantly more under personalized recommendations to capture this additional value. They impose a high idiosyncratic threshold, uniformly raising average prices and absorbing essentially all of the newly generated match surplus (see Theorem \ref{thm:value-personalization-pricing}(a)). 

The dynamics change sharply when $\gamma\in(0,1)$ and demand is on the short side (excess supply). Under {\sf No Information}, fierce competition among undifferentiated sellers collapses all prices to zero, allowing buyers to capture the entire expected surplus. Under {\sf Only Quality Information}, sellers become vertically differentiated, and equilibrium prices take a ``cutoff'' form. The $(m+1)$-st quality item acts as the marginal alternative that pins the agents' expected utility at $(1-\rho)q_{m+1}+\rho/2$. Each of the top $m$ sellers charges exactly its quality advantage relative to this cutoff, $p_y^\star=(1-\rho)(q_y-q_{m+1})$, while lower-quality sellers price at $0$ and do not sell. 
When qualities are made public, rankings induce two distinct effects. First, an \emph{allocative effect}: trade optimally concentrates on the top-$m$ highest quality items, increasing overall social welfare. Second, a \emph{pricing effect}: rankings create vertical differentiation, relaxing price competition and allowing top-tier sellers to sustain positive markups proportional to their quality advantage over the marginal unsold item. These forces push buyer welfare in opposite directions. Consequently, public rankings \emph{reduce} agent welfare when the market imbalance is mild ($\gamma>1/2$) because the pricing effect dominates, but \emph{increase} agent welfare when excess supply is large ($\gamma<1/2$) because the allocative gains dominate (Theorem \ref{thm:value-public-rankings-pricing}(b)). Notably, as the market approaches balance ($\gamma \to 1$), the social welfare under {\sf Only Quality Information} smoothly approaches $1/2$, seamlessly connecting to the balanced market limit. Finally, under {\sf Full Information} in excess supply, buyers can perfectly optimize over idiosyncratic heterogeneity across the competitive fringe. Under {\sf Full Information} and fixed $\gamma\in(0,1)$, Proposition~\ref{prop:FI-general}(b) shows that the same cutoff profile as in the Only Quality Information regime remains an $\mathcal{O}(\sqrt{\log n / n})$-approximate equilibrium. Because demand is on the short side, intense competition among the excess items keeps prices rigidly anchored to the quality cutoff of the marginal unsold item. Sellers cannot raise prices to extract the idiosyncratic fit without losing their sale to a competitor in the fringe. As a result, markups remain stable, and buyers utilize personalized recommendations to find better idiosyncratic fits without facing higher prices, allowing agents to capture the entirety of the $\rho/2$ surplus expansion (see Theorem \ref{thm:value-personalization-pricing}(b)).

\section{Conclusion}
\label{sec:conclusion}
This paper quantifies the incremental welfare value of two ubiquitous information tools—{\it public rankings} and {\it personalized recommendations}—and shows that their usefulness depends fundamentally on whether supply is unconstrained (digital content) or scarce (admissions seats, rentals, limited inventory).
We study a parsimonious model in which utility decomposes into an item-level {\it common} quality component and an agent–item {\it idiosyncratic} fit component, with heterogeneity governed by $\rho$.
Rankings reveal (possibly noisy) common quality; personalization reveals additional (possibly noisy) information about fit.

Our main message is that {\it scarcity changes what information can improve average welfare}.
In uncapacitated settings, information raises welfare through extreme-value selection: better signals help agents pick better options as the choice set grows, so both rankings and personalization can generate large-market gains.
In capacitated one-to-one settings, we identify a conservation property for the common component: because each item is allocated at most once, revealing common quality reshuffles {\it who gets which qualities} but cannot increase the {\it average} common-value contribution.
Accordingly, in our benchmark with independent fit values, public rankings have zero effect on {\it average} agent welfare in the capacitated setting (indeed, exactly zero in the baseline model), while personalization can deliver first-order gains by improving idiosyncratic fit and mitigating correlated demand for the same few “top” options.

We show that these conclusions are robust beyond the Gaussian benchmark.
Across a broad range of tail behaviors (Pareto, exponential, Gaussian-like, bounded), welfare gains inherit the relevant extreme-value scaling, but the capacitated irrelevance of rankings for {\it average} welfare persists, whereas personalization remains valuable through the idiosyncratic component.
We further stress-test the mechanism in institutionally motivated extensions: with supply-side priorities (stable matchings with own-group priorities) the same aggregate contrast continues to hold, and with prices, improved information can raise total surplus while shifting surplus toward sellers in tight markets.

These results offer a simple lens for interpreting empirical findings in school and college choice: non-personalized quality information can change {individual} choices yet produce muted improvements in {\it average} outcomes under binding capacity constraints, while tools that improve preference discovery and fit can generate aggregate gains.
Future work should relax key simplifications and incorporate aspects like correlation in idiosyncratic values, distributional/fairness objectives, strategic behavior, privacy/consent costs, and applicant-program compatibility in qualities,  to better understand the tradeoffs in rankings versus personalization in capacity constrained markets.

\newpage

\bibliographystyle{plainnat}
\bibliography{references}

\newpage
\appendix

\pagenumbering{arabic}
\renewcommand{\thepage}{App-\arabic{page}}
\renewcommand{\theequation}{\thesection.\arabic{equation}}
\renewcommand{\thelemma}{\thesection.\arabic{corollary}}
\renewcommand{\theproposition}{\thesection.\arabic{proposition}}
\renewcommand{\thecorollary}{\thesection.\arabic{corollary}}
\renewcommand{\thetheorem}{\thesection.\arabic{corollary}}
\renewcommand{\theremark}{\thesection.\arabic{corollary}}
\renewcommand{\theconjecture}{\thesection.\arabic{conjecture}}

\setcounter{proposition}{0}
\setcounter{equation}{0}
\setcounter{theorem}{0}
\setcounter{conjecture}{0}

\counterwithin{equation}{section}

\begin{center}
 {\Large \textbf{Electronic Companion: 
\\ Impact of Rankings and Personalized Recommendations in Marketplaces \\}
\medskip
\ifx\blind\undefined
\fi}
\end{center}

\vspace{-4em}

\addcontentsline{toc}{section}{Appendix} 

\part{Appendix} 
\setstretch{1}
\parttoc 
\newpage


\section{Proof of Theorem \ref{thm:noisy-estimate-agent-welfare}}
\label{app:proof-base-model-results}

\subsection{Proof of Theorem \ref{thm:noisy-estimate-uncapacitated-agent-welfare}}

\begin{proof}[\underline{Proof of Theorem \ref{thm:noisy-estimate-uncapacitated-agent-welfare}}]
    In the uncapacitated case, we can focus on a single agent. Given the noisy common $(\tilde{q}_y)$ and idiosyncratic $(\tilde{\varphi}_{1y})$ terms, the agent chooses the item with the highest {\it perceived} utility as defined in \eqref{eq:perceived-utility}.
    Note that $\tilde{q}_y \sim \mathcal{N}(\mu_q, \sigma_{q}^2 + \sigma_{\varepsilon}^2)$ and $\tilde{\varphi}_{1y} \sim \mathcal{N}(\mu_\varphi, \sigma_{\varphi}^2 + \sigma_{\delta}^2)$.
    Using \eqref{eq:common-term-posterior-mean-variance} and \eqref{eq:idiosyncratic-term-posterior-mean-variance}, we have that
    \begin{align*}
        \hat{q}_y \sim \mathcal{N}\left(\mu_q, \frac{\sigma_{q}^4}{\sigma_{q}^2 + \sigma_{\varepsilon}^2} \right), \quad \text{and}\quad
        \hat{\varphi}_{1y} \sim \mathcal{N}\left(\mu_\varphi, \frac{\sigma_{\varphi}^4}{\sigma_{\varphi}^2 + \sigma_{\delta}^2} \right).
    \end{align*}
    Define
    \begin{align*}
        \bar{\mu} &= (1 - \rho) \mu_q + \rho \mu_\varphi, \\
        \bar{\sigma}^2 &= (1 - \rho)^2 \frac{\sigma_{q}^4}{\sigma_q^2 + \sigma_{ \varepsilon}^2} + \rho^2 \frac{\sigma_{\varphi}^4}{\sigma_{\varphi}^2 + \sigma_{\delta}^2}.
    \end{align*}
    Using the fact that $\hat{u}_{1y} = (1 - \rho) \hat{q}_y + \rho \hat{\varphi}_{1y}$, we have that $\hat{u}_{1y} \sim \mathcal{N}(\bar{\mu}, \bar{\sigma}^2)$.
    Let $\pi = \argmax_{y \in \mathcal{Y}} \hat{u}_{1y}$.
    Since $\E[u_{1\pi} \mid (\tilde{q}_y)_{y = 1}^n, (\tilde{\varphi}_{1y})_{y = 1}^n] = \hat{u}_{1\pi}$, we have
    \begin{align*}
        {\sf AW}^{\sf uncap}(n)
        &= \mathbb{E}[u_{1\pi}]
        = \mathbb{E}\big[\mathbb{E}[u_{1\pi} \mid (\tilde{q}_y)_{y = 1}^n , (\tilde{\varphi}_{1y})_{y = 1}^n]\big]
        = \mathbb{E}[\hat{u}_{1\pi}]
        = \mathbb{E}\Big[\max_{y \in \mathcal{Y}} \hat{u}_{1y}\Big].
    \end{align*}
    Using the Fisher--Tippett--Gnedenko theorem for Gaussian maxima, we have that
    $\lim_{n \to \infty} {\sf AW}^{\sf uncap}(n) / \sqrt{2 \ln n} = \bar{\sigma}$, which completes the proof.
\end{proof}

\subsection{Proof of Theorem \ref{thm:noisy-estimate-capacitated-agent-welfare}}

\begin{proof}[\underline{Proof of Theorem \ref{thm:noisy-estimate-capacitated-agent-welfare}}]
Let $\mathcal{R}_k$ denote the set of remaining items when the $k$-th agent arrives to make the selection.
Given the noisy common and idiosyncratic terms, the $k$-th agent chooses the item with the highest {\it perceived} utility $\hat{u}_{ky}$ defined in \eqref{eq:perceived-utility}.
Let $\pi(k)$ denote the index of the item selected by the $k$-th agent. We have $\pi(k) = \argmax_{y \in \mathcal{R}_k} \hat{u}_{ky}$.

Define
\begin{align*}
    \kappa_q \triangleq \frac{\sigma_q^2}{\sigma_q^2+\sigma_{\varepsilon}^2},
\qquad
\kappa_\varphi \triangleq \frac{\sigma_{\varphi}^2}{\sigma_{\varphi}^2 + \sigma_{\delta}^2}.
\end{align*}

The agent welfare in the capacitated case is
\begin{align}
    \nonumber
    {\sf AW}^{\sf cap}(n) &\stackrel{(a)}= n^{-1} \sum_{k = 1}^n \mathbb{E}[u_{k \pi(k)}] \\
    \nonumber
    &\stackrel{(b)}= n^{-1} \sum_{k = 1}^n \mathbb{E}\Big[\mathbb{E}[u_{k\pi(k)} \mid (\tilde{q}_y)_{y \in \mathcal{R}_k}, (\tilde{\varphi}_{ky})_{y \in \mathcal{R}_k}]\Big] \\
    \label{eq:gaussian-noisy-estimate-average-welfare}
    &\stackrel{(c)}= n^{-1} \sum_{k = 1}^n \mathbb{E}[\hat{u}_{k \pi(k)}] \\
    \nonumber
    &\stackrel{(d)}= n^{-1} \sum_{k = 1}^n \mathbb{E}\big[(1 - \rho) \cdot \hat{q}_{\pi(k)} + \rho \cdot \hat{\varphi}_{k \pi(k)}\big] \\
    \nonumber
    &\stackrel{(e)}= n^{-1} (1 - \rho) \cdot \mathbb{E}\left[\sum_{y \in \mathcal{Y}} \hat{q}_y \right]
      + n^{-1} \rho \cdot \mathbb{E}\left[\sum_{k = 1}^n \Big\{ \kappa_\varphi \tilde{\varphi}_{k \pi(k)} + (1 - \kappa_\varphi) \mu_{\varphi} \Big\} \right] \\
    \label{eq:noisy-gaussian-agent-welfare-capacitated}
    &\stackrel{(f)}= (1 - \rho) \mu_q + \rho (1 - \kappa_\varphi) \mu_\varphi
      + n^{-1} \cdot \rho \cdot \kappa_{\varphi} \cdot \sum_{k = 1}^n \mathbb{E}[\tilde{\varphi}_{k \pi(k)}],
\end{align}
where (a) follows from the definition of ${\sf AW}^{\sf cap}(n)$, (b) follows from the law of iterated expectations, (c) follows from $\mathbb{E}[u_{k\pi(k)} \mid \cdot] = \hat{u}_{k\pi(k)}$, (d) follows from the definition of perceived utility, (e) follows from \eqref{eq:idiosyncratic-term-posterior-mean-variance}, and (f) follows from $\mathbb{E}[\sum_{y \in \mathcal{Y}} \hat{q}_y] = n \mathbb{E}[\hat{q}_y] = n\mu_q$.

\paragraph{\bf Upper bound on agent welfare.}
We upper bound $\sum_{k = 1}^n \mathbb{E}[\tilde{\varphi}_{k \pi(k)}]$.
By the principle of deferred decisions, for each agent $k$, we can view $\{\tilde{\varphi}_{ky}: y \in \mathcal{R}_k \}$ as a fresh i.i.d.\ draw from $\mathcal{N}(\mu_\varphi, \sigma_\varphi^2 + \sigma_{\delta}^2)$.
Therefore,
\begin{align}
    \label{eq:upper-bound-noisy-idiosyncratic-sum}
    \sum_{k = 1}^n \mathbb{E}[\tilde{\varphi}_{k \pi(k)}]
    \leq \sum_{k = 1}^n \mathbb{E}\Big[\max_{y \in \mathcal{R}_k} \tilde{\varphi}_{ky}\Big]
    \leq  n\mu_{\varphi} + \sqrt{\sigma_{\varphi}^2 + \sigma_{\delta}^2}\sum_{k = 1}^n \sqrt{2\ln (n - k + 1)}.
\end{align}
Using \eqref{eq:noisy-gaussian-agent-welfare-capacitated} and \eqref{eq:upper-bound-noisy-idiosyncratic-sum}, and the fact that
$n^{-1}\sum_{k=1}^n \sqrt{2\ln(n-k+1)} \leq \sqrt{2\ln n}$, we obtain
\[
{\sf AW}^{\sf cap}(n)
\leq (1-\rho)\mu_q + \rho \mu_\varphi
+ \rho \kappa_\varphi \sqrt{\sigma_\varphi^2 + \sigma_{\delta}^2}\,\sqrt{2\ln n}.
\]
Since $\kappa_\varphi \sqrt{\sigma_\varphi^2 + \sigma_{\delta}^2} = \sigma_\varphi^2 / \sqrt{\sigma_\varphi^2 + \sigma_{\delta}^2}$, we have
\begin{align}
    \label{eq:limsup-agent-welfare-noisy-gaussian-capacitated}
    \limsup_{n \to \infty} \frac{{\sf AW}^{\sf cap}(n) }{ \sqrt{2 \ln n}}
    \leq \rho \cdot \frac{\sigma_{\varphi}^2}{\sqrt{\sigma_{\varphi}^2 + \sigma_{\delta}^2}}.
\end{align}

\paragraph{\bf Lower bound on agent welfare.}
Let $m = n - k + 1$ denote the number of remaining items at step $k$.
Fix $\gamma \in (0,1)$ and let $i = m^\gamma$.
Let $\mathcal{Q}_k$ denote the set of the $i$ largest $\tilde{q}_y$ among the $m$ remaining items.
Then
\begin{align}
    \nonumber
    \hat{u}_{k\pi(k)}
    &= \max_{y \in \mathcal{R}_k} \{(1 - \rho) \cdot \hat{q}_y + \rho \cdot \hat{\varphi}_{ky}\} \\
    \nonumber
    &\geq \max_{y \in \mathcal{Q}_k} \{(1 - \rho) \cdot \hat{q}_y + \rho \cdot \hat{\varphi}_{ky} \} \\
    \nonumber
    &\geq (1 - \rho) \cdot \min_{y \in \mathcal{Q}_k} \hat{q}_y + \rho \cdot \max_{y \in \mathcal{Q}_k} \hat{\varphi}_{ky} \\
    \label{eq:noisy-estimate-agent-i-welfare-lower-bound}
    &\geq (1 - \rho) \cdot \hat{q}_{(m - i + 1:n)} + \rho \cdot \max_{y \in \mathcal{Q}_k} \hat{\varphi}_{ky} \\
    \nonumber
    &= (1 - \rho) \cdot \hat{q}_{(m - i + 1:n)}
       + \rho \cdot \max_{y \in \mathcal{Q}_k} \{\kappa_\varphi \cdot \tilde{\varphi}_{ky} + (1 - \kappa_\varphi) \cdot \mu_{\varphi}\} \\
    \label{eq:noisy-estimate-agent-i-welfare-lower-bound-2}
    &= (1 - \rho) \cdot \hat{q}_{(m - i + 1:n)} + \rho \cdot (1 - \kappa_\varphi) \cdot \mu_{\varphi} + \rho \cdot \kappa_\varphi \cdot \max_{y \in \mathcal{Q}_k} \tilde{\varphi}_{ky}.
\end{align}
Since $\{\tilde{\varphi}_{ky}: y \in \mathcal{Q}_k\}$ are i.i.d.\ draws from $\mathcal{N}(\mu_\varphi, \sigma_{\varphi}^2 + \sigma_{\delta}^2)$ and $|\mathcal{Q}_k| = m^{\gamma}$, we have
\begin{align}
    \label{eq:noisy-gaussian-maximum-n-lower-bound}
    \mathbb{E}[\max_{y \in \mathcal{Q}_k} \tilde{\varphi}_{ky}] - \mu_\varphi
    \geq \sqrt{\sigma_\varphi^2 + \sigma_{\delta}^2}\,\sqrt{2 \ln (m^\gamma)} - o(\sqrt{2\ln m})
    = \sqrt{\sigma_{\varphi}^2 + \sigma_{\delta}^2 } \sqrt{2 \gamma \ln m} - o(\sqrt{2 \ln m}).
\end{align}

Now,
\begin{align*}
    {\sf AW}^{\sf cap}(n)
    &\geq n^{-1} (1 - \rho) \mathbb{E}\Big[\sum_{m = 1}^n \hat{q}_{(\lfloor m -m^\gamma + 1\rfloor:n)}\Big]
      + n^{-1} \rho \kappa_\varphi \sum_{m = 1}^n \mathbb{E}\big[\max_{y \in \mathcal{Q}_m} \tilde{\varphi}_{my}\big] \\
    &\geq  - n^{\gamma - 1} (1 - \rho)\,\kappa_q \sqrt{\sigma_q^2 + \sigma_{\varepsilon}^2}\,\sqrt{2 \ln n}
      + \rho \kappa_\varphi \sqrt{\sigma_{\varphi}^2 + \sigma_{\delta}^2}\,\sqrt{2\gamma \ln n}
      + o(\sqrt{2 \ln n}).
\end{align*}
Therefore,
\begin{align}
    \label{eq:liminf-agent-welfare-noisy-gaussian-capacitated}
    \sqrt{\gamma} \cdot  \rho \cdot \frac{\sigma_{\varphi}^2}{\sqrt{\sigma_{\varphi}^2 + \sigma_{\delta}^2}}
    \leq
    \liminf_{n \to \infty} \frac{{\sf AW}^{\sf cap}(n)}{\sqrt{2\ln n}}.
\end{align}
Combining \eqref{eq:limsup-agent-welfare-noisy-gaussian-capacitated} and \eqref{eq:liminf-agent-welfare-noisy-gaussian-capacitated} yields
\[
    \sqrt{\gamma}\cdot \rho \cdot \frac{\sigma_{\varphi}^2}{\sqrt{\sigma_{\varphi}^2 + \sigma_{\delta}^2}}
    \leq
    \liminf_{n \to \infty} \frac{{\sf AW}^{\sf cap}(n)}{\sqrt{2\ln n}}
    \leq
    \limsup_{n \to \infty} \frac{{\sf AW}^{\sf cap}(n)}{\sqrt{2\ln n}}
    \leq
    \rho \cdot \frac{\sigma_{\varphi}^2}{\sqrt{\sigma_{\varphi}^2 + \sigma_{\delta}^2}}.
\]
Since this holds for any $\gamma \in (0,1)$, this completes the proof.
\end{proof}

\section{Welfare Gains For Different Tail Distributions}
\label{app:proof-main-theorems}

\subsection{Helper Results}
\label{app:helper-results}
In this section we will provide some useful results along with proofs which will be used to prove the main results.

\subsubsection{Characterizing the expected value of maximum of $n$ i.i.d RVs}

\begin{proposition}
    \label{prop:expected-maximum-n}
    Let $X$ be a random variable with distribution $P$. Assume that $X \geq 0$ and $\mathbb{E}[X] < \infty$. Let $X_1, \dots, X_n$ be i.i.d copies of $X$ and define $X_{(n:n)} = \max_{1 \leq k \leq n} X_k$.
    \begin{enumerate}[label = $(\alph*)$]
        \item \textup{(Pareto Tail)} \label{prop:expected-maximum-n-pareto} Fix $c > 0, \alpha > 1$. Assume $P$ has a Pareto tail with parameters $(c, \alpha)$. We have that $\lim_{n \to \infty} \mathbb{E}[X_{(n:n)}] / (c\Gamma(1 - 1 /\alpha) \cdot n^{1 / \alpha}) = 1$.
        
        \item \textup{(Exponential Tail)} \label{prop:expected-maximum-n-exponential} Fix $c > 0, \lambda > 0$. Assume $P$ has an exponential tail with parameters $(c, \lambda)$. We have that $\lim_{n \to \infty} \mathbb{E}[X_{(n:n)}] / (\ln n / \lambda) = 1.$

         \item \label{prop:expected-maximum-n-gaussian} \textup{(Gaussian-like Tail)}  Fix $c > 0, \lambda > 0$ and $\beta \geq 0$. Assume $P$ has a Gaussian-like tail with parameters $(c, \lambda, \beta)$. We have that $\lim_{n \to \infty} \mathbb{E}[X_{(n:n)}] / (\sqrt{\ln n} / \lambda) = 1.$
        
        \item \textup{(Bounded Tail)} \label{thm:uncapacitated-ranking-value-bounded-tail} Fix $0 \leq a < b < \infty$, $\kappa > 0$ and $\beta \geq 1$. Assume $P$ has a continuous distribution over $[a, b]$ and bounded tail with parameter $(b, \kappa, \beta)$. We have that $\lim_{n \to \infty} \mathbb{E}[X_{(n:n)}] = b$.
    \end{enumerate}
\end{proposition}

\begin{proof}[\underline{Proof of Proposition~\ref{prop:expected-maximum-n}}]
Let $\bar F(x):=\mathbb P(X>x)$ and write $M_n:=X_{(n:n)}=\max_{1\le k\le n}X_k$.
We repeatedly use the following two identities: for every $x\ge 0$,
\begin{align}
\mathbb P(M_n>x)
&=1-\mathbb P(X_1\le x,\dots,X_n\le x)
=1-(1-\bar F(x))^n, \label{eq:max-tail}\\
\mathbb E[M_n]
&=\int_0^\infty \mathbb P(M_n>x)\,dx. \label{eq:tailsum}
\end{align}
Moreover, for every $y\in[0,1]$ we have the bounds
\begin{equation}\label{eq:1miny-power}
1-(1-y)^n\le ny,
\qquad
1-(1-y)^n\ge 1-e^{-ny},
\end{equation}
and the map $y\mapsto 1-(1-y)^n$ is nondecreasing on $[0,1]$.

Throughout, fix $\varepsilon\in(0,1/2)$. In each part, by the corresponding tail definition, there exists
$x_0=x_0(\varepsilon)$ such that the required two-sided bound on $\bar F(x)$ holds for all $x\ge x_0$.

\paragraph{(a) Pareto tail.}
Assume $F$ has a Pareto tail with parameters $(c,\alpha)$, i.e.\ $\bar F(x)\sim (c/x)^\alpha$ as $x\to\infty$,
with $c>0$ and $\alpha>1$. Then there exists $x_0$ such that for all $x\ge x_0$,
\begin{equation}\label{eq:pareto-twosided}
(1-\varepsilon)\Big(\frac{c}{x}\Big)^\alpha \le \bar F(x)\le (1+\varepsilon)\Big(\frac{c}{x}\Big)^\alpha.
\end{equation}
Define $\bar c:=(1+\varepsilon)^{1/\alpha}c$ and $\underline c:=(1-\varepsilon)^{1/\alpha}c$, so that
$(1+\varepsilon)(c/x)^\alpha=(\bar c/x)^\alpha$ and $(1-\varepsilon)(c/x)^\alpha=(\underline c/x)^\alpha$.

{\it An integral identity.}
For any $d>0$, define
\[
I_n(d):=\int_{d}^{\infty}\Big(1-\big(1-(d/x)^\alpha\big)^n\Big)\,dx,
\]
where the integrand is well-defined since $(d/x)^\alpha\in(0,1]$ for $x\ge d$.
We compute $I_n(d)$ explicitly. Let $U(x)=x$ and $V(x)=1-(1-(d/x)^\alpha)^n$ for $x\ge d$.
Using \eqref{eq:1miny-power} with $y=(d/x)^\alpha$, we have for all $x\ge d$,
\[
0\le V(x)\le n(d/x)^\alpha
\quad\Rightarrow\quad
0\le xV(x)\le n d^\alpha x^{1-\alpha}\xrightarrow[x\to\infty]{}0
\qquad(\alpha>1).
\]
Also, $V(d)=1-(1-1)^n=1$. Moreover, differentiating,
\[
V'(x)
= -n\Big(1-(d/x)^\alpha\Big)^{n-1}\cdot \frac{d}{dx}\Big(1-(d/x)^\alpha\Big)
= -n\alpha d^\alpha \Big(1-(d/x)^\alpha\Big)^{n-1}x^{-\alpha-1}.
\]
Integration by parts gives
\begin{align}
I_n(d)
&=\int_d^\infty V(x)\,dU(x)
=\big[U(x)V(x)\big]_{d}^{\infty}-\int_d^\infty U(x)\,dV(x) \notag\\
&=\big[ xV(x)\big]_{d}^{\infty}-\int_d^\infty xV'(x)\,dx \notag\\
&=-d+n\alpha\int_d^\infty \Big(1-(d/x)^\alpha\Big)^{n-1}\Big(\frac{d}{x}\Big)^\alpha\,dx. \label{eq:In-parts}
\end{align}
With the change of variables $u=(d/x)^\alpha$ (so $x=du^{-1/\alpha}$ and $dx=-(d/\alpha)u^{-1/\alpha-1}\,du$),
the integral in \eqref{eq:In-parts} becomes
\begin{align}
n\alpha\int_d^\infty \Big(1-(d/x)^\alpha\Big)^{n-1}\Big(\frac{d}{x}\Big)^\alpha\,dx
&=n\alpha\int_{u=1}^{0}(1-u)^{n-1}\,u\cdot\Big(-\frac{d}{\alpha}u^{-1/\alpha-1}\Big)\,du \notag\\
&=nd\int_0^1 (1-u)^{n-1}u^{-1/\alpha}\,du \notag\\
&=nd\int_0^1 (1-u)^{n-1}u^{(1-1/\alpha)-1}\,du \notag\\
&=nd\,B\!\Big(1-\frac1\alpha,n\Big)
=nd\,\frac{\Gamma(1-1/\alpha)\Gamma(n)}{\Gamma(n+1-1/\alpha)}. \label{eq:In-beta}
\end{align}
Combining \eqref{eq:In-parts}--\eqref{eq:In-beta} yields
\begin{equation}\label{eq:In-closed}
I_n(d)
=-d+nd\,\frac{\Gamma(1-1/\alpha)\Gamma(n)}{\Gamma(n+1-1/\alpha)}.
\end{equation}

{\it Upper bound.}
Using \eqref{eq:tailsum} and $\mathbb P(M_n>x)\le 1$,
\begin{align*}
\mathbb E[M_n]
&=\int_0^{\max\{x_0,\bar c\}}\mathbb P(M_n>x)\,dx+\int_{\max\{x_0,\bar c\}}^\infty \mathbb P(M_n>x)\,dx\\
&\le \max\{x_0,\bar c\}+\int_{\max\{x_0,\bar c\}}^\infty \mathbb P(M_n>x)\,dx.
\end{align*}
For $x\ge \max\{x_0,\bar c\}$, we have $x\ge x_0$ and by \eqref{eq:pareto-twosided},
$\bar F(x)\le (\bar c/x)^\alpha\in(0,1]$. Since $y\mapsto 1-(1-y)^n$ is nondecreasing on $[0,1]$, \eqref{eq:max-tail} gives
\[
\mathbb P(M_n>x)=1-(1-\bar F(x))^n \le 1-\Big(1-(\bar c/x)^\alpha\Big)^n.
\]
Hence, using that the integrand is nonnegative on $[\bar c,\infty)$,
\begin{align}
\mathbb E[M_n]
&\le \max\{x_0,\bar c\}+\int_{\max\{x_0,\bar c\}}^\infty \Big(1-\big(1-(\bar c/x)^\alpha\big)^n\Big)\,dx \notag\\
&\le \max\{x_0,\bar c\}+\int_{\bar c}^\infty \Big(1-\big(1-(\bar c/x)^\alpha\big)^n\Big)\,dx \notag\\
&=\max\{x_0,\bar c\}+I_n(\bar c)
=(x_0-\bar c)_+ + n\bar c\,\frac{\Gamma(1-1/\alpha)\Gamma(n)}{\Gamma(n+1-1/\alpha)}. \label{eq:pareto-upper}
\end{align}

{\it Lower bound.}
Similarly, for $x\ge \max\{x_0,\underline c\}$ we have $\bar F(x)\ge (\underline c/x)^\alpha\in(0,1]$, hence
\[
\mathbb P(M_n>x)=1-(1-\bar F(x))^n \ge 1-\Big(1-(\underline c/x)^\alpha\Big)^n.
\]
Therefore,
\begin{align}
\mathbb E[M_n]
&=\int_0^\infty \mathbb P(M_n>x)\,dx
\ge \int_{\max\{x_0,\underline c\}}^\infty \mathbb P(M_n>x)\,dx \notag\\
&\ge \int_{\max\{x_0,\underline c\}}^\infty \Big(1-\big(1-(\underline c/x)^\alpha\big)^n\Big)\,dx \notag\\
&= \int_{\underline c}^\infty \Big(1-\big(1-(\underline c/x)^\alpha\big)^n\Big)\,dx
 - \int_{\underline c}^{\max\{x_0,\underline c\}} \Big(1-\big(1-(\underline c/x)^\alpha\big)^n\Big)\,dx \notag\\
&\ge I_n(\underline c) - (x_0-\underline c)_+ \notag\\
&=-(x_0-\underline c)_+ - \underline c
+ n\underline c\,\frac{\Gamma(1-1/\alpha)\Gamma(n)}{\Gamma(n+1-1/\alpha)}. \label{eq:pareto-lower}
\end{align}

Now use the standard Gamma-ratio asymptotic (e.g.\ via Stirling's formula): since $\alpha>1$,
\begin{equation}\label{eq:gamma-ratio}
\lim_{n\to\infty}\frac{n\,\Gamma(n)}{\Gamma(n+1-1/\alpha)\,n^{1/\alpha}}=1.
\end{equation}
Divide \eqref{eq:pareto-upper} and \eqref{eq:pareto-lower} by $c\,\Gamma(1-1/\alpha)\,n^{1/\alpha}$ and send $n\to\infty$,
using \eqref{eq:gamma-ratio} and that $(x_0-\bar c)_+/n^{1/\alpha}\to 0$ and $(x_0-\underline c)_+/n^{1/\alpha}\to 0$. This yields
\[
\limsup_{n\to\infty}\frac{\mathbb E[M_n]}{c\,\Gamma(1-1/\alpha)\,n^{1/\alpha}}\le \frac{\bar c}{c}=(1+\varepsilon)^{1/\alpha},
\qquad
\liminf_{n\to\infty}\frac{\mathbb E[M_n]}{c\,\Gamma(1-1/\alpha)\,n^{1/\alpha}}\ge \frac{\underline c}{c}=(1-\varepsilon)^{1/\alpha}.
\]
Since $\varepsilon\in(0,1/2)$ was arbitrary, letting $\varepsilon\downarrow 0$ proves
\[
\lim_{n\to\infty}\frac{\mathbb E[M_n]}{c\,\Gamma(1-1/\alpha)\,n^{1/\alpha}}=1.
\]

\paragraph{(b) Exponential tail.}
Assume $F$ has an exponential tail with parameters $(c,\lambda)$, i.e.\ $\bar F(x)\sim c e^{-\lambda x}$ as $x\to\infty$,
with $c,\lambda>0$. Then there exists $x_0$ such that for all $x\ge x_0$,
\begin{equation}\label{eq:exp-twosided}
(1-\varepsilon)c e^{-\lambda x}\le \bar F(x)\le (1+\varepsilon)c e^{-\lambda x}.
\end{equation}
Define $\bar c:=(1+\varepsilon)c$, $\underline c:=(1-\varepsilon)c$, and
\[
\bar s:=\frac{\ln \bar c}{\lambda},\qquad \underline s:=\frac{\ln \underline c}{\lambda}.
\]
Note that for all $x\ge \bar s$, $\bar c e^{-\lambda x}\in(0,1]$, and for all $x\ge \underline s$, $\underline c e^{-\lambda x}\in(0,1]$.

{\it Upper bound.}
By \eqref{eq:tailsum} and $\mathbb P(M_n>x)\le 1$,
\[
\mathbb E[M_n]
\le \max\{x_0,\bar s\}+\int_{\max\{x_0,\bar s\}}^\infty \mathbb P(M_n>x)\,dx.
\]
For $x\ge \max\{x_0,\bar s\}$ we have $x\ge x_0$ and $\bar F(x)\le \bar c e^{-\lambda x}\in(0,1]$, hence by monotonicity and \eqref{eq:max-tail},
\[
\mathbb P(M_n>x)=1-(1-\bar F(x))^n \le 1-\big(1-\bar c e^{-\lambda x}\big)^n.
\]
Therefore, using nonnegativity of the integrand on $[\bar s,\infty)$,
\begin{align}
\mathbb E[M_n]
&\le \max\{x_0,\bar s\}+\int_{\max\{x_0,\bar s\}}^\infty \Big(1-\big(1-\bar c e^{-\lambda x}\big)^n\Big)\,dx \notag\\
&\le \max\{x_0,\bar s\}+\int_{\bar s}^\infty \Big(1-\big(1-\bar c e^{-\lambda x}\big)^n\Big)\,dx. \label{eq:exp-upper-int}
\end{align}
Make the change of variables $u=\bar c e^{-\lambda x}$, so that $dx=-(1/(\lambda u))\,du$ and $u$ runs from $1$ to $0$ as $x$ runs from $\bar s$ to $\infty$:
\begin{align*}
\int_{\bar s}^\infty \Big(1-\big(1-\bar c e^{-\lambda x}\big)^n\Big)\,dx
&=\frac1\lambda\int_0^1 \frac{1-(1-u)^n}{u}\,du.
\end{align*}
Using the (finite) geometric-series identity
\[
\frac{1-(1-u)^n}{u}=\sum_{k=0}^{n-1}(1-u)^k,
\]
we obtain
\begin{align*}
\int_0^1 \frac{1-(1-u)^n}{u}\,du
&=\sum_{k=0}^{n-1}\int_0^1 (1-u)^k\,du
=\sum_{k=0}^{n-1}\frac{1}{k+1}
=H_n,
\end{align*}
where $H_n:=\sum_{k=1}^n \frac1k$ is the $n$th harmonic number. Plugging into \eqref{eq:exp-upper-int},
\begin{equation}\label{eq:exp-upper}
\mathbb E[M_n]\le \max\{x_0,\bar s\}+\frac{H_n}{\lambda}.
\end{equation}
Finally, the bounds $\ln(n+1)\le H_n\le 1+\ln n$ imply $H_n/\ln n\to 1$, hence
\[
\limsup_{n\to\infty}\frac{\mathbb E[M_n]}{\ln n/\lambda}\le 1.
\]

{\it Lower bound.}
For $x\ge x_0$, \eqref{eq:exp-twosided} gives $\bar F(x)\ge \underline c e^{-\lambda x}$. Hence for $x\ge \max\{x_0,\underline s\}$,
$\underline c e^{-\lambda x}\in(0,1]$ and
\[
\mathbb P(M_n>x)=1-(1-\bar F(x))^n \ge 1-\big(1-\underline c e^{-\lambda x}\big)^n.
\]
Therefore,
\begin{align}
\mathbb E[M_n]
&\ge \int_{x_0}^\infty \mathbb P(M_n>x)\,dx
\ge \int_{x_0}^\infty \Big(1-\big(1-\underline c e^{-\lambda x}\big)^n\Big)\,dx \notag\\
&\ge \int_{\underline s}^\infty \Big(1-\big(1-\underline c e^{-\lambda x}\big)^n\Big)\,dx
     - \int_{\underline s}^{x_0} \Big(1-\big(1-\underline c e^{-\lambda x}\big)^n\Big)\,dx \notag\\
&\ge \int_{\underline s}^\infty \Big(1-\big(1-\underline c e^{-\lambda x}\big)^n\Big)\,dx
     - (x_0-\underline s)_+. \label{eq:exp-lower-int}
\end{align}
The same substitution $u=\underline c e^{-\lambda x}$ (so that $u=1$ when $x=\underline s$) yields
\[
\int_{\underline s}^\infty \Big(1-\big(1-\underline c e^{-\lambda x}\big)^n\Big)\,dx
=\frac1\lambda\int_0^1 \frac{1-(1-u)^n}{u}\,du
=\frac{H_n}{\lambda}.
\]
Thus \eqref{eq:exp-lower-int} becomes
\begin{equation}\label{eq:exp-lower}
\mathbb E[M_n]\ge \frac{H_n}{\lambda}-(x_0-\underline s)_+.
\end{equation}
Dividing by $\ln n/\lambda$ and letting $n\to\infty$ gives $\liminf_{n\to\infty}\mathbb E[M_n]/(\ln n/\lambda)\ge 1$.
Together with the upper bound, this proves
\[
\lim_{n\to\infty}\frac{\mathbb E[M_n]}{\ln n/\lambda}=1.
\]

\paragraph{(c) Gaussian-like tail.}
Assume $F$ has a Gaussian-like tail with parameters $(c,\lambda,\beta)$, i.e.\
$\bar F(x)\sim c x^{-\beta}e^{-\lambda^2 x^2}$ as $x\to\infty$, where $c>0$, $\lambda>0$, and $\beta\ge 0$.
Then there exists $x_0$ such that for all $x\ge x_0$,
\begin{equation}\label{eq:gauss-twosided}
(1-\varepsilon)c\,x^{-\beta}e^{-\lambda^2 x^2}
\le \bar F(x)\le
(1+\varepsilon)c\,x^{-\beta}e^{-\lambda^2 x^2}.
\end{equation}
Define $a_n:=\sqrt{\ln n}/\lambda$ and fix $\delta\in(0,1/2)$. Set
\[
\bar t_n:=(1+\delta)a_n,\qquad \underline t_n:=(1-\delta)a_n.
\]
Since $a_n\to\infty$, for all $n$ large enough we have $\underline t_n\ge x_0$ and $\bar t_n\ge x_0$; we work with such $n$ below.

We also use the standard Gaussian tail bound: for any $t>0$,
\begin{equation}\label{eq:gauss-int-bound}
\int_t^\infty e^{-\lambda^2 x^2}\,dx
\le \frac{e^{-\lambda^2 t^2}}{2\lambda^2 t}.
\end{equation}
Indeed, for $x\ge t$ we have $x/t\ge 1$, so
\[
\int_t^\infty e^{-\lambda^2 x^2}\,dx
\le \int_t^\infty \frac{x}{t}e^{-\lambda^2 x^2}\,dx
= \Big[-\frac{e^{-\lambda^2 x^2}}{2\lambda^2 t}\Big]_{x=t}^{x=\infty}
=\frac{e^{-\lambda^2 t^2}}{2\lambda^2 t}.
\]

{\it Upper bound.}
By \eqref{eq:tailsum},
\[
\mathbb E[M_n]\le \bar t_n + \int_{\bar t_n}^\infty \mathbb P(M_n>x)\,dx.
\]
For $x\ge \bar t_n\ge x_0$, \eqref{eq:max-tail} and \eqref{eq:1miny-power} yield
$\mathbb P(M_n>x)=1-(1-\bar F(x))^n\le n\bar F(x)$, hence by \eqref{eq:gauss-twosided},
\[
\mathbb P(M_n>x)\le n(1+\varepsilon)c\,x^{-\beta}e^{-\lambda^2 x^2},\qquad x\ge \bar t_n.
\]
Therefore,
\begin{align*}
\int_{\bar t_n}^\infty \mathbb P(M_n>x)\,dx
&\le n(1+\varepsilon)c \int_{\bar t_n}^\infty x^{-\beta}e^{-\lambda^2 x^2}\,dx
\le n(1+\varepsilon)c\,\bar t_n^{-\beta}\int_{\bar t_n}^\infty e^{-\lambda^2 x^2}\,dx\\
&\le n(1+\varepsilon)c\,\bar t_n^{-\beta}\cdot \frac{e^{-\lambda^2 \bar t_n^2}}{2\lambda^2 \bar t_n}
=\frac{(1+\varepsilon)c}{2\lambda^2}\, \Big(n e^{-\lambda^2 \bar t_n^2}\Big)\,\bar t_n^{-(\beta+1)}.
\end{align*}
Since $\lambda^2 \bar t_n^2=(1+\delta)^2\ln n$, we have
$n e^{-\lambda^2 \bar t_n^2}=n^{1-(1+\delta)^2}=n^{-2\delta-\delta^2}\to 0$, and $\bar t_n\asymp \sqrt{\ln n}$, hence
\[
\int_{\bar t_n}^\infty \mathbb P(M_n>x)\,dx = o(a_n).
\]
Thus
\[
\mathbb E[M_n]\le \bar t_n + o(a_n)=(1+\delta)a_n+o(a_n),
\]
so $\limsup_{n\to\infty}\mathbb E[M_n]/a_n\le 1+\delta$.

{\it Lower bound.}
Using \eqref{eq:max-tail} and $1-y\le e^{-y}$, for any $t\ge 0$,
\[
\mathbb P(M_n\le t)=(1-\bar F(t))^n \le \exp(-n\bar F(t)).
\]
For $t=\underline t_n\ge x_0$, the lower bound in \eqref{eq:gauss-twosided} gives
\[
\bar F(\underline t_n)\ge (1-\varepsilon)c\,\underline t_n^{-\beta}e^{-\lambda^2 \underline t_n^2}.
\]
Therefore,
\begin{align*}
\mathbb P(M_n\le \underline t_n)
&\le \exp\!\Big(-n(1-\varepsilon)c\,\underline t_n^{-\beta}e^{-\lambda^2 \underline t_n^2}\Big).
\end{align*}
Now $\lambda^2\underline t_n^2=(1-\delta)^2\ln n$, so
\[
n e^{-\lambda^2\underline t_n^2}=n^{1-(1-\delta)^2}=n^{2\delta-\delta^2},
\qquad
\underline t_n^{-\beta}=\big((1-\delta)a_n\big)^{-\beta}
=(1-\delta)^{-\beta}\lambda^\beta(\ln n)^{-\beta/2}.
\]
Hence
\[
n\,\underline t_n^{-\beta}e^{-\lambda^2 \underline t_n^2}
=(1-\delta)^{-\beta}\lambda^\beta\, n^{2\delta-\delta^2}(\ln n)^{-\beta/2}\xrightarrow[n\to\infty]{}\infty,
\]
so $\mathbb P(M_n\le \underline t_n)\to 0$, i.e.\ $\mathbb P(M_n>\underline t_n)\to 1$.
Since $x\mapsto \mathbb P(M_n>x)$ is nonincreasing,
\[
\mathbb E[M_n]=\int_0^\infty \mathbb P(M_n>x)\,dx
\ge \int_0^{\underline t_n}\mathbb P(M_n>x)\,dx
\ge \underline t_n\,\mathbb P(M_n>\underline t_n)
=(1-o(1))\,\underline t_n.
\]
Thus $\liminf_{n\to\infty}\mathbb E[M_n]/a_n\ge 1-\delta$.
Letting $\delta\downarrow 0$ and combining with the upper bound yields
\[
\lim_{n\to\infty}\frac{\mathbb E[M_n]}{\sqrt{\ln n}/\lambda}=1.
\]

\paragraph{(d) Bounded tail.}
Assume $X\in[a,b]$ a.s., $0\le a<b<\infty$, $F$ is continuous on $[a,b]$, and $F$ has a bounded tail with parameters $(b,\kappa,\beta)$,
i.e.\ $\bar F(x)\sim \kappa(b-x)^\beta$ as $x\uparrow b$, with $\kappa>0$ and $\beta\ge 1$.
Fix $\varepsilon\in(0,1/2)$. By the definition of bounded tail, there exists $\eta_0\in(0,b-a)$ such that for every $\eta\in(0,\eta_0]$,
\begin{equation}\label{eq:bounded-tail-eta}
(1-\varepsilon)\kappa \eta^\beta \le \mathbb P(X>b-\eta)\le (1+\varepsilon)\kappa \eta^\beta.
\end{equation}

Fix any $\eta\in(0,\eta_0]$ and set $p_\eta:=\mathbb P(X>b-\eta)>0$. Then
\begin{equation}\label{eq:bounded-prob}
\mathbb P(M_n\le b-\eta)
=\mathbb P(X_1\le b-\eta,\dots,X_n\le b-\eta)
=(1-p_\eta)^n
\le e^{-np_\eta}
\le \exp\!\big(-(1-\varepsilon)\kappa\eta^\beta n\big)\xrightarrow[n\to\infty]{}0,
\end{equation}
where the last inequality uses the left-hand side of \eqref{eq:bounded-tail-eta}.
Thus $M_n\xrightarrow{p} b$.

To conclude convergence of expectations via bounded convergence, we strengthen to almost sure convergence.
For $m\in\mathbb N$, set $\eta_m:=\min\{\eta_0,1/m\}$ and $p_m:=\mathbb P(X>b-\eta_m)>0$.
By independence,
\[
\sum_{n=1}^\infty \mathbb P(X_n>b-\eta_m)=\sum_{n=1}^\infty p_m=\infty,
\]
so by the second Borel--Cantelli lemma, $\mathbb P(X_n>b-\eta_m\ \text{i.o.})=1$.
Therefore, with probability $1$, for each $m$ there exists a (random) $N_m<\infty$ such that for all $n\ge N_m$,
\[
M_n=\max_{1\le k\le n}X_k \ge b-\eta_m.
\]
It follows that $\liminf_{n\to\infty}M_n \ge b-\eta_m$ for every $m$, hence $\liminf_{n\to\infty}M_n\ge b$.
Since $M_n\le b$ always, we conclude $M_n\to b$ almost surely.

Finally, since $0\le M_n\le b$ for all $n$ and $M_n\to b$ a.s., the bounded convergence theorem yields
\[
\lim_{n\to\infty}\mathbb E[M_n]=\mathbb E\big[\lim_{n\to\infty}M_n\big]=b.
\]
This completes the proof of all four parts.
\end{proof}

\subsubsection{Characterizing the behavior of convex combination of RVs}

\begin{proposition}
    \label{prop:tail-convex-combination}
    Let $X$ and $Y$ be two random variables with nonnegative support and finite means $\mu_X < \infty$ and $\mu_Y < \infty$ respectively. Fix $\rho \in [0,1]$. Define $Z = (1 - \rho) X + \rho Y$. Then we have that $Z \geq 0$ and $\mathbb{E}[Z] = \mu_Z = (1 - \rho) \mu_X + \rho \mu_Y < \infty$. Furthermore, we have that 
    \begin{enumerate}[label = (\alph*)]
        \item \textup{(Pareto Tail)} If $X$ and $Y$ have Pareto tail with parameters $(c_X, \alpha_X)$ and $(c_Y, \alpha_Y)$ respectively. Then $Z$ has a Pareto tail with parameters $(c_Z, \alpha_Z)$ given as
        \begin{itemize}
            \item \underline{$\alpha_X < \alpha_Y$.} We have $c_Z = (1 - \rho)c_X$ and $\alpha_Z = \alpha_X$.
            \item \underline{$\alpha_X > \alpha_Y$.} We have $c_Z = \rho c_Y$ and $\alpha_Z = \alpha_Y$.
            \item \underline{$\alpha_X = \alpha_Y := \alpha$.} We have $c_Z = (((1 - \rho) c_X)^{\alpha} + (\rho c_Y)^\alpha)^{1 / \alpha}$ and $\alpha_Z = \alpha$.
        \end{itemize}
        \item \textup{(Exponential Tail)} If $X$ and $Y$ have Exponential tail with parameters $(c_X, \lambda_X)$ and $(c_Y, \lambda_Y)$ respectively. 
        Define $\underline{\lambda} = \min\{(1 - \rho)^{-1} \lambda_X, \rho^{-1}\lambda_Y \}$ and $\bar{\lambda} = \max\{(1 - \rho)^{-1} \lambda_X, \rho^{-1} \lambda_Y\}$. Then $Z$ has an exponential-like tail, i.e., $\lim_{t \to \infty} t^{-1} \ln \mathbb{P}(Z > t) = -\underline{\lambda}$. In particular, for sufficiently large $t$, there exists constants $c,C < \infty$ such that $c \cdot \exp(-\underline{\lambda}t) \leq \mathbb{P}(Z > t) \leq C \cdot t^{1 + \bar{\lambda} / \underline{\lambda}} \cdot \exp(-\underline{\lambda} t)$.
        \item \textup{(Gaussian-like Tail)} If $X$ and $Y$ have Gaussian-like tail with parameters $(c_X, \lambda_X, \beta_X)$ and $(c_Y, \lambda_Y, \beta_Y)$ respectively. Define ${\lambda}_{\tilde{X}} = \lambda_X / ( 1 - \rho)$ and ${\lambda}_{\tilde{Y}} = \lambda_Y / \rho$. Define $\lambda_Z = {\lambda}_{\tilde{X}} \cdot {\lambda}_{\tilde{Y}} / \sqrt{{\lambda}_{\tilde{X}}^2 + {\lambda}_{\tilde{Y}}^2}$. Then $Z$ has a Gaussian-like tail, i.e., $\lim_{t \to \infty} t^{-1} \ln \mathbb{P}(Z > t) = -\lambda_Z^2$. In particular, for sufficiently large $t$, there exists constants $\beta,c,C < 0$ such that $c \cdot t^{\beta_X + \beta_Y} \cdot \exp(-\lambda_{Z}^2t^2) \leq \mathbb{P}(Z > t) \leq C \cdot t^{\beta} \cdot \exp(-\lambda_Z^2 t^2)$. 
        \item \textup{(Bounded Tail)} If $X$ and $Y$ have bounded tails with parameters $(b_X,\kappa_X,\beta_X)$ and $(b_Y,\kappa_Y,\beta_Y)$ respectively, then $Z$ is bounded above by $b_Z = (1-\rho)b_X + \rho b_Y.$ Moreover, if $\rho \in (0,1)$, then $Z$ has a bounded tail with parameters $(b_Z,\kappa_Z,\beta_Z)$ given by
            \[
            \beta_Z = \beta_X + \beta_Y,
            \qquad
            \kappa_Z = \kappa_X \kappa_Y (1-\rho)^{-\beta_X}\rho^{-\beta_Y} \cdot \frac{\Gamma(\beta_X+1)\Gamma(\beta_Y+1)}{\Gamma(\beta_X+\beta_Y+1)}.\]
        Equivalently,
\[
\lim_{t\to\infty}\frac{\mathbb{P}\!\left(Z > b_Z - t^{-1}\right)}{\kappa_Z t^{-\beta_Z}} = 1.
\]
    \end{enumerate}
\end{proposition}

\begin{proof}[\underline{Proof of Proposition \ref{prop:tail-convex-combination}}]
    From linearity of expectations, it trivially follows that $\mu_Z = (1 - \rho) \cdot \mu_X + \rho \cdot \mu_Y$. Next we will characterize the tail behavior of $Z$ under different tail behaviors of $X$ and $Y$.
    
    \paragraph{(a) Pareto Tail.} Since $X$ has Pareto tail with parameters $c_X > 0$ and $\alpha_X > 1$, we have that $(1 - \rho) X$ has a pareto tail with parameters $(1 - \rho) c_X > 0$ and $\alpha_{X} > 1$. This is because
    \begin{align*}
        1 = \lim_{x \to \infty } \frac{\mathbb{P}(X > x)}{(c_X / x)^{\alpha_X}} = \lim_{x \to \infty} \frac{\mathbb{P}(X > x / (1 - \rho))}{(c_X / (x / (1 - \rho)))^{\alpha_X}} = \lim_{x \to \infty} \frac{\mathbb{P}((1 - \rho) X > x)}{((1 - \rho) c_X / x)^{\alpha_X}}
    \end{align*}
    Similarly we have that $\rho Y$ has a pareto tail with parameters $\rho c_Y > 0 $ and $\alpha_Y > 1$. Let us denote $\tilde{X} = (1 - \rho) X$ and $\tilde{Y} = \rho Y$, then $Z = \tilde{X} + \tilde{Y}$ and we want to characterize the tail behavior of $Z$.

    \paragraph{\underline{Lower Bound on $\mathbb{P}(Z > t)$.}} We will begin by providing a lower bound on the tail of $Z$. We have that 
    \begin{align}
        \nonumber
        \mathbb{P}(Z > t) &\stackrel{(a)}= \mathbb{P}(\tilde{X} + \tilde{Y} > t), \\
        \nonumber
        &\stackrel{(b)}\geq \mathbb{P}(\max\{\tilde{X}, \tilde{Y}\} > t), \\
        \nonumber
        &\stackrel{(c)}= 1 - (1 - \mathbb{P}(\tilde{X} > t)) ( 1- \mathbb{P}(\tilde{Y} > t)), \\
        \label{eq:tail-Z-lower-bound}
        &\stackrel{(d)}= \mathbb{P}(\tilde{X} > t) + \mathbb{P}(\tilde{Y} > t) - \mathbb{P}(\tilde{X} > t) \mathbb{P}(\tilde{Y} > t),
    \end{align}
    where $(a)$ follows from the definition of $Z$, $(b)$ follows from the fact that $\{\max\{\tilde{X}, \tilde{Y}\} > t\} \implies \{Z > t\}$, $(c)$ follows from the fact that $\mathbb{P}(\max\{\tilde{X}, \tilde{Y}\} > t) = 1 - \mathbb{P}(\tilde{X} < t) \mathbb{P}(\tilde{Y} < t)$ since $\tilde{X}$ and $\tilde{Y}$ are independent, $(d)$ follows trivially. 
    
    \paragraph{\underline{Upper Bound on $\mathbb{P}(Z > t)$.}} Next we will provide an upper bound on the tail of $Z$. Fix a $\delta \in (0,1/2)$. We have that 
    \begin{align}
        \label{eq:tail-Z-upper-bound}
        \mathbb{P}(Z > t) &= \mathbb{P}(\tilde{X} + \tilde{Y} > t)
        \stackrel{(a)}\leq \mathbb{P}(\tilde{X} > (1 - \delta) t) + \mathbb{P}(\tilde{Y} > (1 - \delta) t) + \mathbb{P}(\tilde{X} > \delta t) \mathbb{P}(\tilde{Y} > \delta t),
    \end{align}
    where $(a)$ follows from the fact that the event $\{\tilde{X} + \tilde{Y} > t\}$ implies the event $\{\tilde{X} > (1 - \delta)t \} \cup \{\tilde{Y} > (1 - \delta) t\} \cup \{\tilde{X} > \delta t, \tilde{Y} > \delta t\}$. Next we will consider following cases:
    \begin{enumerate}[label = $(\alph*)$]
        \item \underline{$\alpha_X < \alpha_Y$.} Using \eqref{eq:tail-Z-lower-bound}, we have that $\liminf_{t \to \infty} \mathbb{P}(Z > t) / \mathbb{P}(\tilde{X} > t) \geq 1$ and using \eqref{eq:tail-Z-upper-bound}, we have that $\limsup_{t  \to \infty} \mathbb{P}(Z > t) / \mathbb{P}(\tilde{X} > t) \leq \limsup_{t \to \infty} \mathbb{P}(\tilde{X} > (1 - \delta)t) / \mathbb{P}(\tilde{X} > t) = (1 - \delta)^{-\alpha_X}$. Since the upper bound holds for all $\delta \in (0,1/2)$, we have that $\lim_{t \to \infty} \mathbb{P}(Z > t) / \mathbb{P}(\tilde{X} > t) = 1$. Therefore we have that $Z$ has a pareto tail with parameters $c_Z = (1 - \rho) c_X$ and $\alpha_Z  = \alpha_X$.
        \item \underline{$\alpha_X > \alpha_Y$.} This is completely analogous to the case above.
        \item \underline{$\alpha_X = \alpha_Y = \alpha$.} Note that 
        \begin{align*}
            \lim_{t \to \infty } \frac{\mathbb{P}(\tilde{X} > t) + \mathbb{P}(\tilde{Y} > t)}{(c_Z / t)^\alpha} = 1, \quad \text{where } c_Z = (((1 - \rho) c_X)^\alpha + (\rho c_Y)^{\alpha})^{1 / \alpha}
        \end{align*}
        Using \eqref{eq:tail-Z-lower-bound}, we have that $\liminf_{t \to \infty} \mathbb{P}(Z > t) / (\mathbb{P}(\tilde{X} > t) + \mathbb{P}(\tilde{Y} > t)) \geq 1$ and using \eqref{eq:tail-Z-upper-bound}, we have that $\limsup_{t  \to \infty} \mathbb{P}(Z > t) / (\mathbb{P}(\tilde{X} > t) + \mathbb{P}(\tilde{Y} > t)) \leq \limsup_{t \to \infty} \mathbb{P}(\tilde{X} > (1 - \delta)t) / (\mathbb{P}(\tilde{X} > t) + \mathbb{P}(\tilde{Y} > t)) = (1 - \delta)^{-\alpha}$. Since the upper bound holds for all $\delta \in (0,1/2)$, we have that $\lim_{t \to \infty} \mathbb{P}(Z > t) / (\mathbb{P}(\tilde{X} > t) + \mathbb{P}(\tilde{Y} > t)) = 1$. Therefore we have that $Z$ has a pareto tail with parameters $c_Z = (((1 - \rho) c_X)^\alpha + (\rho c_Y)^{\alpha})^{1 / \alpha}$ and $\alpha_Z  = \alpha$.
    \end{enumerate}
    This completes the proof for the Pareto tails.

    \paragraph{(b) Exponential Tails.} Fix $\epsilon > 0$ and let $\rho \in (0,1)$. Define $\tilde{X} \triangleq (1 - \rho) X$ and $\tilde{Y} \triangleq \rho Y$. We have that $\tilde{X}$ and $\Tilde{Y}$ have exponential tails with parameters $(c_X, \lambda_X / (1 - \rho))$ and $(c_Y, \lambda_Y / \rho)$ respectively. This is because
    \begin{align*}
        1 &= \lim_{x \to \infty} \frac{\mathbb{P}(X > x)}{c_X \exp(-\lambda_X x)} = \lim_{x \to \infty} \frac{\mathbb{P}(X > x /(1 - \rho))}{c_X \exp(-\lambda_X (x / (1 - \rho)))} =  \lim_{x \to \infty} \frac{\mathbb{P}(\tilde{X} > x)}{c_X \exp(-(\lambda_X / (1 - \rho)) x)} \\
        1 &= \lim_{y \to \infty} \frac{\mathbb{P}(Y > y)}{c_Y \exp(-\lambda_Y y)} = \lim_{y \to \infty} \frac{\mathbb{P}(Y > y / \rho)}{c_Y \exp(-\lambda_Y (y / \rho))} =  \lim_{y \to \infty} \frac{\mathbb{P}(\tilde{Y} > y)}{c_Y \exp(-(\lambda_Y / \rho) y)}
    \end{align*}
    Since $\tilde{X}$ and $\tilde{Y}$ have an exponential tail with parameters $(c_X, \lambda_{X} / (1 - \rho))$ and $(c_Y, \lambda_Y / \rho)$ respectively, there exists constants $\tilde{x}_0, \tilde{y}_0$ such that for all $t \geq t_0 = \max\{\tilde{x}_0, \tilde{y}_0\}$, we have that 
    \begin{align}
        \label{eq:tail-bound-exponential-X}
        (1 - \epsilon) c_X \exp(-(\lambda_X / (1 - \rho)) t) &\leq \mathbb{P}(X > t) \leq (1 + \epsilon) c_X \exp( - (\lambda_X / (1 - \rho)) t) \\
        \label{eq:tail-bound-exponential-Y}
        (1 - \epsilon) c_Y \exp( -(\lambda_Y /\rho) t) &\leq \mathbb{P}(Y > t) \leq (1 + \epsilon) c_Y \exp(- (\lambda_Y /\rho) t)
    \end{align}
    Define $\lambda_{\tilde{X}} \triangleq \lambda_X / (1 - \rho)$ and $\lambda_{\tilde{Y}} \triangleq \lambda_Y / \rho$. Furthermore, we define $\underline{\lambda} \triangleq \min\{\lambda_{\tilde{X}}, \lambda_{\tilde{Y}}\}$ and $\bar{\lambda} \triangleq \max\{\lambda_{\tilde{X}}, \lambda_{\tilde{Y}}\}$. 
    
    \paragraph{\underline{Upper Bound on $\mathbb{P}(Z > t)$}}
    Next we want to provide an upper bound on the tail of $Z$. Choose an $s \in (0, \underline{\lambda})$. We will optimize for $s$ later. Then we have that for $t > 1$,
    \begin{align*}
        \mathbb{P}(Z > t) &\stackrel{(a)}= \mathbb{P}(\exp(sZ) > \exp(st)) \stackrel{(b)}\leq \mathbb{E}[\exp(s Z)] \cdot \exp(-st) \stackrel{(c)}= \mathbb{E}[\exp(s\Tilde{X})] \mathbb{E}[\exp(s\Tilde{Y}) ] \exp(- st),
    \end{align*}
    where $(a)$ follows from the fact that $\exp(sx)$ is strictly increasing, $(b)$ follows from Markov's inequality and $(c)$ follows from the fact that $Z = \tilde{X} + \tilde{Y}$ and  $\tilde{X}$ and $\tilde{Y}$ are independent.

    We will now show that there exists constants $c^\prime_X = c(\epsilon, t_0, c_X)$ and $c^\prime_Y = c(\epsilon, t_0, c_Y)$ such that 
    \begin{align}
        \label{eq:mgf-upper-bound}
        \mathbb{E}\left[ \exp(s \tilde{X})\right] \leq c_X^\prime \left(1 - \frac{s}{\lambda_{\tilde{X}}} \right)^{-1}, \ \ \mathbb{E}\left[ \exp(s \tilde{Y}) \right] \leq c_Y^\prime \left( 1 - \frac{s}{\lambda_{\tilde{Y}}}\right)^{-1}.
    \end{align}
    We will show this for $\mathbb{E}[\exp(s \tilde{X})]$ and the steps for $\mathbb{E}[\exp(s \tilde{Y})]$ will follow analogously. Further, we will assume that $t_0 > \max\{1, (1 + \epsilon)c_X\}$. If $t_0 < \max\{1, (1 + \epsilon) c_X\}$, the analysis will require trivial modifications. We have that 
    
    \begin{align*}
        \mathbb{E}[\exp(s\tilde{X})] &\stackrel{(a)}= \int_{0}^\infty \mathbb{P}(\exp(s \tilde{X}) > t) dt, \\
        &\stackrel{(b)}= 1 + \int_{1}^\infty \mathbb{P}( \tilde{X} > \ln t / s) dt, \\
        &\stackrel{(c)}\leq 1 + \int_{1}^{t_0} 1 dt + \int_{t_0}^\infty (1 + \epsilon) c_{X} \exp( - (\lambda_{X} / s(1 - \rho)) \ln t) dt , \\
        &\stackrel{(d)}\leq  t_0 + (1 + \epsilon) c_{X} \int_{1}^\infty \exp(- (\lambda_X / s(1 - \rho)) \ln t) dt, \\
        &\stackrel{(e)}=  t_0 + (1 + \epsilon) c_X \frac{1}{(\lambda_X / (s (1 - \rho))) - 1}, \\
        &\stackrel{(f)}\leq \frac{c^\prime}{(\lambda_X / (s (1 - \rho))) - 1},
    \end{align*}
    where $(a)$ follows from tail sum formula for expectation \citep{durrett2019probability}, $(b)$ follows from the fact that $\mathbb{P}(\exp(s\Tilde{X}) > t) = \mathbb{P}(\tilde{X} > \ln t / s)$, $(c)$ follows from \ref{eq:tail-bound-exponential-X}, $(d)$ follows trivally, $(e)$ follows from the fact that $\int_{1}^\infty \exp(- (\lambda_X / s(1 - \rho)) \ln t) dt = \frac{s(1 - \rho)}{\lambda_X - s(1 - \rho)}$, $(f)$ follows from an appropriate choice of $c^\prime$.
    
    Therefore, we have that
    \begin{align*}
        \mathbb{P}(Z > t) &\stackrel{(a)}\leq c_{X}^\prime c_{Y}^\prime \exp\left(-st - \ln \left(1 - \frac{s }{\lambda_{\tilde{X}}} \right) - \ln \left( 1 - \frac{s}{\lambda_{\tilde{Y}}} \right) \right), \\
        &\stackrel{(b)}\leq c_X^\prime c_Y^\prime \exp\left( -st - \left( \frac{\underline{\lambda}}{\lambda_{\tilde{X}}} + \frac{\underline{\lambda}}{\lambda_{\tilde{Y}}} \right) \ln \left( 1 - \frac{s}{\underline{\lambda}}\right)\right), \\
        &\stackrel{(c)}= \exp(\underline{\lambda}) c_X^\prime c_Y^\prime t^{1 + (\bar{\lambda} / \underline{\lambda})} \exp(-\underline{\lambda} t),
    \end{align*}
    where $(a)$ follows from \eqref{eq:mgf-upper-bound}, $(b)$ follows from the fact that $-\ln\left(1 - \frac{s}{\lambda_{\tilde{X}}} \right) \leq - \frac{\underline{\lambda}}{\lambda_{\tilde{X}}} \ln \left( 1 - \frac{s}{\underline{\lambda}}\right)$ and $-\ln\left(1 - \frac{s}{\lambda_{\tilde{Y}}} \right) \leq - \frac{\underline{\lambda}}{\lambda_{\tilde{Y}}} \ln \left( 1 - \frac{s}{\underline{\lambda}}\right)$ and $(c)$ follows from choosing $s = (1 - 1/t) \underline{\lambda}$. Define $C = \exp(\underline{\lambda}) c_X^\prime c_Y^\prime$. Then we have that $\mathbb{P}(Z > t) \leq C t^{1 + (\bar{\lambda} / \underline{\lambda})} \exp(-\underline{\lambda} t)$.

    \paragraph{\underline{Lower Bound on $\mathbb{P}(Z > t)$}}
    Next we want to provide a lower bound on the tail of $Z$. 
    Define $L = \tilde{X} \mathbbm{1}\{\lambda_{\tilde{X}} < \lambda_{\tilde{Y}}\} + \tilde{Y} \mathbbm{1}\{\lambda_{\tilde{X}} > \lambda_{\tilde{Y}}\}$. Let $\underline{c} = \min\{c_X, c_Y\}$.
    For all $t \geq t_0$, we have that, 
    \begin{align*}
        \mathbb{P}(Z > t) &\stackrel{(a)}\geq \mathbb{P}(L > t) \stackrel{(b)}\geq (1 - \epsilon) \underline{c} \exp(- \underline{\lambda} t) := c \exp(-\underline{\lambda} t),
    \end{align*}
    where $(a)$ follows from the fact that $\{L > t\}$ implies $\{Z > t\}$ and $(b)$ follows from \eqref{eq:tail-bound-exponential-X} and \eqref{eq:tail-bound-exponential-Y}.
    This completes the proof for the exponential tails.

    \paragraph{(c) Gaussian-like Tails.} Fix $\epsilon > 0$ and let $\rho \in (0,1)$. Define $\tilde{X} = (1 - \rho) X$ and $\tilde{Y} = \rho Y$. Define $\lambda_{\tilde{X}} \triangleq \lambda_X / (1 - \rho)$ and $\lambda_{\tilde{Y}} \triangleq \lambda_Y / \rho$. Furthermore, define $c_{\tilde{X}} = c_X \cdot (1 - \rho)^{\beta_X}$ and $c_{\tilde{Y}} = c_Y \cdot \rho^{\beta_Y}$.
    We have that $\tilde{X}$ and $\tilde{Y}$ have Gaussian-like tails with parameters $(c_{\tilde{X}}, \beta_X, \lambda_{\tilde{X}})$ and $(c_{\tilde{Y}}, \beta_Y, \lambda_{\tilde{Y}})$ respectively. This is because
    \begin{align*}
        1 &= \lim_{y \to \infty} \frac{\mathbb{P}(Y > y)}{c_Y y^{-\beta_Y} \exp(-\lambda_Y^2 y^2)} = \lim_{y \to \infty} \frac{\mathbb{P}(Y > y/\rho)}{c_Y (y / \rho)^{-\beta_Y} \exp(-\lambda_Y^2 (y / \rho)^2)} = \lim_{y \to \infty} \frac{\mathbb{P}(\tilde{Y} > y)}{c_{\tilde{Y}} \cdot y^{-\beta_Y} \cdot \exp(-\lambda_{\tilde{Y}}^2 y^2)}
    \end{align*}
    The case of $\tilde{X}$ follows analogously. Since $\tilde{X}$ and $\tilde{Y}$ have a Gaussian-like tail with parameters $(c_{\tilde{X}}, \beta_X, \lambda_{\tilde{X}})$ and $(c_{\tilde{Y}}, \beta_Y, \lambda_{\tilde{Y}})$ respectively, there exists constants $\tilde{x}_0, \tilde{y}_0$ such that for all $t \geq t_0 = \max\{\tilde{x}_0, \tilde{y}_0\}$, we have that
    \begin{align}
        \label{eq:tilde-X-gaussian-tail}
        (1 - \epsilon) c_{\tilde{X}} x^{-\beta_X} \exp(-\lambda_{\tilde{X}}^2 x^2 ) &\leq \mathbb{P}(\tilde{X} > x) \leq (1 + \epsilon) c_{\tilde{X}} x^{-\beta_X} \exp(-\lambda_{\tilde{X}}^2 x^2) \\
        \label{eq:tilde-Y-gaussian-tail}
        (1 - \epsilon) c_{\tilde{Y}} y^{-\beta_Y} \exp(-\lambda_{\tilde{Y}}^2 y^2 ) &\leq \mathbb{P}(\tilde{Y} > y) \leq (1 + \epsilon) c_{\tilde{Y}} y^{-\beta_Y} \exp(-\lambda_{\tilde{Y}}^2 y^2) 
    \end{align}

    \paragraph{\underline{Upper Bound on $\mathbb{P}(Z > t)$}} Next we will provide an upper bound on the tail of $Z$. Choose an $s  > 0$. We will optimize for $s$ later. Then we have that for $t > 1$,
    \begin{align*}
        \mathbb{P}(Z > t) &\stackrel{(a)}= \mathbb{P}(\exp(sZ) > \exp(st)) \stackrel{(b)}\leq \exp(-st) \cdot \mathbb{E}[\exp(sZ)] \stackrel{(c)}= \mathbb{E}[\exp(s\tilde{X})] \mathbb{E}[\exp(s\tilde{Y})] \exp(-st),
    \end{align*}
    where $(a)$ follows from the fact that $\exp(sx)$ is strictly increasing, $(b)$ follows from Markov's inequality and $(c)$ follows from the fact that $Z = \tilde{X} + \tilde{Y}$ and $\tilde{X}$ and $\tilde{Y}$ are independent.

    We have that
    \begin{align*}
        \mathbb{E}[\exp(s \tilde{X})] &\stackrel{(a)}= \int_{0}^\infty \mathbb{P}(\exp(s\tilde{X}) > t) dt, \\
        &\stackrel{(b)}= 1 + \int_{1}^\infty \mathbb{P}(\tilde{X} > \ln t / s) dt, \\
        &\stackrel{(c)}= 1 + s \int_0^{\infty} \exp(su) \mathbb{P}(\tilde{X} > u) du, \\ 
        &\stackrel{(d)}\leq \exp(s t_0) + s \int_{t_0}^\infty  (1 + \epsilon) u^{-\beta_X} \exp(-\lambda_{\tilde{X}}^2 u^2 + su) du, 
    \end{align*}
    where $(a)$ follows from the tail sum formula for expectation \cite{durrett2019probability}, $(b)$ follows from the fact that $\mathbb{P}(\exp(s\tilde{X}) > t) = \mathbb{P}(\tilde{X} > \ln t / s)$, $(c)$ follows from a change of variable $t = \exp(su)$ and $(d)$ follows from the fact that for $u \in (0,t_0)$, $\mathbb{P}(\tilde{X} > u) \leq 1$ and for $u > t_0$, we have that $\mathbb{P}(\tilde{X} > u) \leq (1 + \epsilon) u^{\beta_X}\exp(\lambda_{\tilde{X}}^2 u^2)$.
    
    Define $I(s) = \int_{t_0}^\infty u^{-\beta_X} \exp(-\lambda_{\tilde{X}}^2 u^2 + su)du$ and $I^\prime(s) = s(1 + \epsilon) I(s)$. Note that we can write $su - \lambda_{\tilde{X}}^2 u^2 = - \lambda_{\tilde{X}}^2(u - \lambda_{\tilde{X}}^{-2}s/2)^2 + \lambda_{\tilde{X}}^{-2}s^2 / 4$. Define $u_0 = \lambda_{\tilde{X}}^{-2}s/2$ and assume that $s$ is large enough. Then we have that 
    \begin{align*}
        I(s) &\stackrel{(a)}= \exp(\lambda_{\tilde{X}}^{-2} s^2 / 4) \int_{t_0}^{\infty} u^{-\beta_{X}} \exp(-\lambda_{\tilde{X}}^2 (u - u_0)^2) du, \\
        &\stackrel{(b)}\leq \exp(\lambda_{\tilde{X}}^{-2} s^2 / 4) t_{0}^{-\beta_X} \int_{t_0}^\infty \exp(-\lambda_{\tilde{X}}^{2} (u - u_0)^2) du, \\
        &\stackrel{(c)}\leq \exp(\lambda_{\tilde{X}}^{-2} s^2/4) t_0^{-\beta_X} \int_{-\infty}^{\infty} \exp(-\lambda_{\tilde{X}}^2 (u - u_0)^2) du, \\
        &\stackrel{(d)}= \exp(\lambda_{X}^{-2} s^2 / 4) t_0^{-\beta_X} \lambda_{\tilde{X}}^{-1} \sqrt{\pi},
    \end{align*}
    where $(a)$ follows from completing the square, $(b)$ follows from the fact that $u^{-\beta_X} \leq t_0^{-\beta_X}$ for all $u \geq t_0$, $(c)$ follows trivially, $(d)$ follows from the fact that $\int_{-\infty}^\infty \exp(-\lambda_{\tilde{X}}^2 (u - u_0)^2) du = \sqrt{\pi / \lambda_{\tilde{X}}^2}$. 
    
    Therefore, there are constants $C_{\tilde{X}}, C_{\tilde{X}}^\prime$ such that for large enough $s$, we have that 
    \begin{align}
        \label{eq:gaussian-mgf-tilde-X-upper-bound}
        \mathbb{E}[\exp(s \tilde{X})] \leq \exp(st_0) + C_{\tilde{X}} s \exp(\lambda_{\tilde{X}}^{-2} s^2 / 4) \leq C_{\tilde{X}}^\prime \cdot s \cdot \exp(\lambda_{\tilde{X}}^{-2} s^2 / 4)
    \end{align}
    
    Similarly, there are constants $C_{\tilde{Y}}, C_{\tilde{Y}}^\prime$ such that for large enough $s$, we have that 
    \begin{align}
        \label{eq:gaussian-mgf-tilde-Y-upper-bound}
        \mathbb{E}[\exp(s \tilde{Y})] \leq \exp(s t_0) + C_{\tilde{Y}} s \exp(\lambda_{\tilde{Y}}^{-2} s^2 / 4) \leq C_{\tilde{Y}}^\prime \cdot s \cdot \exp(\lambda_{\tilde{X}}^{-2} s^2 / 4)
    \end{align}
    Using Markov's inequality,  \eqref{eq:gaussian-mgf-tilde-X-upper-bound} and \eqref{eq:gaussian-mgf-tilde-Y-upper-bound}, we have that,
    \begin{align*}
        \mathbb{P}(Z > t) \leq C \cdot s^2 \cdot \exp(-st + (\lambda_{\tilde{X}}^{-2} + \lambda_{\tilde{Y}}^{-2}) s^2 / 4) \leq C \cdot t^2 \cdot \exp(-\lambda_{Z}^2 t^2),
    \end{align*}
    where the last inequality follows from choosing $s = 2\lambda_{Z}^2t$ where recall that $\lambda_{Z}^2 = \lambda_{\tilde{X}}^2 \lambda_{\tilde{Y}}^2 / (\lambda_{\tilde{X}}^2 + \lambda_{\tilde{Y}}^2)$.

    \paragraph{\underline{Lower Bound on $\mathbb{P}(Z > t)$}} Next we will provide a lower bound on the tail of $Z$. For any $t$, consider the event $E_t = \{\tilde{X} > (\lambda_{\tilde{Y}}^2 / (\lambda_{\tilde{X}}^2 + \lambda_{\tilde{Y}}^2)) \cdot  t\} \cap \{\tilde{Y} > (\lambda_{\tilde{X}}^2 / (\lambda_{\tilde{X}}^2 + \lambda_{\tilde{Y}}^2)) \cdot t\}$. Now the event $E_t$ implies the event $\{Z > t\}$ since $Z = \tilde{X} + \tilde{Y}$. Therefore we have that,
    \begin{align*}
        \mathbb{P}(Z > t) &\stackrel{(a)}\geq \mathbb{P}(\tilde{X} > (\lambda_{\tilde{Y}}^2 / (\lambda_{\tilde{X}}^2 + \lambda_{\tilde{Y}}^2)) \cdot t) \cdot \mathbb{P}(\tilde{Y} > (\lambda_{\tilde{X}}^2 / (\lambda_{\tilde{X}}^2 + \lambda_{\tilde{Y}}^2)) \cdot t), \\
        &\stackrel{(b)}\geq c \cdot  t^{-\beta_{X} - \beta_{Y}} \cdot \exp\left(-\lambda_{\tilde{X}}^2  \cdot \frac{\lambda_{\tilde{Y}}^4}{(\lambda_{\tilde{X}}^2 + \lambda_{\tilde{Y}}^2)^2} \cdot t^2 \right) \cdot \exp\left( - \lambda_{\tilde{Y}}^2 \cdot \frac{\lambda_{\tilde{X}}^4}{(\lambda_{\tilde{X}}^2 + \lambda_{\tilde{Y}}^2)^2} \cdot t^2 \right), \\
        &\stackrel{(c)}= c \cdot t^{-\beta_{x} - \beta_{Y}} \cdot \exp\left( - \frac{\lambda_{\tilde{X}}^2 \lambda_{\tilde{Y}}^2}{\lambda_{\tilde{X}}^2 + \lambda_{\tilde{Y}}^2} \cdot t \right) = c \cdot t^{-\beta_{X} - \beta_{Y}} \cdot \exp(-\lambda_{Z}^2 t^2),
    \end{align*}
    where $(a)$ follows from the fact that $\tilde{X}$ and $\tilde{Y}$ are independent, $(b)$ follows from \eqref{eq:tilde-X-gaussian-tail} and \eqref{eq:tilde-Y-gaussian-tail} and $(c)$ follows trivially and the last equality follows from the definition of $\lambda_{Z}.$
    This completes the proof for the Gaussian-like tails.

    \paragraph{(d) Bounded Tails.} Fix $\epsilon>0$ and let $\rho\in(0,1)$. Define
\[
\tilde{X} \triangleq (1-\rho)X,
\qquad
\tilde{Y} \triangleq \rho Y,
\qquad
Z=\tilde{X}+\tilde{Y}.
\]
Since $X\le b_X$ and $Y\le b_Y$ almost surely, we have $\tilde{X}\le (1-\rho)b_X$ and $\tilde{Y}\le \rho b_Y$ almost surely. Hence $Z$ is bounded above by
\[
b_Z \triangleq (1-\rho)b_X + \rho b_Y.
\]

Next we characterize the behavior of $\mathbb{P}(Z>b_Z-t^{-1})$ as $t\to\infty$.
Define the {\it gaps to the upper endpoints}
\[
U \triangleq (1-\rho)b_X - \tilde{X} = (1-\rho)(b_X-X),
\qquad
V \triangleq \rho b_Y - \tilde{Y} = \rho(b_Y-Y).
\]
Then $U,V\ge 0$, $U$ and $V$ are independent, and
\[
b_Z - Z = U+V.
\]
Therefore, for any $t>0$,
\begin{equation}
\label{eq:bounded-tail-Z-gap}
\mathbb{P}\!\left(Z > b_Z - t^{-1}\right)
=
\mathbb{P}\!\left(U+V < t^{-1}\right).
\end{equation}

\paragraph{\underline{Step 1: Tail behavior of $U$ and $V$ near $0$.}}
Since $X$ has a bounded tail with parameters $(b_X,\kappa_X,\beta_X)$, there exists $t_{X,0}$ such that for all $s\ge t_{X,0}$,
\begin{equation}
\label{eq:bounded-tail-X}
(1-\epsilon)\kappa_X s^{-\beta_X}
\le
\mathbb{P}\!\left(X>b_X-s^{-1}\right)
\le
(1+\epsilon)\kappa_X s^{-\beta_X}.
\end{equation}
Similarly, since $Y$ has a bounded tail with parameters $(b_Y,\kappa_Y,\beta_Y)$, there exists $t_{Y,0}$ such that for all $s\ge t_{Y,0}$,
\begin{equation}
\label{eq:bounded-tail-Y}
(1-\epsilon)\kappa_Y s^{-\beta_Y}
\le
\mathbb{P}\!\left(Y>b_Y-s^{-1}\right)
\le
(1+\epsilon)\kappa_Y s^{-\beta_Y}.
\end{equation}

Now note that for any $t>0$,
\[
\mathbb{P}\!\left(U < t^{-1}\right)
=
\mathbb{P}\!\left(\tilde{X}>(1-\rho)b_X - t^{-1}\right)
=
\mathbb{P}\!\left(X>b_X - ((1-\rho)t)^{-1}\right).
\]
Applying \eqref{eq:bounded-tail-X} with $s=(1-\rho)t$, we obtain that for all $t$ large enough,
\begin{equation}
\label{eq:bounded-tail-U}
(1-\epsilon)\kappa_X (1-\rho)^{-\beta_X} t^{-\beta_X}
\le
\mathbb{P}\!\left(U<t^{-1}\right)
\le
(1+\epsilon)\kappa_X (1-\rho)^{-\beta_X} t^{-\beta_X}.
\end{equation}
Define
\[
\kappa_{\tilde{X}} \triangleq \kappa_X (1-\rho)^{-\beta_X}.
\]
Similarly,
\[
\mathbb{P}\!\left(V < t^{-1}\right)
=
\mathbb{P}\!\left(Y>b_Y-(\rho t)^{-1}\right),
\]
so by \eqref{eq:bounded-tail-Y}, for all $t$ large enough,
\begin{equation}
\label{eq:bounded-tail-V}
(1-\epsilon)\kappa_Y \rho^{-\beta_Y} t^{-\beta_Y}
\le
\mathbb{P}\!\left(V<t^{-1}\right)
\le
(1+\epsilon)\kappa_Y \rho^{-\beta_Y} t^{-\beta_Y}.
\end{equation}
Define
\[
\kappa_{\tilde{Y}} \triangleq \kappa_Y \rho^{-\beta_Y}.
\]

Equivalently, letting $x=t^{-1}$ and taking $x$ sufficiently small, there exists $x_0>0$ such that for all $x\in(0,x_0]$,
\begin{equation}
\label{eq:small-x-FU}
(1-\epsilon)\kappa_{\tilde{X}} x^{\beta_X}
\le
F_U(x)
\le
(1+\epsilon)\kappa_{\tilde{X}} x^{\beta_X},
\qquad
F_U(x)\triangleq \mathbb{P}(U<x),
\end{equation}
and
\begin{equation}
\label{eq:small-x-FV}
(1-\epsilon)\kappa_{\tilde{Y}} x^{\beta_Y}
\le
F_V(x)
\le
(1+\epsilon)\kappa_{\tilde{Y}} x^{\beta_Y},
\qquad
F_V(x)\triangleq \mathbb{P}(V<x).
\end{equation}

\paragraph{\underline{Step 2: Convolution asymptotics for $\mathbb{P}(U+V<x)$ as $x\downarrow 0$.}}
Fix $x\in(0,x_0]$. By independence and the standard convolution identity,
\begin{equation}
\label{eq:convolution}
\mathbb{P}(U+V<x)
=
\int_{0}^{x} F_U(x-u)\, dF_V(u),
\end{equation}
where the integral is a Riemann--Stieltjes integral. Using \eqref{eq:small-x-FU}, we obtain
\begin{equation}
\label{eq:convolution-bounds-FU}
(1-\epsilon)\kappa_{\tilde{X}}
\int_0^x (x-u)^{\beta_X}\, dF_V(u)
\le
\mathbb{P}(U+V<x)
\le
(1+\epsilon)\kappa_{\tilde{X}}
\int_0^x (x-u)^{\beta_X}\, dF_V(u).
\end{equation}

We now analyze the integral
\[
I(x)\triangleq \int_0^x (x-u)^{\beta_X}\, dF_V(u).
\]
Let $g(u)=(x-u)^{\beta_X}$ on $[0,x]$. Since $g$ is continuously differentiable on $[0,x]$ and $g(x)=0$, an integration-by-parts identity for Riemann--Stieltjes integrals yields
\[
I(x)
=
g(x)F_V(x)-g(0)F_V(0)-\int_0^x F_V(u)\, dg(u)
=
\beta_X\int_0^x F_V(u)\,(x-u)^{\beta_X-1}\, du,
\]
where we used that $F_V(0)=\mathbb{P}(V<0)=0$. Applying \eqref{eq:small-x-FV} inside the integral gives
\begin{equation}
\label{eq:I-bounds}
\beta_X(1-\epsilon)\kappa_{\tilde{Y}}
\int_0^x u^{\beta_Y}(x-u)^{\beta_X-1}\, du
\le
I(x)
\le
\beta_X(1+\epsilon)\kappa_{\tilde{Y}}
\int_0^x u^{\beta_Y}(x-u)^{\beta_X-1}\, du.
\end{equation}

The remaining integral can be evaluated via the change of variables $u=xr$:
\begin{align}
\label{eq:beta-integral}
\int_0^x u^{\beta_Y}(x-u)^{\beta_X-1}\, du
&=
x^{\beta_X+\beta_Y}\int_0^1 r^{\beta_Y}(1-r)^{\beta_X-1}\, dr \nonumber\\
&=
x^{\beta_X+\beta_Y}\,\mathrm{B}(\beta_Y+1,\beta_X),
\end{align}
where $\mathrm{B}(\cdot,\cdot)$ denotes the Beta function. Using the identity
\[
\mathrm{B}(\beta_Y+1,\beta_X)
=
\frac{\Gamma(\beta_Y+1)\Gamma(\beta_X)}{\Gamma(\beta_X+\beta_Y+1)},
\]
and $\beta_X\Gamma(\beta_X)=\Gamma(\beta_X+1)$, we conclude that
\begin{equation}
\label{eq:I-asymptotic-constant}
I(x)
=
\left(1+o_\epsilon(1)\right)\kappa_{\tilde{Y}}
\cdot
\frac{\Gamma(\beta_X+1)\Gamma(\beta_Y+1)}{\Gamma(\beta_X+\beta_Y+1)}
\cdot x^{\beta_X+\beta_Y},
\qquad x\downarrow 0,
\end{equation}
in the sense of two-sided bounds with multiplicative factors $(1\pm\epsilon)$ from \eqref{eq:I-bounds}.

Combining \eqref{eq:convolution-bounds-FU}, \eqref{eq:I-bounds}, and \eqref{eq:beta-integral} yields: for all $x\in(0,x_0]$,
\begin{align}
\label{eq:UV-small-sum-bounds}
(1-\epsilon)^2 \kappa_{\tilde{X}}\kappa_{\tilde{Y}}
\frac{\Gamma(\beta_X+1)\Gamma(\beta_Y+1)}{\Gamma(\beta_X+\beta_Y+1)}
x^{\beta_X+\beta_Y}
\le
\mathbb{P}(U+V<x)
\le
(1+\epsilon)^2 \kappa_{\tilde{X}}\kappa_{\tilde{Y}}
\frac{\Gamma(\beta_X+1)\Gamma(\beta_Y+1)}{\Gamma(\beta_X+\beta_Y+1)}
x^{\beta_X+\beta_Y}.
\end{align}

\paragraph{\underline{Step 3: Translate back to $Z$.}}
Set $x=t^{-1}$ in \eqref{eq:UV-small-sum-bounds} and use \eqref{eq:bounded-tail-Z-gap}. For all $t$ sufficiently large,
\begin{align*}
(1-\epsilon)^2 \kappa_{\tilde{X}}\kappa_{\tilde{Y}}
\frac{\Gamma(\beta_X+1)\Gamma(\beta_Y+1)}{\Gamma(\beta_X+\beta_Y+1)}
t^{-(\beta_X+\beta_Y)}
\le
\mathbb{P}\!\left(Z>b_Z-t^{-1}\right)
\le
(1+\epsilon)^2 \kappa_{\tilde{X}}\kappa_{\tilde{Y}}
\frac{\Gamma(\beta_X+1)\Gamma(\beta_Y+1)}{\Gamma(\beta_X+\beta_Y+1)}
t^{-(\beta_X+\beta_Y)}.
\end{align*}
Since $\epsilon>0$ was arbitrary, we conclude that
\[
\lim_{t\to\infty}
\frac{\mathbb{P}\!\left(Z>b_Z-t^{-1}\right)}
{\kappa_Z t^{-(\beta_X+\beta_Y)}}
=1,
\]
where
\[
\kappa_Z
=
\kappa_{\tilde{X}}\kappa_{\tilde{Y}}
\frac{\Gamma(\beta_X+1)\Gamma(\beta_Y+1)}{\Gamma(\beta_X+\beta_Y+1)}
=
\kappa_X\kappa_Y(1-\rho)^{-\beta_X}\rho^{-\beta_Y}
\frac{\Gamma(\beta_X+1)\Gamma(\beta_Y+1)}{\Gamma(\beta_X+\beta_Y+1)}.
\]
Thus $Z$ has a bounded tail with parameters $(b_Z,\kappa_Z,\beta_Z)$ where $b_Z=(1-\rho)b_X+\rho b_Y$ and $\beta_Z=\beta_X+\beta_Y$.
This completes the proof for the bounded tails.
\end{proof}

\begin{proposition}
    \label{prop:maximum-n-convex-combination}
    Let $X$ and $Y$ be two random variables with non-negative support and finite means $\mu_X < \infty$ and $\mu_Y < \infty$ respectively. Fix $\rho \in [0,1]$. Define $Z = (1 - \rho) X + \rho Y$. Let $Z_1, Z_2, \dots, Z_n$ be $n$ i.i.d copies of $Z$ and let $Z_{(n:n)} = \max_{1 \leq k \leq n } Z_k$. Then we have that,
    \begin{enumerate}[label = $(\alph*)$]
        \item \textup{(Pareto Tail)} If $X$ and $Y$ have Pareto tail with parameters $(c_X, \alpha_X)$ and $(c_Y, \alpha_Y)$ respectively. Then we have that 
        \begin{itemize}
            \item \underline{$\alpha_X < \alpha_Y$}. We have that $\lim_{n \to \infty} \mathbb{E}[Z_{(n:n)}] / ( c_{X} \cdot (1 - \rho) \cdot \Gamma(1 - 1 /\alpha_X) \cdot n^{1 / \alpha_X}) = 1$.
            \item \underline{$\alpha_X > \alpha_Y$.} We have that $\lim_{n \to \infty} \mathbb{E}[Z_{(n:n)}] / (c_{Y} \cdot \rho \cdot \Gamma(1 - 1 / \alpha_Y) \cdot n^{1 / \alpha_Y}) = 1$.
            \item \underline{$\alpha_X = \alpha_Y = \alpha$.} Define $c_Z = (((1 - \rho) c_X)^\alpha + (\rho c_Y)^\alpha)^{1 / \alpha}$. We have that $\lim_{n \to \infty} \mathbb{E}[Z_{(n:n)}] / (c_Z \cdot \Gamma(1 - 1 / \alpha) \cdot n^{1 / \alpha}) = 1.$
        \end{itemize}
        
        \item \textup{(Exponential Tail)} If $X$ and $Y$ have exponential tail with parameters $(c_X, \lambda_X)$ and $(c_Y, \lambda_Y)$ respectively. Define $\lambda_Z = \min\{(1 - \rho)^{-1} \lambda_X, \rho^{-1} \lambda_Y \}$. We have that $\lim_{n \to \infty} \mathbb{E}[Z_{(n:n)}] / (\ln n / \lambda_Z) = 1.$

        \item \textup{(Gaussian-like Tail)} If $X$ and $Y$ have Gaussian-like tail with parameters $(c_X, \beta_X, \lambda_X)$ and $(c_Y, \beta_Y, \lambda_Y)$ respectively. Define $\lambda_{\tilde{X}} = \lambda_X / (1 - \rho)$, $\lambda_{\tilde{Y}} = \lambda_{Y} / \rho$ and $\lambda_Z = \lambda_{\tilde{X}} \lambda_{\tilde{Y}} / \sqrt{\lambda_{\tilde{X}}^2 + \lambda_{\tilde{Y}}^2}$.
        We have that $\lim_{n \to \infty} \mathbb{E}[Z_{(n:n)}] / (\sqrt{\ln n } / \lambda_Z) = 1.$

        \item \textup{(Bounded Tail)} If $X$ and $Y$ have bounded tail with parameters $(b_X, \kappa_X, \beta_X)$ and $(b_Y, \kappa_Y, \beta_Y)$. Define $b_Z = (1 - \rho) b_X + \rho b_Y$. We have that $\lim_{n \to \infty} \mathbb{E}[Z_{(n:n)}] = b_Z$.
    \end{enumerate}
\end{proposition}

\begin{proof}[\underline{Proof of Proposition \ref{prop:maximum-n-convex-combination}}] We provide the proof for different tail distributions of $X$ and $Y$.
\paragraph{(a) Pareto Tail.} The result follows from Proposition \ref{prop:expected-maximum-n} and \ref{prop:tail-convex-combination}.

\paragraph{(b) Exponential Tail.} Let $\bar{\lambda} = \max\{(1 - \rho)^{-1} \lambda_X, \rho^{-1} \lambda_Y\}$. Using Proposition \ref{prop:tail-convex-combination} and \eqref{eq:max-tail}, for $t$ large enough, we have that
\begin{align}
        \label{eq:maximum-tail-upper-lower-bound-exponential-tail}
        1 - (1 - c \exp(-{\lambda_Z}t))^n \leq \mathbb{P}(Z_{(n:n)} > t) \leq 1 - (1 - Ct^{1 + (\bar{\lambda} / {\lambda_Z})} \exp(-{\lambda_Z}t ))^n 
    \end{align}
    Choose a $\delta \in (0, \lambda_Z)$. There exists $t_0^\prime$ such for all $t \geq t_0^\prime$, we have that $t^{1 + \bar{\lambda} / \lambda_{Z}} \leq \exp(\delta t)$. Therefore, we have that 
    \begin{align}
        \label{eq:maximum-tail-upper-lower-bound-exponential-tail-2}
        1 - (1 - c \exp(-{\lambda_Z}t))^n \leq \mathbb{P}(Z_{(n:n)} > t) \leq 1 - (1 - C \exp(-({\lambda_Z} - \delta) t))^n 
    \end{align}
    Using the analysis in the proof of Proposition \ref{prop:expected-maximum-n}(b), we have that
    \begin{align*}
        \frac{1}{{\lambda}_Z} \leq \liminf_{n \to \infty} \frac{\mathbb{E}[Z_{(n:n)}]}{\ln n} \leq \limsup_{n \to \infty} \frac{\mathbb{E}[Z_{(n:n)}]}{\ln n} \leq \frac{1}{{\lambda}_Z - \delta}
    \end{align*}
    Since this set of inequalities is true for all $\delta \in (0,{\lambda}_Z)$, we have that $\lim_{n \to \infty} {\mathbb{E}[Z_{(n:n)}]} / {\ln n} = {1} / {\lambda_Z}$.

\paragraph{(c) Gaussian-like Tail.}
The cases $\rho=0$ and $\rho=1$ are immediate (then $Z=X$ or $Z=Y$), so assume $\rho\in(0,1)$.
Recall $\lambda_{\tilde{X}}=\lambda_X/(1-\rho)$, $\lambda_{\tilde{Y}}=\lambda_Y/\rho$, and
\[
\lambda_Z=\frac{\lambda_{\tilde{X}}\lambda_{\tilde{Y}}}{\sqrt{\lambda_{\tilde{X}}^2+\lambda_{\tilde{Y}}^2}}.
\]
By Proposition~\ref{prop:tail-convex-combination}\textup{(c)}, there exist constants
$t_0<\infty$, $c,C>0$ and exponents $\gamma_1,\gamma_2\in\mathbb{R}$ such that for all $t\ge t_0$,
\begin{equation}\label{eq:Z-gaussian-two-sided}
c\, t^{\gamma_1}e^{-\lambda_Z^2 t^2}\ \le\ \mathbb P(Z>t)\ \le\ C\, t^{\gamma_2}e^{-\lambda_Z^2 t^2}.
\end{equation}
Combining \eqref{eq:Z-gaussian-two-sided} with \eqref{eq:max-tail}, for all $t\ge t_0$,
\begin{equation}\label{eq:max-tail-gaussian-Z}
1-\Big(1-c\,t^{\gamma_1}e^{-\lambda_Z^2 t^2}\Big)^n
\ \le\ \mathbb P(Z_{(n:n)}>t)\ \le\
1-\Big(1-C\,t^{\gamma_2}e^{-\lambda_Z^2 t^2}\Big)^n .
\end{equation}
Define
\[
a_n:=\frac{\sqrt{\ln n}}{\lambda_Z},
\]
and fix $\delta\in(0,1/2)$.

{\it Upper bound.}
Let $\bar t_n=(1+\delta)a_n$. By \eqref{eq:tailsum},
\[
\mathbb E[Z_{(n:n)}]
=\int_0^{\bar t_n}\mathbb P(Z_{(n:n)}>t)\,dt+\int_{\bar t_n}^\infty\mathbb P(Z_{(n:n)}>t)\,dt
\le \bar t_n+\int_{\bar t_n}^\infty\mathbb P(Z_{(n:n)}>t)\,dt.
\]
Using $1-(1-y)^n\le ny$ from \eqref{eq:1miny-power} and the upper bound in \eqref{eq:Z-gaussian-two-sided},
for $n$ large enough so that $\bar t_n\ge t_0$,
\[
\mathbb P(Z_{(n:n)}>t)\le n\,\mathbb P(Z>t)\le nC\,t^{\gamma_2}e^{-\lambda_Z^2 t^2},
\qquad t\ge \bar t_n.
\]
Hence
\[
\int_{\bar t_n}^\infty\mathbb P(Z_{(n:n)}>t)\,dt
\le nC\int_{\bar t_n}^\infty t^{\gamma_2}e^{-\lambda_Z^2 t^2}\,dt.
\]
An integration-by-parts bound gives that for some constant $K<\infty$ and all sufficiently large $x$,
\[
\int_x^\infty t^{\gamma_2}e^{-\lambda_Z^2 t^2}\,dt\le K\,x^{\gamma_2-1}e^{-\lambda_Z^2 x^2}.
\]
Applying this with $x=\bar t_n$ yields
\[
\int_{\bar t_n}^\infty\mathbb P(Z_{(n:n)}>t)\,dt
\le K n\,\bar t_n^{\gamma_2-1}e^{-\lambda_Z^2 \bar t_n^2}.
\]
But $e^{-\lambda_Z^2\bar t_n^2}=e^{-(1+\delta)^2\ln n}=n^{-(1+\delta)^2}$, so
\[
K n\,\bar t_n^{\gamma_2-1}e^{-\lambda_Z^2 \bar t_n^2}
=K\,\bar t_n^{\gamma_2-1}\,n^{1-(1+\delta)^2}
=K\,\bar t_n^{\gamma_2-1}\,n^{-2\delta-\delta^2}
=o(a_n),
\]
since $\bar t_n$ grows only polynomially in $\ln n$ while $n^{-2\delta-\delta^2}\to 0$.
Therefore,
\[
\limsup_{n\to\infty}\frac{\mathbb E[Z_{(n:n)}]}{a_n}\le 1+\delta.
\]

{\it Lower bound.}
Let $\underline t_n=(1-\delta)a_n$. Using $(1-y)^n\le e^{-ny}$ from \eqref{eq:1miny-power} and the lower
bound in \eqref{eq:Z-gaussian-two-sided},
\begin{align*}
\mathbb P(Z_{(n:n)}\le \underline t_n)
&=\big(1-\mathbb P(Z>\underline t_n)\big)^n\\
&\le \exp\big(-n\,\mathbb P(Z>\underline t_n)\big)\\
&\le \exp\!\left(-n c\,\underline t_n^{\gamma_1}e^{-\lambda_Z^2\underline t_n^2}\right)
= \exp\!\left(-c\,\underline t_n^{\gamma_1}\,n^{1-(1-\delta)^2}\right)
\to 0,
\end{align*}
since $1-(1-\delta)^2=2\delta-\delta^2>0$. Hence $\mathbb P(Z_{(n:n)}>\underline t_n)\to 1$ and, by
monotonicity of $t\mapsto \mathbb P(Z_{(n:n)}>t)$,
\[
\mathbb E[Z_{(n:n)}]\ge \int_0^{\underline t_n}\mathbb P(Z_{(n:n)}>t)\,dt
\ge \underline t_n\,\mathbb P(Z_{(n:n)}>\underline t_n)
=(1-o(1))\,\underline t_n.
\]
Thus $\liminf_{n\to\infty}\mathbb E[Z_{(n:n)}]/a_n\ge 1-\delta$. Letting $\delta\downarrow 0$ gives
\[
\lim_{n\to\infty}\frac{\mathbb E[Z_{(n:n)}]}{\sqrt{\ln n}/\lambda_Z}=1.
\]

\paragraph{(d) Bounded Tail.}
The cases $\rho=0$ and $\rho=1$ are immediate, so assume $\rho\in(0,1)$.
Since $X\le b_X$ and $Y\le b_Y$ almost surely, we have
\[
0\le Z=(1-\rho)X+\rho Y\le (1-\rho)b_X+\rho b_Y=:b_Z\qquad\text{a.s.}
\]
Hence $0\le Z_{(n:n)}\le b_Z$ for all $n$.
Fix $\eta>0$ and define
\[
E_\eta := \left\{X> b_X-\frac{\eta}{2(1-\rho)}\right\}\cap
\left\{Y> b_Y-\frac{\eta}{2\rho}\right\}.
\]
On $E_\eta$ we have
\[
Z=(1-\rho)X+\rho Y
> (1-\rho)\left(b_X-\frac{\eta}{2(1-\rho)}\right)
+\rho\left(b_Y-\frac{\eta}{2\rho}\right)
=b_Z-\eta,
\]
so $E_\eta\subseteq\{Z>b_Z-\eta\}$. By independence,
\[
\mathbb P(Z>b_Z-\eta)\ \ge\ \mathbb P(E_\eta)
=\mathbb P\!\left(X> b_X-\frac{\eta}{2(1-\rho)}\right)\,
\mathbb P\!\left(Y> b_Y-\frac{\eta}{2\rho}\right)
=:p_\eta.
\]
Under the bounded-tail assumptions, each factor is strictly positive for every $\eta>0$, hence $p_\eta>0$.
Therefore,
\[
\mathbb P\big(Z_{(n:n)}\le b_Z-\eta\big)
=\big(1-\mathbb P(Z>b_Z-\eta)\big)^n
\le (1-p_\eta)^n \xrightarrow[n\to\infty]{} 0,
\]
so $Z_{(n:n)}\xrightarrow{p} b_Z$. Since $0\le Z_{(n:n)}\le b_Z$ for all $n$, bounded convergence yields
\[
\lim_{n\to\infty}\mathbb E[Z_{(n:n)}]=b_Z.
\]
    
\end{proof}

\subsection{Proof of Theorem \ref{thm:uncapacitated-ranking-value}}
\label{app:proof-thm-uncapacitated-ranking-value}

\begin{proof}[\underline{Proof of Theorem \ref{thm:uncapacitated-ranking-value}}]
    In the uncapacitated setting, there is a single agent and $n$ items and in the {\sf No Information} regime, the agent chooses a random item since the agent has no information about the common terms $(q_y)$ or the idiosyncratic terms $(\varphi_{xy})$. Recall that $\pi_{\emptyset}(1)$ denotes the index of the item chosen by the agent. Therefore, the social welfare ${\sf AW}_{\emptyset}^{\sf uncap}(n)$ is given as 
    \begin{align}
        \label{eq:sw-no-info-uncap-pareto}
        {\sf AW}_{\emptyset}^{\sf uncap}(n) = \mathbb{E}\left[ (1 - \rho) q_{\pi_{\emptyset}(1)} + \rho \varphi_{1\pi_{\emptyset}(1)}\right] = (1 - \rho) \mu_q + \rho \mu_\varphi,
    \end{align}
    where the last equality follows from the fact that index $\pi_\emptyset(1)$ is uniformly random in $\{1,2,\dots, n\}$.

    In the {\sf Only Quality Information} regime, the agent chooses the item with highest common term value. Therefore, the social welfare ${\sf AW}_{q}^{\sf uncap}(n)$ is given as 
    \begin{align}
        \label{eq:sw-quality-info-uncap-pareto}
        {\sf AW}_{q}^{\sf uncap}(n) = \mathbb{E}\left[ ( 1 - \rho) q_{\pi_{q}(1)} + \rho \varphi_{1 \pi_{q}(1)} \right] = \mathbb{E}\left[(1 - \rho) q_{(n:n)} + \rho \varphi_{1 \pi_{q}(1)} \right] = (1 - \rho) \mathbb{E}[q_{(n:n)}] + \rho \mu_{\varphi},
    \end{align}
    where the last equality follows from the fact that index $\sigma_q(1)$ is uniformly random in $\{1,2,\dots, n\}$. Using \eqref{eq:sw-no-info-uncap-pareto}, \eqref{eq:sw-quality-info-uncap-pareto} and Proposition \ref{prop:expected-maximum-n}, we get the required result.
\end{proof}

\subsection{Proof of Theorem \ref{thm:uncapacitated-personalized-recos-value}}
\label{app:proof-thm-uncapacitated-personalized-recos-value}
\begin{proof}[\underline{Proof of Theorem \ref{thm:uncapacitated-personalized-recos-value}}]
Since there are no capacity constraints on the supply side, we will drop the agent index $i$ and we have the agent utility for item $j$ is given as $u_j = (1 - \rho) \cdot q_j + \rho \cdot \varphi_j$. In the {\sf Full Information} setting, we have that ${\sf AW}_{u}^{\sf uncap}(n) = \mathbb{E}[u_{(n:n)}]$. Therefore we have that $\Delta_{q \to u}^{\sf uncap}(n) = \mathbb{E}[u_{(n:n)}] - \left( (1 - \rho) \cdot \mathbb{E}[q_{(n:n)}] + \rho \cdot \mu_{\varphi} \right)$. Now the results for different tail distributions follows from Proposition  \ref{prop:expected-maximum-n} and \ref{prop:maximum-n-convex-combination}. 
\end{proof}

\subsection{Proof of Theorem \ref{thm:capacitated-ranking-value}}
\label{app:proof-thm-capacitated-ranking-value}

\begin{proof}{\underline{Proof of Theorem \ref{thm:capacitated-ranking-value}}}
    In the {\sf No Information} regime, since the agents do not have information about the common term or the idiosyncratic term, they randomly choose an item from the remaining set of items. Recall that $\pi_\emptyset(k)$ denotes the index of the item chosen by agent $k$ in the {\sf No Information} regime. Therefore we have that,
\begin{align}
    \label{eq:sw-cap-no-info}
    {\sf AW}^{\sf cap}_{\emptyset}(n) 
    &\stackrel{(a)}= \frac{1}{n}\mathbb{E}\left[\sum_{k = 1}^n (1 - \rho) q_{\pi_{\emptyset}(k)} + \rho \varphi_{k \pi_{\emptyset}(k)} \right]
    \stackrel{(b)}= (1 - \rho) \frac{1}{n} \mathbb{E}\left[ \sum_{k = 1}^n q_{k} \right] + \rho \frac{1}{n} \sum_{k = 1}^n \mathbb{E}[\varphi_{k \pi_{\emptyset}(k)}],  
\end{align}
where $(a)$ follows from definition of $u_{k\pi_{\empty}(k)}$, $(b)$ follows from the fact that $\sum_{k = 1}^n q_{\pi_{\emptyset}(k)} = \sum_{k = 1}^n q_k$.

In the {\sf Only Quality Information} regime, the agents base their decisions solely on the common term $(q_y)$. Therefore, we have that the agent $k$ will choose the item with common term value $q_{(k:n)}$ (recall that $X_{(k:n)}$ denotes the $k$-th smallest value of $n$ i.i.d copies of $X$). 
Recall that $\pi_{q}(k)$ denotes the index of the item chosen by agent $k$ in the {\sf Only Quality Information} regime. This means that $q_{\pi_q(k)} = q_{(k:n)}$. Therefore we have that,
\begin{align}
    \label{eq:sw-cap-quality-info}
    {\sf AW}_{q}^{\sf cap}(n) &\stackrel{(a)}= \frac{1}{n}\mathbb{E}\left[\sum_{k = 1}^n (1 - \rho) q_{\pi_{q}(k)} + \rho \varphi_{k \pi_{q}(k)} \right]
    \stackrel{(b)}= (1 - \rho) \frac{1}{n} \mathbb{E}\left[ \sum_{k = 1}^n q_{k} \right] + \rho \frac{1}{n} \sum_{k = 1}^n \mathbb{E}[\varphi_{k \pi_{q}(k)}], 
\end{align}
where $(a)$ follows from definition of $u_{k\pi_{q}(k)}$, $(b)$ follows from the fact that $\sum_{k = 1}^n q_{\pi_{\emptyset}(k)} = \sum_{k = 1}^n q_{(k:n)} = \sum_{k = 1}^n q_k$.
Note that the index $\pi_{\emptyset}(k)$ and $\pi_{q}(k)$ are random and hence we have that $\varphi_{k \pi_{\emptyset}(k)} \stackrel{d}= \varphi_{k \pi_q(k)}$ (have the same distribution) and therefore $\mathbb{E}[\varphi_{k \pi_{\emptyset}(k)}] = \mathbb{E}[\varphi_{k \pi_q(k)}]$. Comparing \eqref{eq:sw-cap-no-info} and \eqref{eq:sw-cap-quality-info}, we have that $\Delta_{\emptyset \to q}^{\sf cap}(n) = 0$.
\end{proof}

\subsection{Proof of Theorem \ref{thm:capacitated-personalized-recos-value}}

\begin{proof}[\underline{Proof of Theorem~\ref{thm:capacitated-personalized-recos-value}}]
Fix $\rho\in[0,1]$. If $\rho=0$ then $u_{xy}=q_y$ for all $(x,y)$, hence the two regimes coincide and
$\Delta^{\sf cap}_{q\to u}(n)=0$ for all $n$, so all limits hold trivially.  Assume henceforth that
$\rho\in(0,1]$.

\smallskip
\noindent\textbf{Step 0: A policy-invariance identity for the $q$--part.}
In the capacitated model each item is chosen exactly once, so under {\it any} sequential policy $\pi$
the chosen indices $\{\sigma_\pi(x)\}_{x=1}^n$ form a permutation of $\{1,\dots,n\}$ and therefore
\begin{equation}\label{eq:q-invariance}
\sum_{x=1}^n q_{\sigma_\pi(x)}=\sum_{y=1}^n q_y \qquad\text{(pathwise)}.
\end{equation}
Taking expectations and dividing by $n$ gives
\begin{equation}\label{eq:q-invariance-exp}
\frac1n\sum_{x=1}^n \mathbb{E}[q_{\sigma_\pi(x)}]=\mathbb{E}[q_1]=\mu_q.
\end{equation}

\smallskip
\noindent\textbf{Step 1: Rewrite $\Delta^{\sf cap}_{q\to u}(n)$ in terms of the chosen idiosyncratic draws.}
By definition,
\[
{\sf AW}^{\sf cap}_r(n)=\frac1n \mathbb{E}\!\Big[\sum_{x=1}^n \big((1-\rho)q_{\sigma_r(x)}+\rho\,\varphi_{x,\sigma_r(x)}\big)\Big],
\qquad r\in\{u,q\}.
\]
Using \eqref{eq:q-invariance-exp} for $r=u$ and $r=q$ we obtain the exact identity
\begin{equation}\label{eq:Delta-idiosyncratic}
\Delta^{\sf cap}_{q\to u}(n)
=
\rho\Bigg(\frac1n\sum_{x=1}^n \mathbb{E}[\varphi_{x,\sigma_u(x)}]-\frac1n\sum_{x=1}^n \mathbb{E}[\varphi_{x,\sigma_q(x)}]\Bigg).
\end{equation}
In the {\it only-quality} regime, $\sigma_q(x)$ is measurable with respect to $(q_y)_{y\le n}$ and is independent of
$(\varphi_{xy})_{x,y\le n}$; by symmetry, $\varphi_{x,\sigma_q(x)}\stackrel{d}= \varphi_{11}$ and hence
\begin{equation}\label{eq:only-q-phi-mean}
\frac1n\sum_{x=1}^n \mathbb{E}[\varphi_{x,\sigma_q(x)}]=\mu_\varphi.
\end{equation}
Combining \eqref{eq:Delta-idiosyncratic}--\eqref{eq:only-q-phi-mean} yields
\begin{equation}\label{eq:Delta-basic}
\Delta^{\sf cap}_{q\to u}(n)=
\rho\Bigg(\frac1n\sum_{x=1}^n \mathbb{E}[\varphi_{x,\sigma_u(x)}]-\mu_\varphi\Bigg).
\end{equation}

\smallskip
\noindent\textbf{Step 2: A universal upper--lower envelope via row-maxima (used for parts (a)--(c)).}
For $m\in\{1,\dots,n\}$ let $\varphi_{(m:m)}$ denote the maximum of $m$ i.i.d.\ samples from $P_\varphi$, and define
\[
M_m:=\mathbb{E}[\varphi_{(m:m)}],
\qquad
\Phi_n:=\frac1n\sum_{m=1}^n M_m.
\]
Consider agent $x$ when there are $m=n-x+1$ items remaining. Conditional on the history, the vector of remaining
idiosyncratic draws $\{\varphi_{x y}: y\ \text{remaining}\}$ is i.i.d.\ from $P_\varphi$ and independent of the history
(principle of deferred decisions). Hence
\[
\varphi_{x,\sigma_u(x)}
\le
\max\{\varphi_{x y}: y\ \text{remaining}\}\ \stackrel{d}= \ \varphi_{(m:m)}.
\]
Summing over $x$ and taking expectations gives the upper bound
\begin{equation}\label{eq:phi-upper-by-Phi}
\frac1n\sum_{x=1}^n \mathbb{E}[\varphi_{x,\sigma_u(x)}]\le \Phi_n.
\end{equation}
For a lower bound, use $q_y\ge 0$:
\[
u_{x,\sigma_u(x)}
=
\max_{y\ \text{remaining}}\big((1-\rho)q_y+\rho\varphi_{xy}\big)
\ge
\rho\,\max_{y\ \text{remaining}}\varphi_{xy},
\]
so $\mathbb{E}[u_{x,\sigma_u(x)}]\ge \rho\,M_m$. Averaging over $x$ yields
\begin{equation}\label{eq:AWu-lower-by-Phi}
{\sf AW}^{\sf cap}_u(n)\ge \rho\,\Phi_n.
\end{equation}
Similarly, by \eqref{eq:q-invariance-exp} and \eqref{eq:phi-upper-by-Phi},
\begin{equation}\label{eq:AWu-upper-by-Phi}
{\sf AW}^{\sf cap}_u(n)
=
(1-\rho)\mu_q + \rho\cdot \frac1n\sum_{x=1}^n \mathbb{E}[\varphi_{x,\sigma_u(x)}]
\le
(1-\rho)\mu_q+\rho\,\Phi_n.
\end{equation}
Finally, using \eqref{eq:q-invariance-exp} and \eqref{eq:only-q-phi-mean} we have
\begin{equation}\label{eq:AWq-exact}
{\sf AW}^{\sf cap}_q(n)=(1-\rho)\mu_q+\rho\mu_\varphi.
\end{equation}
Subtracting \eqref{eq:AWq-exact} from \eqref{eq:AWu-lower-by-Phi} and \eqref{eq:AWu-upper-by-Phi} gives, for all $n$,
\begin{equation}\label{eq:Delta-sandwich-Phi}
\rho\,\Phi_n-\big((1-\rho)\mu_q+\rho\mu_\varphi\big)
\ \le\
\Delta^{\sf cap}_{q\to u}(n)
\ \le\
\rho(\Phi_n-\mu_\varphi).
\end{equation}
Dividing by $\Phi_n$ (when $\Phi_n>0$) yields
\begin{equation}\label{eq:Delta-over-Phi-sandwich}
\rho-\frac{(1-\rho)\mu_q+\rho\mu_\varphi}{\Phi_n}
\ \le\
\frac{\Delta^{\sf cap}_{q\to u}(n)}{\Phi_n}
\ \le\
\rho-\frac{\rho\mu_\varphi}{\Phi_n}.
\end{equation}

\smallskip
\noindent\textbf{Parts (a)--(c): Unbounded tails.}
In each of the tail assumptions (Pareto / Exponential / Gaussian-like), the expected maximum satisfies a standard
extreme-value asymptotic: for every $\varepsilon>0$ there exists $m_0$ such that for all $m\ge m_0$,
\begin{align}
\label{eq:EVT-Pareto}
\text{Pareto:}\quad
&(1-\varepsilon)\,c_\varphi\Gamma\!\Big(1-\frac1{\alpha_\varphi}\Big)m^{1/\alpha_\varphi}
\le
M_m
\le
(1+\varepsilon)\,c_\varphi\Gamma\!\Big(1-\frac1{\alpha_\varphi}\Big)m^{1/\alpha_\varphi},
\\
\label{eq:EVT-Exp}
\text{Exponential:}\quad
&(1-\varepsilon)\,\frac{\ln m}{\lambda_\varphi}
\le
M_m
\le
(1+\varepsilon)\,\frac{\ln m}{\lambda_\varphi},
\\
\label{eq:EVT-Gauss}
\text{Gaussian-like:}\quad
&(1-\varepsilon)\,\frac{\sqrt{\ln m}}{\lambda_\varphi}
\le
M_m
\le
(1+\varepsilon)\,\frac{\sqrt{\ln m}}{\lambda_\varphi}.
\end{align}

We now compute $\Phi_n=\frac1n\sum_{m=1}^n M_m$ by splitting the sum at $m_0$ and comparing the tail sum to an integral.

\smallskip
{\it (a) Pareto tail.}
Let $\alpha_\varphi>1$ and set
\[
C_\varphi:=c_\varphi\Big(\frac{\alpha_\varphi}{\alpha_\varphi+1}\Big)\Gamma\!\Big(1-\frac1{\alpha_\varphi}\Big).
\]
Using $M_m\le m\mu_\varphi$ for $m<m_0$ and \eqref{eq:EVT-Pareto} for $m\ge m_0$,
\[
\Phi_n
=
\frac1n\sum_{m=1}^{m_0-1}M_m+\frac1n\sum_{m=m_0}^n M_m
\le
O\!\Big(\frac1n\Big)
+\frac{(1+\varepsilon)c_\varphi\Gamma(1-1/\alpha_\varphi)}{n}\sum_{m=1}^n m^{1/\alpha_\varphi}.
\]
Since $x\mapsto x^{1/\alpha_\varphi}$ is increasing,
\[
\sum_{m=1}^n m^{1/\alpha_\varphi}
\le
\int_0^n x^{1/\alpha_\varphi}\,dx + n^{1/\alpha_\varphi}
=
\frac{\alpha_\varphi}{\alpha_\varphi+1}n^{1+1/\alpha_\varphi}+n^{1/\alpha_\varphi}.
\]
Dividing by $n$ yields
\[
\limsup_{n\to\infty}\frac{\Phi_n}{C_\varphi n^{1/\alpha_\varphi}}\le 1+\varepsilon.
\]
A matching lower bound follows from \eqref{eq:EVT-Pareto} and $\sum_{m=m_0}^n m^{1/\alpha_\varphi}\ge \int_{m_0}^n x^{1/\alpha_\varphi}\,dx$,
giving
\[
\liminf_{n\to\infty}\frac{\Phi_n}{C_\varphi n^{1/\alpha_\varphi}}\ge 1-\varepsilon.
\]
As $\varepsilon>0$ is arbitrary,
\begin{equation}\label{eq:Phi-asymp-Pareto}
\frac{\Phi_n}{C_\varphi n^{1/\alpha_\varphi}}\to 1.
\end{equation}
In particular $\Phi_n\to\infty$, so \eqref{eq:Delta-over-Phi-sandwich} implies
$\Delta^{\sf cap}_{q\to u}(n)/\Phi_n\to \rho$. Combining with \eqref{eq:Phi-asymp-Pareto} yields
\[
\frac{\Delta^{\sf cap}_{q\to u}(n)}{C_\varphi n^{1/\alpha_\varphi}}
=
\frac{\Delta^{\sf cap}_{q\to u}(n)}{\Phi_n}\cdot \frac{\Phi_n}{C_\varphi n^{1/\alpha_\varphi}}
\ \longrightarrow\ \rho.
\]

\smallskip
{\it (b) Exponential tail.}
Using $M_m\le m\mu_\varphi$ for $m<m_0$ and \eqref{eq:EVT-Exp} for $m\ge m_0$,
\[
\Phi_n
\le
O\!\Big(\frac1n\Big)
+\frac{1+\varepsilon}{\lambda_\varphi}\cdot\frac1n\sum_{m=1}^n \ln m.
\]
Since $x\mapsto \ln x$ is increasing and $\int_1^n \ln x\,dx = n\ln n-n+1$,
\[
\int_1^n \ln x\,dx
\ \le\
\sum_{m=1}^n \ln m
\ \le\
\ln n + \int_1^n \ln x\,dx,
\]
hence $\frac1n\sum_{m=1}^n \ln m = \ln n + O(1)$ and therefore
\begin{equation}\label{eq:Phi-asymp-Exp}
\frac{\Phi_n}{(\ln n)/\lambda_\varphi}\to 1.
\end{equation}
Again $\Phi_n\to\infty$, so \eqref{eq:Delta-over-Phi-sandwich} gives $\Delta^{\sf cap}_{q\to u}(n)/\Phi_n\to\rho$ and thus,
by \eqref{eq:Phi-asymp-Exp},
\[
\frac{\Delta^{\sf cap}_{q\to u}(n)}{(\ln n)/\lambda_\varphi}\to \rho.
\]

\smallskip
{\it (c) Gaussian-like tail.}
Using $M_m\le m\mu_\varphi$ for $m<m_0$ and \eqref{eq:EVT-Gauss} for $m\ge m_0$,
\[
\Phi_n
\le
O\!\Big(\frac1n\Big)
+\frac{1+\varepsilon}{\lambda_\varphi}\cdot\frac1n\sum_{m=2}^n \sqrt{\ln m}.
\]
For $f(x):=\sqrt{\ln x}$ (increasing on $[e,\infty)$), we have the integral comparison
\[
\int_1^n f(x)\,dx \ \le\ \sum_{m=2}^n f(m)\ \le\ f(n)+\int_1^n f(x)\,dx.
\]
A standard change of variables $x=e^{t^2}$ gives
\[
\int_1^n \sqrt{\ln x}\,dx
=
n\sqrt{\ln n}-\frac{n}{2\sqrt{\ln n}}+O\!\Big(\frac{n}{(\ln n)^{3/2}}\Big),
\]
so $\frac1n\sum_{m=2}^n \sqrt{\ln m}\sim \sqrt{\ln n}$. Consequently,
\begin{equation}\label{eq:Phi-asymp-Gauss}
\frac{\Phi_n}{(\sqrt{\ln n})/\lambda_\varphi}\to 1.
\end{equation}
Since $\Phi_n\to\infty$, \eqref{eq:Delta-over-Phi-sandwich} implies $\Delta^{\sf cap}_{q\to u}(n)/\Phi_n\to\rho$ and then,
by \eqref{eq:Phi-asymp-Gauss},
\[
\frac{\Delta^{\sf cap}_{q\to u}(n)}{(\sqrt{\ln n})/\lambda_\varphi}\to \rho.
\]

\smallskip
\noindent\textbf{Part (d): Bounded tail.}
Assume now that $P_\varphi$ is continuous on $[a_\varphi,b_\varphi]$ with $b_\varphi<\infty$ (in particular,
$\varphi_{xy}\le b_\varphi$ a.s.).  We prove
\[
\Delta^{\sf cap}_{q\to u}(n)\to \rho\,(b_\varphi-\mu_\varphi),
\qquad\text{equivalently}\qquad
\frac{\Delta^{\sf cap}_{q\to u}(n)}{b_\varphi-\mu_\varphi}\to \rho.
\]

By \eqref{eq:Delta-basic}, it suffices to show
\begin{equation}\label{eq:goal-bounded}
\frac1n\sum_{x=1}^n \mathbb{E}[\varphi_{x,\sigma_u(x)}]\to b_\varphi.
\end{equation}
The upper bound is immediate: $\varphi_{x,\sigma_u(x)}\le b_\varphi$ a.s.\ for each $x$, hence
\[
\limsup_{n\to\infty}\frac1n\sum_{x=1}^n \mathbb{E}[\varphi_{x,\sigma_u(x)}]\le b_\varphi.
\]
We establish the matching lower bound using a top-$k$ comparison.

Fix $\gamma\in(0,1)$ and, at a step with $m$ remaining items, define $k(m):=\lceil m^\gamma\rceil$.
At agent $x$ (so $m=n-x+1$), let $\mathcal Q_x$ be the subset of the $k(m)$ {\it largest} qualities among the
remaining items, and define
\[
q^{\min}_{\mathcal Q_x}:=\min_{y\in\mathcal Q_x} q_y,
\qquad
\varphi^\star_{\mathcal Q_x}:=\max_{y\in\mathcal Q_x}\varphi_{xy}.
\]
Since $q_y\ge q^{\min}_{\mathcal Q_x}$ for all $y\in\mathcal Q_x$ and $\sigma_u(x)$ maximizes $u_{xy}$ over all remaining $y$,
we have the deterministic inequality
\begin{equation}\label{eq:topk-ineq}
(1-\rho)q_{\sigma_u(x)}+\rho\varphi_{x,\sigma_u(x)}
\ \ge\
(1-\rho)q^{\min}_{\mathcal Q_x}+\rho\varphi^\star_{\mathcal Q_x},
\end{equation}
and hence
\begin{equation}\label{eq:topk-rearranged}
\varphi_{x,\sigma_u(x)}
\ \ge\
\varphi^\star_{\mathcal Q_x}
+\frac{1-\rho}{\rho}\big(q^{\min}_{\mathcal Q_x}-q_{\sigma_u(x)}\big).
\end{equation}
Averaging \eqref{eq:topk-rearranged} over $x$ and taking expectations gives
\begin{equation}\label{eq:avg-topk}
\frac1n\sum_{x=1}^n \mathbb{E}[\varphi_{x,\sigma_u(x)}]
\ \ge\
\frac1n\sum_{x=1}^n \mathbb{E}[\varphi^\star_{\mathcal Q_x}]
+\frac{1-\rho}{\rho}\left(
\frac1n\sum_{x=1}^n \mathbb{E}[q^{\min}_{\mathcal Q_x}]
-\frac1n\sum_{x=1}^n \mathbb{E}[q_{\sigma_u(x)}]
\right).
\end{equation}
By \eqref{eq:q-invariance-exp}, $\frac1n\sum_{x=1}^n \mathbb{E}[q_{\sigma_u(x)}]=\mu_q$, so the second bracket in
\eqref{eq:avg-topk} is $\frac1n\sum_x\mathbb{E}[q^{\min}_{\mathcal Q_x}]-\mu_q$.

\smallskip
{\it (d.1) The $\varphi$-term tends to $b_\varphi$.}
Conditional on the history and on $\mathcal Q_x$, the random variables $\{\varphi_{xy}:y\in\mathcal Q_x\}$ are i.i.d.\ from $P_\varphi$
and independent of $\mathcal Q_x$ (deferred decisions). Hence
\[
\mathbb{E}[\varphi^\star_{\mathcal Q_x}\mid \mathcal Q_x]=M_{k(m)}
\quad\Rightarrow\quad
\mathbb{E}[\varphi^\star_{\mathcal Q_x}]=M_{k(m)}.
\]
Therefore
\begin{equation}\label{eq:avg-phi-star}
\frac1n\sum_{x=1}^n \mathbb{E}[\varphi^\star_{\mathcal Q_x}]
=
\frac1n\sum_{m=1}^n M_{k(m)}.
\end{equation}
Because $\varphi_{(k:k)}\uparrow b_\varphi$ almost surely as $k\to\infty$ and $0\le \varphi_{(k:k)}\le b_\varphi$,
dominated convergence implies $M_k\to b_\varphi$ as $k\to\infty$.
Since $k(m)\to\infty$ as $m\to\infty$ and $M_k$ is increasing in $k$, a standard split-sum argument yields
\begin{equation}\label{eq:avg-phi-star-limit}
\frac1n\sum_{m=1}^n M_{k(m)}\to b_\varphi.
\end{equation}

\smallskip
{\it (d.2) The $q$-correction is asymptotically negligible from below.}
We show
\begin{equation}\label{eq:qmin-avg-close}
\frac1n\sum_{x=1}^n \E[q^{\min}_{\mathcal Q_x}]
\ \ge\ \mu_q-o(1).
\end{equation}
Let $q_{(1:n)}\le\cdots\le q_{(n:n)}$ be the order statistics of $\{q_y\}_{y=1}^n$.
At a step with $m$ remaining items and $|\mathcal Q_x|=k(m)$, the minimum of the top $k(m)$ remaining qualities is
pathwise lower bounded by the corresponding global order statistic:
\begin{equation}\label{eq:qmin-coupling}
q^{\min}_{\mathcal Q_x}\ \ge\ q_{(m-k(m)+1:n)} \qquad\text{(pathwise)}.
\end{equation}
Now set $r_n:=\max_{1\le m\le n}k(m)=k(n)=\lceil n^\gamma\rceil=o(n)$. Since $k(m)\le r_n$,
\[
m-k(m)+1 \ \ge\ \max\{1,m-r_n+1\},
\quad\text{hence}\quad
q_{(m-k(m)+1:n)}\ge q_{(\max\{1,m-r_n+1\}:n)}.
\]
Define
\[
A_n(r):=\frac1n\sum_{m=1}^n \E\big[q_{(\max\{1,m-r+1\}:n)}\big].
\]
Then \eqref{eq:qmin-coupling} implies
\begin{equation}\label{eq:qmin-to-An}
\frac1n\sum_{x=1}^n \E[q^{\min}_{\mathcal Q_x}] \ \ge\ A_n(r_n).
\end{equation}
We claim $A_n(r_n)\to\mu_q$.  To see this, note first that (pathwise)
\[
\sum_{m=1}^n q_{(m:n)}=\sum_{y=1}^n q_y,
\qquad\text{so}\qquad
\frac1n\sum_{m=1}^n \E[q_{(m:n)}]=\mu_q.
\]
Also, for any integer $r\ge 1$, by monotonicity of order statistics,
\[
0\le \frac1n\sum_{m=1}^n q_{(m:n)}-\frac1n\sum_{m=1}^n q_{(\max\{1,m-r+1\}:n)}
\le
\frac1n\sum_{t=n-r+2}^{n} q_{(t:n)},
\]
hence
\begin{equation}\label{eq:mu-minus-An}
0\le \mu_q-A_n(r)\le \mathbb{E}\!\left[\frac1n\sum_{t=n-r+2}^{n} q_{(t:n)}\right].
\end{equation}
To bound the expectation on the right, fix a threshold $T>0$ and use that for any realization,
\[
\sum_{t=n-r+2}^{n} q_{(t:n)}
\le
T(r-1)+\sum_{y=1}^n q_y\,\mathbf 1\{q_y>T\}.
\]
Taking expectations and dividing by $n$ gives
\begin{equation}\label{eq:top-order-bound}
\mathbb{E}\!\left[\frac1n\sum_{t=n-r+2}^{n} q_{(t:n)}\right]
\le
T\frac{r}{n}+\mathbb{E}[q_1\,\mathbf 1\{q_1>T\}].
\end{equation}
Now choose $T=T_n:=\sqrt{n/r_n}$. Since $r_n=o(n)$, we have $T_n\to\infty$ and
$T_n r_n/n=\sqrt{r_n/n}\to 0$. Moreover, $\mathbb{E}[q_1\,\mathbf 1\{q_1>T_n\}]\to 0$ by dominated convergence
(using $\mathbb{E}[q_1]<\infty$ and $q_1\mathbf 1\{q_1>T_n\}\downarrow 0$ a.s.).
Applying \eqref{eq:top-order-bound} with $(r,T)=(r_n,T_n)$ and then \eqref{eq:mu-minus-An} yields
$\mu_q-A_n(r_n)\to 0$, i.e.\ $A_n(r_n)\to\mu_q$. Combining with \eqref{eq:qmin-to-An} proves \eqref{eq:qmin-avg-close}.

\smallskip
{\it (d.3) Conclude \eqref{eq:goal-bounded}.}
Insert \eqref{eq:avg-phi-star}--\eqref{eq:avg-phi-star-limit} and \eqref{eq:qmin-avg-close} into \eqref{eq:avg-topk}:
\[
\liminf_{n\to\infty}\frac1n\sum_{x=1}^n \mathbb{E}[\varphi_{x,\sigma_u(x)}]
\ \ge\
\liminf_{n\to\infty}\frac1n\sum_{m=1}^n M_{k(m)}
+\frac{1-\rho}{\rho}\cdot \liminf_{n\to\infty}\Big(\frac1n\sum_x \mathbb{E}[q^{\min}_{\mathcal Q_x}]-\mu_q\Big)
\ \ge\ b_\varphi+0=b_\varphi.
\]
Together with the earlier $\limsup\le b_\varphi$, this proves \eqref{eq:goal-bounded}. Returning to \eqref{eq:Delta-basic},
we obtain
\[
\Delta^{\sf cap}_{q\to u}(n)\to \rho(b_\varphi-\mu_\varphi),
\qquad\text{hence}\qquad
\frac{\Delta^{\sf cap}_{q\to u}(n)}{b_\varphi-\mu_\varphi}\to \rho.
\]
This completes the proof of the theorem.
\end{proof}

\section{Value of information provisioning tools with supply-side priorities}
\label{app:supply-side-priority-proofs}

\subsection{Proof of Proposition~\ref{prop:gpsd-stable-arbitrary}}
\label{app:proof-gpsd-stable-all}

\begin{proof}[\underline{Proof of Proposition~\ref{prop:gpsd-stable-arbitrary}}]
Fix the item priorities, the priority order $\pi$, and \emph{any arbitrary profile of strict agent preferences} over the items (let $\succ_x$ denote the strict preference of agent $x$). Let $\mu=\mu^{\sf gpsd}$ be the outcome of Algorithm~\ref{alg:gpsd}. We show that no blocking pair $(x,y)$ exists.

\paragraph{Step 1: No blocking pairs involving a group-2 item.}
Let $y\in\mathcal G_{\mathcal Y}^{2}$.
In GPSD, no group-1 agent ever applies to or takes a group-2 item in Phase~1, so every group-2 item is assigned in Phase~2 to some group-2 agent.
If $x\in\mathcal G_{\mathcal X}^{1}$, then $y$ strictly prefers its group-2 assignee to $x$ (due to own-group priority), so $(x,y)$ cannot block.
If $x\in\mathcal G_{\mathcal X}^{2}$, then Phase~2 processes group-2 agents in order of $\pi$ and each agent picks her most-preferred available item from the remaining set.
Thus, if $y$ was available when $x$ was processed, $x$ cannot strictly prefer $y$ to $\mu(x)$ (because she would have chosen it, meaning $\mu(x) \succ_x y$);
and if $y$ was not available, it was taken by some group-2 agent $x'$ with higher priority ($\pi(x')<\pi(x)$). Because item $y$ ranks group-2 agents exactly by $\pi$, $y$ strictly prefers $x'$ to $x$.
Hence, no blocking pair involves a group-2 item.

\paragraph{Step 2: No blocking pairs involving a group-1 item.}
Let $y\in\mathcal G_{\mathcal Y}^{1}$.
We consider two subcases based on when $y$ is assigned.

\smallskip
\noindent{\it (i) $y$ is assigned in Phase~1 to a group-1 agent.}
Then $\mu^{-1}(y)\in\mathcal G_{\mathcal X}^{1}$.
If $x\in\mathcal G_{\mathcal X}^{2}$, item $y$ strictly prefers any group-1 agent to any group-2 agent, so $y$ does not strictly prefer $x$ to its assignee.
If $x\in\mathcal G_{\mathcal X}^{1}$, then in Phase~1 group-1 agents are processed by $\pi$ and choose their most-preferred available group-1 item.
If $y$ was available when $x$ was processed, $x$ cannot strictly prefer $y$ to $\mu(x)$.
If $y$ was not available, it was taken earlier by some group-1 agent $x'$ with $\pi(x')<\pi(x)$. Because item $y$ ranks group-1 agents by $\pi$, $y$ strictly prefers $x'$ to $x$.
Thus, no blocking pair involves such a $y$.

\smallskip
\noindent{\it (ii) $y$ remains unassigned after Phase~1 and is assigned in Phase~2 to a group-2 agent.}
In this case, $y$ was available throughout Phase~1 and was not chosen by any group-1 agent when they had the opportunity to pick from the available group-1 items.
Hence, for any group-1 agent $x\in\mathcal G_{\mathcal X}^{1}$, $x$ cannot strictly prefer $y$ to $\mu(x)$ (otherwise $x$ would have selected $y$ during Phase~1). Note that even if $x$ hypothetically preferred a group-2 item over all group-1 items, she is restricted to group-1 items in Phase 1; since $y$ is a group-1 item, it was in her choice set, meaning she actively preferred her chosen group-1 item $\mu(x)$ over $y$.
Therefore, no blocking pair can involve a group-1 agent and such an item $y$.

Now consider $x\in\mathcal G_{\mathcal X}^{2}$.
Let $x_y:=\mu^{-1}(y)\in\mathcal G_{\mathcal X}^{2}$ be the group-2 assignee of $y$ chosen in Phase~2.
If $\pi(x)>\pi(x_y)$, then item $y$ ranks group-2 agents by $\pi$ and therefore strictly prefers $x_y$ to $x$, so it does not strictly prefer $x$ to its current assignee.
If $\pi(x)<\pi(x_y)$, then $x$ was processed {\it before} $x_y$ in Phase~2. Since $y$ was ultimately assigned to $x_y$, it was necessarily available when $x$ was processed; hence $x$ cannot strictly prefer $y$ to $\mu(x)$ (otherwise $x$ would have chosen $y$).
Thus, no blocking pair involves such a $y$ either.

All possible cases are covered, and no blocking pair exists. Therefore $\mu^{\sf gpsd}$ yields a stable matching for any arbitrary strict agent preferences. 
\end{proof}

\subsection{Proof of Theorem~\ref{thm:value-public-rankings-supply-side-priority}}
\label{app:proof-public-rankings-supply-side-priority}

\begin{proof}[\underline{Proof of Theorem~\ref{thm:value-public-rankings-supply-side-priority}}]
Let $\mu^{\sf gpsd}_{\emptyset}$ and $\mu^{\sf gpsd}_{q}$ denote the GPSD matchings in the {\sf No Information} and {\sf Only Quality} regimes, respectively.

\paragraph{Common component is conserved.}
For any bijection $\mu$, $\sum_{x} q_{\mu(x)}=\sum_{y} q_y$, hence
\[
\frac{1}{n}\,\mathbb{E}\!\left[\sum_{x}(1-\rho)q_{\mu(x)}\right]
=(1-\rho)\mathbb{E}[q_y]=(1-\rho)\mu_q,
\]
independently of the regime and independently of the selection rule among stable matchings.

\paragraph{Idiosyncratic component under $\emptyset$ and $q$.}
In both regimes $\emptyset$ and $q$, the GPSD outcome depends only on (i) the group labels and priority order $\pi$, (ii) the tie-breaking rule (under $\emptyset$), and (iii) the realized qualities $(q_y)$ (under $q$). In particular, $\mu^{\sf gpsd}_{\emptyset}$ and $\mu^{\sf gpsd}_{q}$ are measurable with respect to $\sigma(q,\text{tie-break})$ and are independent of the idiosyncratic matrix $(\varphi_{xy})$.

Fix any agent $x$. Conditional on $\mu(x)=y$, we have $\mathbb{E}[\varphi_{x,\mu(x)}\mid \mu(x)=y]=\mathbb{E}[\varphi_{xy}]=\mu_\varphi$ because $\varphi_{xy}$ is independent of $\mu$ and identically distributed across $y$.
Therefore $\mathbb{E}[\varphi_{x,\mu(x)}]=\mu_\varphi$, and summing over agents yields
\[
\frac{1}{n}\,\mathbb{E}\!\left[\sum_{x}\rho\,\varphi_{x,\mu(x)}\right]
=\rho\,\mu_\varphi
\]
in both the $\emptyset$ and $q$ regimes. Combining the common and idiosyncratic components gives
${\sf AW}^{\sf cap}_{\emptyset}(n)=(1-\rho)\mu_q+\rho\mu_\varphi
={\sf AW}^{\sf cap}_{q}(n)$,
hence $\Delta_{\emptyset\to q}^{\sf cap}(n)=0$ for all $n$.
\end{proof}

\subsection{Proof of Theorem~\ref{thm:value-personalized-recos-supply-side-priority}}
\label{app:proof-personalized-recos-supply-side-priority}

\begin{proof}[\underline{Proof of Theorem~\ref{thm:value-personalized-recos-supply-side-priority}}]
Let $\mu^{\sf gpsd}_{u}$ be the GPSD matching in the {\sf Full Information} regime.
Let $n_1:=|\mathcal G_{\mathcal X}^{1}|=\gamma_X n$ and $n_2:=|\mathcal G_{\mathcal X}^{2}|=(1-\gamma_X) n$ denote the number of group-1 and group-2 agents respectively. Index the group-1 agents in increasing priority order as $x_{1,1},\dots,x_{1,n_1}$, and similarly index the group-2 agents as $x_{2,1},\dots,x_{2,n_2}$.

\paragraph{Step 1: A lower bound on total welfare.}
Recall that in a balanced market where every item is assigned exactly once, the sum of the common quality components across all matched pairs is deterministically conserved: $\sum_{x} q_{\mu(x)} = \sum_{y} q_y$. Thus, any shifting of high-quality items between group-1 and group-2 agents across different information regimes is perfectly zero-sum in the aggregate.

To safely lower-bound the total welfare without explicitly tracking these complex quality shifts, we use the fact that $P_q$ has non-negative support ($q_y \ge 0$). For any agent $x$ choosing from an available set $S_x$, their chosen utility satisfies:
\[
u_{x,\mu^{\sf gpsd}_u(x)} = \max_{y\in S_x} \big\{(1-\rho)q_y+\rho\varphi_{xy}\big\} \ge \max_{y\in S_x} \{\rho\varphi_{xy}\} = \rho \max_{y\in S_x}\varphi_{xy}.
\]
Summing this over all agents gives a lower bound on the aggregate welfare that depends purely on the idiosyncratic shocks:
\[
{\sf AW}^{\sf cap}_{u}(n)
=\frac{1}{n}\,\mathbb{E}\!\left[\sum_{x \in \mathcal{X}} u_{x,\mu^{\sf gpsd}_{u}(x)}\right]
\ \ge\
\frac{\rho}{n}\,\mathbb{E}\!\left[\sum_{k=1}^{n_1} \max_{y\in S_{1,k}}\varphi_{x_{1,k} y} + \sum_{k=1}^{n_2} \max_{y\in S_{2,k}}\varphi_{x_{2,k} y} \right].
\]

Consider agent $x_{1,k}$ in Phase~1 of GPSD. 
Since Phase~1 assigns only group-1 items and $\gamma_X\le \gamma_Y$, we have $|S_{1,k}|=\gamma_Y n-(k-1)\ge n_1-k+1$. Conditional on $S_{1,k}$, the collection $\{\varphi_{x_{1,k} y}: y\in S_{1,k}\}$ consists of $|S_{1,k}|$ i.i.d.\ draws from $P_\varphi$ and is independent of the history (by row-independence of agent shocks). Letting $\varphi_{(m:m)}$ denote the maximum of $m$ i.i.d.\ draws from $P_\varphi$, we have:
\[
\mathbb{E}\!\left[\max_{y\in S_{1,k}}\varphi_{x_{1,k} y}\right]
\ \ge\
\mathbb{E}[\varphi_{(n_1-k+1:n_1-k+1)}].
\]

Now consider agent $x_{2,k}$ in Phase~2. Exactly $n_1$ items were assigned in Phase~1, and $k-1$ items have been assigned so far in Phase~2, so $|S_{2,k}| = n - n_1 - (k-1) = n_2 - k + 1$. 
While the $q_y$ qualities of the items in $S_{2,k}$ are highly path-dependent based on Group 1's choices, the idiosyncratic shocks $\varphi_{xy}$ are completely independent across all agents. Therefore, the conditional distribution of $\{\varphi_{x_{2,k} y} : y \in S_{2,k}\}$ still consists of $|S_{2,k}|$ fresh i.i.d.\ draws from $P_\varphi$. Applying the exact same deferred decisions principle yields:
\[
\mathbb{E}\!\left[\max_{y\in S_{2,k}}\varphi_{x_{2,k} y}\right]
\ \ge\
\mathbb{E}[\varphi_{(n_2-k+1:n_2-k+1)}].
\]

Summing over all agents in both groups and reindexing $m=n_1-k+1$ and $m=n_2-k+1$ yields
\begin{align*}
{\sf AW}^{\sf cap}_{u}(n)
&\ge
\frac{\rho}{n}\sum_{m=1}^{n_1}\mathbb{E}[\varphi_{(m:m)}] + \frac{\rho}{n}\sum_{m=1}^{n_2}\mathbb{E}[\varphi_{(m:m)}]\\
&=
\rho\,\gamma_X \cdot \Phi_{n_1} + \rho\,(1-\gamma_X) \cdot \Phi_{n_2},
\end{align*}
where we define the ``average-of-maxima'' functional $\Phi_{m} := \frac{1}{m}\sum_{t=1}^{m}\mathbb{E}[\varphi_{(t:t)}]$.

\paragraph{Step 2: Subtract the $q$-regime benchmark and take limits.}
By Theorem~\ref{thm:value-public-rankings-supply-side-priority},
${\sf AW}^{\sf cap}_{q}(n)=(1-\rho)\mu_q+\rho\mu_\varphi$ for all $n$.
Hence
\[
\Delta^{\sf cap}_{q\to u}(n)
=
{\sf AW}^{\sf cap}_u(n)-{\sf AW}^{\sf cap}_q(n)
\ \ge\
\rho\,\big(\gamma_X \Phi_{n_1} + (1-\gamma_X) \Phi_{n_2}\big)
-\big((1-\rho)\mu_q+\rho\mu_\varphi\big).
\]
Since $\mu_q,\mu_\varphi<\infty$, the final subtracted parenthesis is a deterministic $O(1)$ constant. This constant vanishes asymptotically under the relevant extreme-value scalings of $\Phi_n$. Thus it suffices to characterize the asymptotic behavior of $\gamma_X \Phi_{n_1} + (1-\gamma_X) \Phi_{n_2}$.

\paragraph{Step 3: Tail asymptotics for $\Phi_m$.}
The tail assumptions on $P_\varphi$ imply the following standard asymptotics for the average of maxima:
\begin{itemize}
\item \textup{(Pareto tail)} If $P_\varphi$ has a Pareto tail $(c_\varphi,\alpha_\varphi)$ with $\alpha_\varphi>1$, then
$\Phi_m = C_\varphi\, m^{1/\alpha_\varphi}\,(1+o(1))$.
Substituting $n_1=\gamma_X n$ and $n_2=(1-\gamma_X)n$ yields:
\begin{align*}
\gamma_X \Phi_{n_1} + (1-\gamma_X) \Phi_{n_2} 
&= C_\varphi \big( \gamma_X (\gamma_X n)^{1/\alpha_\varphi} + (1-\gamma_X) ((1-\gamma_X)n)^{1/\alpha_\varphi} \big) (1+o(1)) \\
&= C_\varphi \big(\gamma_X^{1+1/\alpha_\varphi} + (1-\gamma_X)^{1+1/\alpha_\varphi}\big) n^{1/\alpha_\varphi} (1+o(1)).
\end{align*}

\item \textup{(Exponential tail)} If $P_\varphi$ has an exponential tail $(c_\varphi,\lambda_\varphi)$, then
$\Phi_m = (\ln m)/\lambda_\varphi + O(1)$.
Substituting $n_1$ and $n_2$ yields:
\begin{align*}
\gamma_X \Phi_{n_1} + (1-\gamma_X) \Phi_{n_2} 
&= \gamma_X \frac{\ln(\gamma_X n)}{\lambda_\varphi} + (1-\gamma_X) \frac{\ln((1-\gamma_X) n)}{\lambda_\varphi} + O(1) \\
&= \frac{\gamma_X \ln n + \gamma_X \ln \gamma_X + (1-\gamma_X) \ln n + (1-\gamma_X) \ln(1-\gamma_X)}{\lambda_\varphi} + O(1) \\
&= \frac{\ln n}{\lambda_\varphi} + O(1).
\end{align*}

\item \textup{(Gaussian-like tail)} If $P_\varphi$ has a Gaussian-like tail $(c_\varphi,\lambda_\varphi,\beta_\varphi)$, then
$\Phi_m = (\sqrt{\ln m})/\lambda_\varphi\,(1+o(1))$.
Substituting $n_1$ and $n_2$ yields:
\begin{align*}
\gamma_X \Phi_{n_1} + (1-\gamma_X) \Phi_{n_2} 
&= \gamma_X \frac{\sqrt{\ln(\gamma_X n)}}{\lambda_\varphi} + (1-\gamma_X) \frac{\sqrt{\ln((1-\gamma_X) n)}}{\lambda_\varphi} (1+o(1)) \\
&= \frac{\sqrt{\ln n}}{\lambda_\varphi} (\gamma_X + 1 - \gamma_X) (1+o(1)) = \frac{\sqrt{\ln n}}{\lambda_\varphi} (1+o(1)).
\end{align*}
\end{itemize}
Dividing by the appropriate normalizers directly yields the stated lower bounds in (a)--(c).
\end{proof}

\section{Value of information provisioning tools with supply side pricing}
\label{app:supply-side-pricing}
\subsection{Equilibrium prices and outcomes by market imbalance}
\label{app:subsec:equilibrium-prices}

Throughout this appendix, let
\[
\gamma_n:=\frac{m}{n}.
\]
When asymptotics are taken along a sequence with $\gamma_n\to\gamma$, we refer to $\gamma$ as the market-imbalance ratio. Since $q_y\sim \mathrm{Unif}([0,1])$ is continuous, realized qualities are distinct almost surely; whenever ranks matter we condition on such a realization and relabel sellers so that
\[
q_1>q_2>\cdots>q_n.
\]

\begin{proposition}[{\sf No Information}]
\label{prop:NI-general}
Fix $m,n\in\mathbb N$ and let $\gamma_n=m/n$.
\begin{enumerate}[label=(\alph*)]
    \item \textbf{Excess demand and balanced markets ($\gamma_n\ge 1$).}
    The unique interim Nash equilibrium is
    \[
    p_y^\star=\frac12 \qquad \forall y\in\mathcal Y.
    \]
    Under this equilibrium,
    \[
    {\sf AW}_{\emptyset}(m,n)=0,\qquad
    {\sf SR}_{\emptyset}(m,n)=\frac12,\qquad
    {\sf SW}_{\emptyset}(m,n)=\frac12.
    \]

    \item \textbf{Excess supply markets ($\gamma_n<1$).}
    The symmetric profile $p_y^\star\equiv 0$ is an interim Nash equilibrium. More generally, a price vector $\mathbf p$ is an interim Nash equilibrium if and only if at least $m+1$ sellers post price $0$. Every such equilibrium yields the same aggregate outcomes:
    \[
    {\sf AW}_{\emptyset}(m,n)=\frac12,\qquad
    {\sf SR}_{\emptyset}(m,n)=0,\qquad
    {\sf SW}_{\emptyset}(m,n)=\frac12.
    \]
\end{enumerate}
Consequently, if $\gamma_n\to\gamma$, then the asymptotic outcomes are
\[
({\sf AW}_{\emptyset},{\sf SR}_{\emptyset},{\sf SW}_{\emptyset})
=
\begin{cases}
(0,\frac12,\frac12), & \gamma\ge 1,\\[3pt]
(\frac12,0,\frac12), & \gamma\in(0,1).
\end{cases}
\]
\end{proposition}

\begin{proof}[\underline{Proof of Proposition \ref{prop:NI-general}}]
Under {\sf No Information}, buyers do not observe $\mathbf q$ and maximize
\[
\mathbb E[u_{xy}\mid \mathbf p]
=
(1-\rho)\mathbb E[q_y]+\rho\mathbb E[\varphi_{xy}]-p_y
=
\frac12-p_y.
\]
Thus the market is a homogeneous-goods Bertrand game with unit capacities and outside option $0$.

\paragraph{Part (a): $\gamma_n\ge 1$.}
Consider the profile $p_y^\star\equiv \frac12$. Then every available item yields expected utility $0$, and tie-breaking in favor of trade implies that each arriving buyer purchases an item whenever one is available. Since $m\ge n$, all $n$ items sell.

Fix seller $y$ and a deviation $p_y'\in[0,1]$.
If $p_y'>\frac12$, then $\frac12-p_y'<0$, so seller $y$ never sells and earns $0$.
If $p_y'<\frac12$, then seller $y$ still sells with probability $1$, but at a strictly lower price.
Hence $p_y^\star=\frac12$ is a best response for every seller.

For uniqueness, suppose $\mathbf p$ is an interim Nash equilibrium.
If some seller posts $p_y>\frac12$, then seller $y$ sells with probability $0$ and can deviate to $p_y'=\frac12$ and obtain strictly positive revenue, a contradiction. Thus $p_y\le \frac12$ for all $y$.
Let $\underline p:=\min_y p_y$ and choose $y$ with $p_y=\underline p$.
If $\underline p<\frac12$, then seller $y$ can raise price slightly to some
\[
p_y'\in\Big(\underline p,\min\Big\{\frac12,p_{(2)}\Big\}\Big),
\]
where $p_{(2)}$ denotes the second-lowest price. This preserves sale probability $1$ (seller $y$ remains among the cheapest available items and all cheap items sell because $m\ge n$), while strictly increasing price. Contradiction. Hence $\underline p=\frac12$, so all prices equal $\frac12$.

Finally, every traded buyer gets expected utility $\frac12-\frac12=0$, and any non-trading buyer (if $m>n$) gets utility $0$. Thus ${\sf AW}_{\emptyset}(m,n)=0$. Total seller revenue is $n\cdot \frac12$, so normalized seller revenue is ${\sf SR}_{\emptyset}(m,n)=\frac12$. Hence ${\sf SW}_{\emptyset}(m,n)=\frac12$.

\paragraph{Part (b): $\gamma_n<1$.}
We first show that any profile with at least $m+1$ zero-priced sellers is an interim Nash equilibrium.
Fix such a profile.
Because there are at least $m+1$ sellers at price $0$, every buyer strictly prefers a zero-priced seller to any seller with positive price. Hence only zero-priced sellers are ever purchased, exactly $m$ of them sell, and total seller revenue is $0$.

Now fix a seller and consider deviations.
If a zero-priced seller deviates to some $p_y'>0$, then there remain at least $m$ other sellers at price $0$, so those $m$ sellers absorb all $m$ buyers and the deviator sells with probability $0$. Her revenue remains $0$.
If a positive-price seller deviates to another positive price, she still sells with probability $0$. If she deviates to $0$, her revenue is still $0$.
Hence no seller can profitably deviate.

We now prove the converse.
Suppose $\mathbf p$ is an interim Nash equilibrium, and let $z(\mathbf p)$ denote the number of sellers pricing at $0$.

First, we claim that the minimum price must equal $0$.
Suppose instead that $p_{\min}>0$, and let $k$ be the number of sellers charging $p_{\min}$.

If $k\le m$, then there exists a seller with price strictly above $p_{\min}$ (since $n>m\ge k$). That seller can deviate to $p_{\min}-\varepsilon>0$ for sufficiently small $\varepsilon$, become the unique cheapest seller, sell with probability $1$, and earn positive revenue. Contradiction.

If $k\ge m+1$, then each seller charging $p_{\min}$ sells with probability at most $m/k<1$. Such a seller can deviate to $p_{\min}-\varepsilon>0$, become strictly cheapest, sell with probability $1$, and earn revenue $p_{\min}-\varepsilon$. For $\varepsilon>0$ small enough,
\[
p_{\min}-\varepsilon > p_{\min}\frac{m}{k},
\]
so the deviation is profitable. Contradiction.

Thus $p_{\min}=0$, i.e.\ $z(\mathbf p)\ge 1$.

Next, suppose $z(\mathbf p)\le m$.
Then the $(m+1)$-st lowest price, denoted $p_{(m+1)}$, is strictly positive.
Pick a zero-priced seller and let
\[
\varepsilon\in\Big(0,\min\Big\{\frac12,p_{(m+1)}\Big\}\Big).
\]
If this seller deviates from $0$ to $\varepsilon$, then at most $m-1$ sellers are priced strictly below $\varepsilon$, so the deviator is among the $m$ cheapest sellers. Since $\varepsilon<\frac12$, buyers weakly prefer trade, and because there are exactly $m$ buyers, every one of the $m$ cheapest sellers sells with probability $1$. The deviator's revenue rises from $0$ to $\varepsilon>0$, a contradiction.
Hence every equilibrium must satisfy $z(\mathbf p)\ge m+1$.

Finally, in every equilibrium there are at least $m+1$ zero-priced sellers and only zero-priced sellers sell. Every purchasing buyer therefore gets expected utility
\[
(1-\rho)\mathbb E[q_y]+\rho\mathbb E[\varphi_{xy}] = \frac12,
\]
so ${\sf AW}_{\emptyset}(m,n)=\frac12$. Seller revenue is $0$, hence ${\sf SR}_{\emptyset}(m,n)=0$ and ${\sf SW}_{\emptyset}(m,n)=\frac12$.
\end{proof}

\begin{proposition}[{\sf Only Quality Information}]
\label{prop:QI-general}
Fix $m,n\in\mathbb N$ and let $\gamma_n=m/n$.
\begin{enumerate}[label=(\alph*)]
    \item \textbf{Excess demand and balanced markets ($\gamma_n\ge 1$).}
    The unique interim Nash equilibrium is
    \[
    p_y^\star=(1-\rho)q_y+\frac{\rho}{2}
    \qquad \forall y\in\mathcal Y.
    \]
    Under this equilibrium,
    \[
    {\sf AW}_{q}(m,n)=0,\qquad
    {\sf SR}_{q}(m,n)=\frac12,\qquad
    {\sf SW}_{q}(m,n)=\frac12.
    \]

    \item \textbf{Excess supply markets ($\gamma_n<1$).}
    After relabeling so that $q_1>\cdots>q_n$, define
    \begin{equation}
    \label{eq:Q-excess-supply-prices}
    p_y^\star=
    \begin{cases}
    (1-\rho)(q_y-q_{m+1}), & y=1,\dots,m,\\
    0, & y=m+1,\dots,n.
    \end{cases}
    \end{equation}
    Then $\mathbf p^\star$ is an interim Nash equilibrium, unique up to the prices of sellers that never sell. The sold set is exactly $\{1,\dots,m\}$, and
    \begin{align*}
    {\sf AW}_{q}(m,n)
    &=\frac{\rho}{2}+(1-\rho)\frac{n-m}{n+1},\\
    {\sf SR}_{q}(m,n)
    &=(1-\rho)\frac{m(m+1)}{2n(n+1)},\\
    {\sf SW}_{q}(m,n)
    &=\frac{\rho}{2}+(1-\rho)\frac{n-m}{n+1}
      +(1-\rho)\frac{m(m+1)}{2n(n+1)}.
    \end{align*}
\end{enumerate}
Consequently, if $\gamma_n\to\gamma$, then
\[
({\sf AW}_{q},{\sf SR}_{q},{\sf SW}_{q})
=
\begin{cases}
(0,\frac12,\frac12), & \gamma\ge 1,\\[4pt]
\big((1-\rho)(1-\gamma)+\frac{\rho}{2},\;
(1-\rho)\frac{\gamma^2}{2},\;
(1-\rho)(1-\gamma+\frac{\gamma^2}{2})+\frac{\rho}{2}\big), & \gamma\in(0,1).
\end{cases}
\]
\end{proposition}

\begin{proof}[\underline{Proof of Proposition \ref{prop:QI-general}}]
Under {\sf Only Quality Information}, conditional on $(\mathbf p,\mathbf q)$ a buyer maximizes
\[
(1-\rho)q_y+\frac{\rho}{2}-p_y.
\]
Write
\[
s_y:=(1-\rho)q_y+\frac{\rho}{2}.
\]

\paragraph{Part (a): $\gamma_n\ge 1$.}
Consider the profile $p_y^\star=s_y$ for all $y$.
Then every available item yields expected utility $0$, so tie-breaking in favor of trade implies that each arriving buyer purchases an item whenever one remains. Since $m\ge n$, all $n$ items sell.

Fix seller $y$ and a deviation $p_y'\in[0,1]$.
If $p_y'>s_y$, then seller $y$ offers negative expected utility and therefore sells with probability $0$.
If $p_y'<s_y$, then seller $y$ still sells with probability $1$ but at a lower price.
Hence $p_y^\star=s_y$ is a best response for every seller.

For uniqueness, if some seller posts $p_y>s_y$, then she sells with probability $0$ but can deviate to $p_y'=s_y$ and obtain positive revenue. If some seller posts $p_y<s_y$, then she can raise price to $s_y$ without affecting sale probability $1$ and thereby increase revenue. Thus every equilibrium must satisfy $p_y=s_y$ for all $y$.

Under this equilibrium, every traded buyer gets expected utility $0$, so ${\sf AW}_q(m,n)=0$. Every seller sells once, hence normalized seller revenue is
\[
{\sf SR}_q(m,n)=\mathbb E[s_y]=(1-\rho)\mathbb E[q_y]+\frac{\rho}{2}
=\frac{1-\rho}{2}+\frac{\rho}{2}
=\frac12.
\]
Therefore ${\sf SW}_q(m,n)=\frac12$.

\paragraph{Part (b): $\gamma_n<1$.}
Fix a realization $\mathbf q$ with distinct qualities and relabel so that
\[
q_1>q_2>\cdots>q_n.
\]
Let $\mathbf p^\star$ be given by \eqref{eq:Q-excess-supply-prices}.

\smallskip
\noindent\emph{Step 1: utilities under $\mathbf p^\star$.}
Define
\[
U_y:=(1-\rho)q_y+\frac{\rho}{2}-p_y^\star.
\]
For $y\le m$,
\[
U_y
=
\Big((1-\rho)q_y+\frac{\rho}{2}\Big)-(1-\rho)(q_y-q_{m+1})
=
(1-\rho)q_{m+1}+\frac{\rho}{2}.
\]
For $y\ge m+1$, $p_y^\star=0$, so
\[
U_y=(1-\rho)q_y+\frac{\rho}{2}\le (1-\rho)q_{m+1}+\frac{\rho}{2},
\]
with equality only at $y=m+1$.
Hence the expected-utility maximizers are exactly $\{1,\dots,m+1\}$, all tied at the same maximal level.
By the quality-favoring tie-break, buyers choose the highest-quality maximizers first, so the sold set is exactly $\{1,\dots,m\}$.

\smallskip
\noindent\emph{Step 2: no profitable deviation.}
Fix seller $y$ and a deviation $p_y'\in[0,1]$.

\emph{Case 1: $y\le m$.}
Under $\mathbf p^\star$, seller $y$ sells with probability $1$ and earns $p_y^\star$.
If $p_y'<p_y^\star$, then seller $y$ still sells with probability $1$ but at a lower price.
If $p_y'>p_y^\star$, then
\[
(1-\rho)q_y+\frac{\rho}{2}-p_y'
<
(1-\rho)q_{m+1}+\frac{\rho}{2},
\]
so seller $y$ falls strictly below the $m$ items
\[
\{1,\dots,m+1\}\setminus\{y\},
\]
all of which still offer the maximal expected utility level. Since there are exactly $m$ buyers, those $m$ items are purchased and seller $y$ does not sell. Hence no profitable deviation exists.

\emph{Case 2: $y=m+1$.}
Under $\mathbf p^\star$, seller $m+1$ is tied at the maximum but loses the quality tie-break and does not sell, so her revenue is $0$.
Any deviation to $p_y'>0$ strictly lowers her expected utility below the maximizer level, so she still does not sell and still earns $0$.

\emph{Case 3: $y\ge m+2$.}
Even at price $0$, seller $y$ offers expected utility
\[
(1-\rho)q_y+\frac{\rho}{2}
<
(1-\rho)q_{m+1}+\frac{\rho}{2},
\]
so seller $y$ never sells.
Since prices are constrained to $[0,1]$, no deviation can raise her expected utility above its value at price $0$. Thus she cannot become chosen under any feasible deviation and her revenue remains $0$.

Hence $\mathbf p^\star$ is an interim Nash equilibrium. Moreover, the prices of sellers $1,\dots,m$ are pinned down by the common marginal-utility level, while sellers $m+1,\dots,n$ never sell. This gives uniqueness up to the prices of never-sold sellers.

\smallskip
\noindent\emph{Step 3: aggregate outcomes.}
Every buyer purchases one of the top $m$ items, and the expected utility of each traded buyer equals the common maximizer level
\[
(1-\rho)q_{m+1}+\frac{\rho}{2}.
\]
Therefore,
\[
{\sf AW}_q(m,n\mid \mathbf q)=(1-\rho)q_{m+1}+\frac{\rho}{2}.
\]
Taking expectations and using
\[
\mathbb E[q_{m+1}]=\frac{n-m}{n+1}
\]
for descending order statistics of $n$ iid $\mathrm{Unif}([0,1])$ draws gives
\[
{\sf AW}_q(m,n)=\frac{\rho}{2}+(1-\rho)\frac{n-m}{n+1}.
\]

Total seller revenue is
\[
\sum_{y=1}^m p_y^\star
=
(1-\rho)\sum_{y=1}^m (q_y-q_{m+1}).
\]
Hence
\[
{\sf SR}_q(m,n)
=
\frac{1-\rho}{n}\,
\mathbb E\!\left[\sum_{y=1}^m (q_y-q_{m+1})\right].
\]
Using
\[
\mathbb E[q_y]=\frac{n-y+1}{n+1}
\qquad (y=1,\dots,n)
\]
for descending order statistics, we obtain
\[
\mathbb E\!\left[\sum_{y=1}^m q_y\right]
=
\sum_{y=1}^m \frac{n-y+1}{n+1}
=
\frac{m(2n-m+1)}{2(n+1)},
\]
and therefore
\[
\mathbb E\!\left[\sum_{y=1}^m (q_y-q_{m+1})\right]
=
\frac{m(2n-m+1)}{2(n+1)}
-\frac{m(n-m)}{n+1}
=
\frac{m(m+1)}{2(n+1)}.
\]
Thus
\[
{\sf SR}_q(m,n)
=
(1-\rho)\frac{m(m+1)}{2n(n+1)}.
\]
Finally, ${\sf SW}_q={\sf AW}_q+{\sf SR}_q$.
\end{proof}

\begin{lemma}[Balanced sale probability under a common threshold]
\label{lem:Sn-balanced-rigorous}
Fix $\rho\in(0,1]$ and consider the balanced market $m=n$.
Let $\tau\in(0,1)$ and suppose every seller uses the price
\[
p_y=(1-\rho)q_y+\rho\tau.
\]
Let $S_t$ be the number of unsold items after the first $t$ buyers, so $S_0=n$. Then
\[
S_t=
\begin{cases}
S_{t-1}, & \text{with probability }\tau^{S_{t-1}},\\
S_{t-1}-1, & \text{with probability }1-\tau^{S_{t-1}}.
\end{cases}
\]
If $\tau=\tau_n=1-\varepsilon_n$ with $\varepsilon_n\downarrow 0$ and $\varepsilon_n n\to\infty$, then
\[
\mathbb E[S_n]=\frac{\ln 2}{\varepsilon_n}\,(1+o(1)),
\qquad
1-\frac{1}{n}\mathbb E[S_n]
=
1-\frac{\ln 2}{\varepsilon_n n}\,(1+o(1)).
\]
\end{lemma}

\begin{proof}[\underline{Proof of Lemma \ref{lem:Sn-balanced-rigorous}}]
For $u=1,\dots,n$, let $G_u$ be the waiting time to move from $u$ unsold items to $u-1$ unsold items. Conditional on being at level $u$, each arriving buyer makes a purchase with probability $1-\tau^u$, independently of the past, so
\[
G_u\sim \mathrm{Geom}(1-\tau^u),
\qquad
\mathbb E[G_u]=\frac{1}{1-\tau^u}.
\]
The random variables $(G_u)_{u=1}^n$ are independent because they are generated by disjoint blocks of buyer arrivals.

For $s\in\{0,\dots,n\}$, let
\[
T(s):=\sum_{u=s+1}^{n} G_u.
\]
Then $T(s)$ is the number of buyers required to reduce inventory from $n$ to $s$, and
\[
S_n\le s
\quad\Longleftrightarrow\quad
T(s)\le n.
\]
Write
\[
\mu(s):=\mathbb E[T(s)]=\sum_{u=s+1}^{n}\frac{1}{1-\tau^u}.
\]
Since $x\mapsto (1-\tau^x)^{-1}$ is decreasing, integral comparison gives
\[
\int_s^n \frac{dx}{1-\tau^x}
\le \mu(s)
\le
\int_{s+1}^{n+1}\frac{dx}{1-\tau^x}.
\]
Let $s^\circ\in(0,n)$ be the unique real solution of
\[
\int_{s^\circ}^{n}\frac{dx}{1-\tau^x}=n.
\]
Evaluating the antiderivative yields
\[
\tau^{s^\circ}=\frac{1}{2-\tau^n}.
\]
Because $\tau_n^n=(1-\varepsilon_n)^n\to 0$, we obtain
\[
\tau_n^{s^\circ}\to \frac12,
\qquad
s^\circ
=
\frac{\ln(1/2)}{\ln \tau_n}
=
\frac{\ln 2}{-\ln(1-\varepsilon_n)}
=
\frac{\ln 2}{\varepsilon_n}\,(1+o(1)).
\]

It remains to show that $S_n$ is concentrated in an $o(\varepsilon_n^{-1})$ window around $s^\circ$.
Fix $L_n=M/\sqrt{\varepsilon_n}$, where $M>0$ is constant.
For all $s\in[s^\circ-L_n,s^\circ+L_n]$ and all sufficiently large $n$, the identity $\tau_n^{s^\circ}\to 1/2$ implies
\[
\tau_n^s\in\Big[\frac13,\frac23\Big].
\]
Hence for every $u\ge s^\circ-L_n$,
\[
1-\tau_n^u\ge \frac13,
\qquad
\Var(G_u)=\frac{\tau_n^u}{(1-\tau_n^u)^2}\le 9\tau_n^u.
\]
Therefore, for every $s\in[s^\circ-L_n,s^\circ+L_n]$,
\[
\Var(T(s))
=
\sum_{u=s+1}^{n}\Var(G_u)
\le
9\sum_{u=s+1}^{n}\tau_n^u
\le
\frac{9\tau_n^s}{1-\tau_n}
\le
\frac{C}{\varepsilon_n}
\]
for a universal constant $C$. Moreover, on this same window,
\[
\mu(s)-\mu(s+1)=\frac{1}{1-\tau_n^{s+1}}\in[1,3].
\]
Thus
\[
\mu(s^\circ-L_n)-n \ge cL_n,
\qquad
n-\mu(s^\circ+L_n)\ge cL_n
\]
for some universal $c>0$ and all large $n$. Applying Bernstein's inequality gives,
\[
\Pr\big(T(s^\circ-L_n)\le n\big)\le 2e^{-c'M^2},
\qquad
\Pr\big(T(s^\circ+L_n)\ge n\big)\le 2e^{-c'M^2}
\]
for some $c'>0$ independent of $n$ and $M$.
Equivalently,
\[
\Pr\big(|S_n-s^\circ|>L_n\big)\le 4e^{-c'M^2}.
\]
Taking $M$ dyadically and summing the resulting tail bound yields
\[
\mathbb E|S_n-s^\circ|=O(\varepsilon_n^{-1/2}).
\]
Since $\varepsilon_n^{-1/2}=o(\varepsilon_n^{-1})$, we conclude that
\[
\mathbb E[S_n]
=
s^\circ+o(\varepsilon_n^{-1})
=
\frac{\ln 2}{\varepsilon_n}\,(1+o(1)).
\]
Dividing by $n$ gives the second claim.
\end{proof}

\begin{proposition}[{\sf Full Information}]
\label{prop:FI-general}\label{prop:FI-balanced}
Let $\gamma_n=m/n$.
\begin{enumerate}[label=(\alph*)]
    \item \textbf{Excess demand and balanced markets ($\gamma_n\ge 1$).}
    Fix $\rho\in(0,1]$ and define
    \[
    a_\rho:=\sqrt{\frac{\ln 2\,(1+\rho)}{2\rho}},
    \qquad
    \tau_n^\star:=1-\frac{a_\rho}{\sqrt n},
    \qquad
    p_y^\dagger:=(1-\rho)q_y+\rho\tau_n^\star .
    \]
    Then for all sufficiently large $n$, the profile $\mathbf p^\dagger$ is an interim $C_\rho n^{-1/2}$-approximate equilibrium for some constant $C_\rho>0$ depending only on $\rho$. Under this profile,
    \[
    {\sf AW}_{u}(m,n)=O_\rho(n^{-1/2}),\qquad
    {\sf SR}_{u}(m,n)=\frac{1+\rho}{2}-\Theta_\rho(n^{-1/2}),\qquad
    {\sf SW}_{u}(m,n)=\frac{1+\rho}{2}-\Theta_\rho(n^{-1/2}).
    \]

    \item \textbf{Excess supply markets with fixed proportional imbalance ($\gamma_n\to\gamma\in(0,1)$).}
    Fix $\gamma\in(0,1)$ and suppose $m=\lfloor \gamma n\rfloor$. Fix $\rho\in(0,1)$ and define the cutoff profile
    \[
    p_y^\dagger(\mathbf q):=
    \begin{cases}
    (1-\rho)(q_y-q_{m+1}), & y=1,\dots,m,\\
    0, & y=m+1,\dots,n.
    \end{cases}
    \]
    There exists an event $\mathcal E_n$ in the quality space with
    $\mathbb{P}(\mathcal E_n)\ge 1-c_{\gamma,\rho}n^{-2}$,
    such that, for every $\mathbf q\in\mathcal E_n$, the profile $\mathbf p^\dagger(\mathbf q)$ is an interim
    $C_{\gamma,\rho}\sqrt{\frac{\log n}{n}}$-approximate equilibrium for some constant $C_{\gamma,\rho}>0$.
    Moreover, under this profile,
    \begin{align*}
    {\sf AW}_{u}(m,n)
    &= (1-\rho)\frac{n-m}{n+1}+\rho
       + O_{\gamma,\rho}\!\left(\sqrt{\frac{\log n}{n}}\right),\\
    {\sf SR}_{u}(m,n)
    &= (1-\rho)\frac{m(m+1)}{2n(n+1)}
       + O_{\gamma,\rho}\!\left(\sqrt{\frac{\log n}{n}}\right),\\
    {\sf SW}_{u}(m,n)
    &= (1-\rho)\frac{n-m}{n+1}
       +(1-\rho)\frac{m(m+1)}{2n(n+1)}
       +\rho
       + O_{\gamma,\rho}\!\left(\sqrt{\frac{\log n}{n}}\right).
    \end{align*}
    Consequently, if $m/n\to\gamma\in(0,1)$, then
    \[
    ({\sf AW}_{u},{\sf SR}_{u},{\sf SW}_{u})
    =
    \big((1-\rho)(1-\gamma)+\rho,\;
    (1-\rho)\frac{\gamma^2}{2},\;
    (1-\rho)(1-\gamma+\frac{\gamma^2}{2})+\rho\big)
    +o(1).
    \]
\end{enumerate}
\end{proposition}

\begin{proof}[\underline{Proof of Proposition \ref{prop:FI-general}}] We have that
\paragraph{Part (a): $\gamma_n\ge 1$.}
Under the profile
\[
p_y^\dagger=(1-\rho)q_y+\rho\tau_n^\star,
\]
buyer utility becomes
\[
u_{xy}=(1-\rho)q_y+\rho\varphi_{xy}-p_y^\dagger
=
\rho(\varphi_{xy}-\tau_n^\star).
\]
Hence the common-value component $q_y$ drops out of buyer choice, and all sellers are ex ante symmetric under $\mathbf p^\dagger$. Let $\pi_{m,n}$ denote the common sale probability of a seller under this profile with $m$ buyers and $n$ sellers.
By Lemma~\ref{lem:Sn-balanced-rigorous}, in the balanced market $m=n$,
\[
\pi_{n,n}
=
1-\frac{\ln 2}{a_\rho\sqrt n}\,(1+o(1))
=
1-O_\rho(n^{-1/2}).
\]
If $m\ge n$, then adding extra buyers can only weakly increase every seller's probability of sale. By coupling the first $n$ buyers in the $m$-buyer market with the entire balanced market and then revealing the remaining $m-n$ buyers, we obtain
\[
\pi_{m,n}\ge \pi_{n,n}=1-O_\rho(n^{-1/2}).
\]

We now bound deviations.
Fix $\mathbf q$, a seller $y$, and any deviation $p_y'\in[0,1]$.
Since $p_y'\le (1-\rho)q_y+\rho$, the deviator's interim revenue is always bounded by
\[
v_{y,u}(p_y';\mathbf p_{-y}^\dagger\mid \mathbf q)\le (1-\rho)q_y+\rho.
\]
Under $\mathbf p^\dagger$, seller $y$'s interim revenue equals
\[
v_{y,u}(\mathbf p^\dagger\mid \mathbf q)
=
\big((1-\rho)q_y+\rho\tau_n^\star\big)\pi_{m,n}.
\]
Hence
\begin{align*}
\sup_{p_y'\in[0,1]}v_{y,u}(p_y';\mathbf p_{-y}^\dagger\mid \mathbf q)
-
v_{y,u}(\mathbf p^\dagger\mid \mathbf q)
&\le
(1-\rho)q_y(1-\pi_{m,n})
+\rho\big(1-\tau_n^\star\pi_{m,n}\big)\\
&\le
(1-\pi_{m,n})+\rho(1-\tau_n^\star)+\rho(1-\pi_{m,n})\\
&=
O_\rho(n^{-1/2}),
\end{align*}
uniformly in $q_y\in[0,1]$.
Thus $\mathbf p^\dagger$ is an interim $C_\rho n^{-1/2}$-approximate equilibrium.

We next compute the induced outcomes.
Every buyer who purchases gets realized utility
\[
\rho\Big(\max_{y\in\mathcal Y_{\mathrm{avail}}}\varphi_{xy}-\tau_n^\star\Big)\in[0,\rho(1-\tau_n^\star)],
\]
so
\[
0\le {\sf AW}_u(m,n)\le \rho(1-\tau_n^\star)=O_\rho(n^{-1/2}).
\]
Because all sellers have the same sale probability under $\mathbf p^\dagger$,
\[
{\sf SR}_u(m,n)
=
\left(\frac1n\sum_{y=1}^n p_y^\dagger\right)\pi_{m,n}
=
\left(\frac{1-\rho}{2}+\rho\tau_n^\star\right)\pi_{m,n}.
\]
Since
\[
\frac{1-\rho}{2}+\rho\tau_n^\star
=
\frac{1+\rho}{2}-\rho\frac{a_\rho}{\sqrt n},
\]
and $\pi_{m,n}=1-O_\rho(n^{-1/2})$, we get
\[
{\sf SR}_u(m,n)=\frac{1+\rho}{2}-\Theta_\rho(n^{-1/2}).
\]
Finally, ${\sf SW}_u={\sf AW}_u+{\sf SR}_u$, so
\[
{\sf SW}_u(m,n)=\frac{1+\rho}{2}-\Theta_\rho(n^{-1/2}).
\]

\paragraph{Part (b): $\gamma_n\to\gamma\in(0,1)$.}
Fix $\gamma\in(0,1)$ and suppose $m=\lfloor\gamma n\rfloor$.
Fix a realization $\mathbf q$ with
\[
q_1>\cdots>q_n.
\]
Define the baseline utility
\[
b_y:=(1-\rho)q_y-p_y^\dagger(\mathbf q).
\]
Then
\[
b_y=(1-\rho)q_{m+1}=:B^\star
\qquad \text{for } y=1,\dots,m+1,
\]
while
\[
b_y=(1-\rho)q_y<B^\star
\qquad \text{for } y\ge m+2.
\]

\smallskip
\noindent\emph{Step 1: a high-probability quality event.}
By the Dvoretzky--Kiefer--Wolfowitz inequality for the empirical cdf of iid $\mathrm{Unif}([0,1])$ draws, there exists a universal constant $c_0>0$ such that the event
\[
\mathcal E_q
:=
\left\{
\max_{1\le y\le n}
\left|q_y-\left(1-\frac{y}{n}\right)\right|
\le
2\sqrt{\frac{\log n}{n}}
\right\}
\]
satisfies
\[
\Pr(\mathcal E_q)\ge 1-c_0n^{-2}.
\]
On $\mathcal E_q$, for every $a<b$,
\begin{equation}
\label{eq:quality-gap-bound}
q_a-q_b
\le
\frac{b-a}{n}+4\sqrt{\frac{\log n}{n}},
\qquad
q_a-q_b
\ge
\frac{b-a}{n}-4\sqrt{\frac{\log n}{n}}.
\end{equation}
We henceforth fix $\mathbf q\in\mathcal E_q$.

\smallskip
\noindent\emph{Step 2: a safe competitive set and an active set.}
Let
\[
K:=\left\lceil \sqrt{n\log n}\right\rceil,
\qquad
\mathcal C:=\{1,\dots,m+K\}.
\]
Since $\gamma<1$ is fixed and $K=o(n)$, for all sufficiently large $n$ we have $m+K\le n$.

Under $\mathbf p^\dagger(\mathbf q)$, every available item gives nonnegative utility, because $b_y\ge 0$ for all $y$.
Hence every buyer purchases, and after the first $t-1$ buyers have arrived, at most $t-1<m$ items have been removed. Therefore the set of still-available sellers in $\mathcal C$, denoted $\mathcal C_t$, satisfies
\[
|\mathcal C_t|\ge |\mathcal C|-(t-1)\ge K+1
\qquad (t=1,\dots,m).
\]

Let $\mathcal F_{t-1}$ be the sigma-field generated by the idiosyncratic shocks of buyers $1,\dots,t-1$.
Conditional on $\mathcal F_{t-1}$, the set $\mathcal C_t$ is fixed and the random variables $(\varphi_{tz})_{z\in\mathcal C_t}$ are iid $\mathrm{Unif}([0,1])$.
Thus
\[
\Pr\!\left(
\max_{z\in\mathcal C_t}\varphi_{tz}
<
1-\frac{4\log n}{K}
\,\middle|\,\mathcal F_{t-1},\mathbf q
\right)
\le
\left(1-\frac{4\log n}{K}\right)^K
\le n^{-4}.
\]
By the tower property and a union bound, the event
\[
\mathcal E_\varphi^\star
:=
\bigg\{
\forall t\in\{1,\dots,m\},
\ \max_{z\in\mathcal C_t}\varphi_{tz}\ge 1-\frac{4\log n}{K}
\bigg\}
\]
satisfies
\[
\Pr(\mathcal E_\varphi^\star\mid \mathbf q)\ge 1-mn^{-4}\ge 1-n^{-3}.
\]

On $\mathcal E_\varphi^\star$, every buyer has an available seller yielding utility at least
\[
V_{\mathrm{safe}}^\star
:=
(1-\rho)q_{m+K}
+
\rho\left(1-\frac{4\log n}{K}\right).
\]
Indeed, every seller in $\mathcal C$ has baseline utility at least $(1-\rho)q_{m+K}$.

We next show that sellers far below the cutoff can never be chosen on $\mathcal E_\varphi^\star$.
Fix $z>m+K$.
Even under the best possible idiosyncratic realization $\varphi_{tz}=1$, seller $z$ provides utility at most
\[
(1-\rho)q_z+\rho.
\]
Thus a necessary condition for seller $z$ to be chosen on $\mathcal E_\varphi^\star$ is
\[
(1-\rho)q_z+\rho\ge V_{\mathrm{safe}}^\star,
\]
equivalently,
\[
(1-\rho)(q_{m+K}-q_z)\le \rho\frac{4\log n}{K}.
\]
Using \eqref{eq:quality-gap-bound}, this implies
\[
(1-\rho)\left(
\frac{z-(m+K)}{n}
-
4\sqrt{\frac{\log n}{n}}
\right)
\le
4\rho\sqrt{\frac{\log n}{n}}.
\]
Hence any seller that can possibly sell on $\mathcal E_\varphi^\star$ must satisfy
\[
z-(m+K)\le
\left(4+\frac{4\rho}{1-\rho}\right)\sqrt{n\log n}.
\]
Define
\[
W:=\left\lceil\left(4+\frac{4\rho}{1-\rho}\right)\sqrt{n\log n}\right\rceil,
\qquad
K':=K+W,
\qquad
\mathcal S:=\{1,\dots,m+K'\}.
\]
Then, on $\mathcal E_\varphi^\star$, no seller outside $\mathcal S$ is ever chosen.

\smallskip
\noindent\emph{Step 3: sale probabilities of the top sellers.}
Under $\mathbf p^\dagger(\mathbf q)$, exactly $m$ buyers arrive and every buyer purchases, so there are exactly $m$ sales.

Because $\Pr(\mathcal E_\varphi^{\star c}\mid \mathbf q)\le n^{-3}$, the expected number of sales outside $\mathcal S$ is at most
\[
m\Pr(\mathcal E_\varphi^{\star c}\mid \mathbf q)\le n^{-2}.
\]
Thus the expected number of sales inside $\mathcal S$ is at least $m-n^{-2}$, and therefore the expected number of unsold sellers inside $\mathcal S$ is at most
\[
|\mathcal S|-(m-n^{-2})\le K'+n^{-2}.
\]

Now note that sellers $1,\dots,m+1$ all have the same baseline utility $B^\star$ and iid idiosyncratic shocks, so they are perfectly symmetric from the buyers' perspective. Let $\pi_{\mathrm{top}}$ denote their common interim sale probability under $\mathbf p^\dagger(\mathbf q)$.
Since the top $m+1$ sellers form a subset of $\mathcal S$, the expected number of unsold sellers among them is at most the expected number of unsold sellers in $\mathcal S$:
\[
(m+1)(1-\pi_{\mathrm{top}})
\le
K'+n^{-2}.
\]
Because $m=\lfloor \gamma n\rfloor$ with $\gamma\in(0,1)$ fixed, this yields
\begin{equation}
\label{eq:pi-top-bound}
1-\pi_{\mathrm{top}}
\le
\frac{K'+n^{-2}}{m+1}
=
O_{\gamma,\rho}\!\left(\sqrt{\frac{\log n}{n}}\right).
\end{equation}

\smallskip
\noindent\emph{Step 4: deviation bounds.}
Fix a seller $y$ and a deviation to some $p_y'\in[0,1]$.
We bound the increase in interim expected revenue conditional on $\mathbf q$.

For the deviation analysis we need a safe alternative that excludes $y$ itself.
Let $\mathcal C_t'$ be the set of sellers in $\mathcal C$ still available when buyer $t$ arrives in the post-deviation process.
Since at most $m-1$ sales have occurred before buyer $t$ arrives,
\[
|\mathcal C_t'\setminus\{y\}|
\ge
|\mathcal C|-(m-1)-1
=
K.
\]
Define
\[
\mathcal E_\varphi^{-y}
:=
\bigg\{
\forall t\in\{1,\dots,m\},
\ \max_{z\in \mathcal C_t'\setminus\{y\}}\varphi_{tz}
\ge
1-\frac{4\log n}{K}
\bigg\}.
\]
The same conditional argument as above gives
\[
\Pr(\mathcal E_\varphi^{-y}\mid \mathbf q)\ge 1-n^{-3}.
\]
On $\mathcal E_\varphi^{-y}$, every buyer has an available alternative \emph{other than $y$} that yields utility at least
\[
V_{\mathrm{safe}}^{-y}
:=
(1-\rho)q_{m+K}
+
\rho\left(1-\frac{4\log n}{K}\right).
\]

\emph{Case 1: $y\le m$ and $p_y'<p_y^\dagger$.}
Let $\pi_y'$ denote seller $y$'s sale probability after the deviation.
Then
\[
\Delta v_y
=
p_y'\pi_y'-p_y^\dagger\pi_{\mathrm{top}}
\le
p_y^\dagger-p_y^\dagger\pi_{\mathrm{top}}
=
p_y^\dagger(1-\pi_{\mathrm{top}}).
\]
Since $p_y^\dagger\le 1-\rho\le 1$, \eqref{eq:pi-top-bound} implies
\[
\Delta v_y
\le
1-\pi_{\mathrm{top}}
=
O_{\gamma,\rho}\!\left(\sqrt{\frac{\log n}{n}}\right).
\]

\emph{Case 2: $y\le m$ and $p_y'=p_y^\dagger+\delta$ with $\delta>0$.}
Seller $y$'s baseline utility drops from $B^\star$ to $B^\star-\delta$.
If seller $y$ sells on $\mathcal E_\varphi^{-y}$, her utility must beat the safe alternative $V_{\mathrm{safe}}^{-y}$, so necessarily
\[
B^\star-\delta+\rho\ge V_{\mathrm{safe}}^{-y}.
\]
Hence
\[
\delta
\le
(1-\rho)(q_{m+1}-q_{m+K})+\rho\frac{4\log n}{K}
=:\delta_{\max}.
\]
Using \eqref{eq:quality-gap-bound},
\[
\delta_{\max}
\le
(1-\rho)\left(\frac{K}{n}+4\sqrt{\frac{\log n}{n}}\right)
+\rho\frac{4\log n}{K}
=
O_\rho\!\left(\sqrt{\frac{\log n}{n}}\right).
\]

If $\delta>\delta_{\max}$, then seller $y$ can sell only on the failure event $\mathcal E_\varphi^{-y\,c}$, so
\[
\Delta v_y
\le
\Pr(\mathcal E_\varphi^{-y\,c}\mid \mathbf q)
\le
n^{-3}.
\]
If $\delta\le \delta_{\max}$, then
\begin{align*}
\Delta v_y
=
(p_y^\dagger+\delta)\pi_y'-p_y^\dagger\pi_{\mathrm{top}}
=
p_y^\dagger(\pi_y'-\pi_{\mathrm{top}})+\delta\pi_y'
\le
p_y^\dagger(1-\pi_{\mathrm{top}})+\delta =
O_{\gamma,\rho}\!\left(\sqrt{\frac{\log n}{n}}\right),
\end{align*}
where we used $\pi_y'\le 1$, \eqref{eq:pi-top-bound}, and $\delta\le\delta_{\max}$.

\emph{Case 3: $y\ge m+1$.}
Under $\mathbf p^\dagger$, seller $y$ earns revenue $0$.
Any deviation has the form $p_y'=\delta\in[0,1]$.
If seller $y$ sells on $\mathcal E_\varphi^{-y}$, then
\[
(1-\rho)q_y-\delta+\rho\ge V_{\mathrm{safe}}^{-y}.
\]
Since $q_y\le q_{m+1}$, this implies
\[
\delta\le (1-\rho)(q_{m+1}-q_{m+K})+\rho\frac{4\log n}{K}=\delta_{\max}.
\]
Therefore, if $\delta>\delta_{\max}$ then seller $y$ can sell only on $\mathcal E_\varphi^{-y\,c}$ and $\Delta v_y\le n^{-3}.$

If $\delta\le\delta_{\max}$, then
\[
\Delta v_y\le \delta\le \delta_{\max}
=
O_\rho\!\left(\sqrt{\frac{\log n}{n}}\right).
\]

Combining the three cases shows that, for every $\mathbf q\in\mathcal E_q$ and every seller $y$,
\[
\sup_{p_y'\in[0,1]}
v_{y,u}(p_y';\mathbf p_{-y}^\dagger(\mathbf q)\mid \mathbf q)
\le
v_{y,u}(\mathbf p^\dagger(\mathbf q)\mid \mathbf q)
+
C_{\gamma,\rho}\sqrt{\frac{\log n}{n}}.
\]
Thus $\mathbf p^\dagger(\mathbf q)$ is an interim $C_{\gamma,\rho}\sqrt{(\log n)/n}$-approximate equilibrium for every $\mathbf q\in\mathcal E_q$. Since $\Pr(\mathcal E_q)\ge 1-c_0n^{-2}$, this proves the high-probability equilibrium claim.

\smallskip
\noindent\emph{Step 5: outcome formulas under the cutoff profile.}
We first bound agent welfare.
For every realization of $(\mathbf q,\boldsymbol\varphi)$, no seller offers baseline utility above $B^\star=(1-\rho)q_{m+1}$, and $\varphi_{xy}\le 1$, so every buyer's realized utility is at most
\[
B^\star+\rho=(1-\rho)q_{m+1}+\rho.
\]
On the other hand, on $\mathcal E_\varphi^\star$ every buyer gets utility at least
\[
V_{\mathrm{safe}}^\star
=
(1-\rho)q_{m+K}
+
\rho\left(1-\frac{4\log n}{K}\right).
\]
Since utilities are always nonnegative under $\mathbf p^\dagger$, it follows that conditional on $\mathbf q\in\mathcal E_q$,
\[
(1-\rho)q_{m+K}
+
\rho\left(1-\frac{4\log n}{K}\right)
-
n^{-3}
\le
{\sf AW}_u(m,n\mid \mathbf q)
\le
(1-\rho)q_{m+1}+\rho.
\]
Using \eqref{eq:quality-gap-bound},
\[
q_{m+1}-q_{m+K}
\le
\frac{K-1}{n}+4\sqrt{\frac{\log n}{n}}
=
O\!\left(\sqrt{\frac{\log n}{n}}\right),
\]
so
\[
{\sf AW}_u(m,n\mid \mathbf q)
=
(1-\rho)q_{m+1}+\rho
+
O_\rho\!\left(\sqrt{\frac{\log n}{n}}\right)
\qquad
(\mathbf q\in\mathcal E_q).
\]
Taking expectations over $\mathbf q$ and using $\Pr(\mathcal E_q^c)=O(n^{-2})$ and $\mathbb E[q_{m+1}]=(n-m)/(n+1)$ yields
\[
{\sf AW}_u(m,n)
=
(1-\rho)\frac{n-m}{n+1}+\rho
+
O_{\gamma,\rho}\!\left(\sqrt{\frac{\log n}{n}}\right).
\]

We next bound seller revenue.
On the event $\mathcal E_\varphi^\star$, all sales occur inside $\mathcal S$.
Since $|\mathcal S|=m+K'$, at most $K'$ sellers in $\mathcal S$ remain unsold.
Only sellers $1,\dots,m$ have positive prices, and each such price is at most $1$.
Therefore, on $\mathcal E_\varphi^\star$,
\[
\sum_{y=1}^m p_y^\dagger - K'
\le
\text{total seller revenue}
\le
\sum_{y=1}^m p_y^\dagger.
\]
On $\mathcal E_\varphi^{\star c}$, total seller revenue is at most $m$.
Dividing by $n$ and taking conditional expectations given $\mathbf q\in\mathcal E_q$,
\[
\frac1n\sum_{y=1}^m p_y^\dagger
-
\frac{K'}{n}
-
n^{-2}
\le
{\sf SR}_u(m,n\mid \mathbf q)
\le
\frac1n\sum_{y=1}^m p_y^\dagger.
\]
Hence
\[
{\sf SR}_u(m,n\mid \mathbf q)
=
\frac1n\sum_{y=1}^m p_y^\dagger
+
O_{\rho}\!\left(\sqrt{\frac{\log n}{n}}\right)
\qquad
(\mathbf q\in\mathcal E_q).
\]
Taking expectations over $\mathbf q$ and using the exact order-statistic calculation from Proposition~\ref{prop:QI-general}(b),
\[
\mathbb E\!\left[\frac1n\sum_{y=1}^m p_y^\dagger\right]
=
(1-\rho)\frac{m(m+1)}{2n(n+1)}.
\]
Therefore
\[
{\sf SR}_u(m,n)
=
(1-\rho)\frac{m(m+1)}{2n(n+1)}
+
O_{\gamma,\rho}\!\left(\sqrt{\frac{\log n}{n}}\right).
\]
Finally,
\[
{\sf SW}_u(m,n)={\sf AW}_u(m,n)+{\sf SR}_u(m,n),
\]
which gives the stated formula.
\end{proof}

\subsection{Proof of Theorem \ref{thm:value-public-rankings-pricing}}

\begin{proof}[\underline{Proof of Theorem \ref{thm:value-public-rankings-pricing}}] The result follows from Propositions \ref{prop:NI-general} and \ref{prop:QI-general}.
\end{proof}

\subsection{Proof of Theorem \ref{thm:value-personalization-pricing}}

\begin{proof}[\underline{Proof of Theorem \ref{thm:value-personalization-pricing}}] The result follows from Propositions \ref{prop:QI-general} and \ref{prop:FI-general}.
\end{proof}

\end{document}